\documentclass[aps,prl,onecolumn,groupedaddress]{revtex4}
\usepackage[pdftex]{graphicx}
\begin{document}
\title{Single-electron Transport in Graphene-like Nanostructures}

\author{K. L. Chiu$^1$, Yang Xu$^2$}

\affiliation{$^1$Department of Physics, Massachusetts Institute of Technology, Cambridge, MA 02139, USA}
\affiliation{$^2$Institute of Microelectronics and Optoelectronics, College of Information Science and Electronic Engineering, Zhejiang University, 310027 P. R. China}

\date{\today}

\begin{abstract}
Two-dimensional (2D) materials for their versatile band structures and strictly 2D nature have attracted considerable attention over the past decade. Graphene is a robust material for spintronics owing to its weak spin-orbit and hyperfine interactions, while monolayer transition metal dichalcogenides (TMDs) possess a Zeeman effect-like band splitting in which the spin and valley degrees of freedom are nondegenerate. The surface states of topological insulators (TIs) exhibit a spin-momentum locking that opens up the possibility of controlling the spin degree of freedom in the absence of an external magnetic field. Nanostructures made of these materials are also viable for use in quantum computing applications involving the superposition and entanglement of individual charge and spin quanta. In this article, we review a selection of transport studies addressing the confinement and manipulation of charges in nanostructures fabricated from various 2D materials. We supply the entry-level knowledge for this field by first introducing the fundamental properties of 2D bulk materials followed by the theoretical background relevant to the physics of nanostructures. Subsequently, a historical review of experimental development in this field is presented, from the early demonstration of graphene nanodevices on SiO$_{2}$ substrate to more recent progress in utilizing hexagonal boron nitride to reduce substrate disorder. In the second part of this article, we extend our discussion to TMDs and TI nanostructures. We aim to outline the current challenges and suggest how future work will be geared towards developing spin qubits in 2D materials.

\end{abstract}

\maketitle
\section{1. INTRODUCTION}

Since the 1960s, the density of components on silicon chips has doubled approximately every 18 months, following a trend known as Moore's law after Intel’s cofounder Gordon Moore, who predicted the phenomenon. Silicon-based transistor manufacturing has now reached the sub-10nm scale, heralding the limit of Moore's law and stimulating the development of alternative switching technologies and host materials for processing and storing bits of information. Quantum bits, or 'qubits', are at the heart of quantum computing, an entirely different paradigm in which information is encoded using the superposition states of individual quanta. Ideally, it is desirable to use single electron transistors (SETs) or quantum dots (QDs) to manipulate single-electron spin as building blocks for more complicated computing devices. To reach this goal, tremendous efforts have been dedicated to studying the transport properties of QDs made from semiconductors, such as GaAs and silicon, and, more recently, graphene and other two-dimensional (2D) materials \cite{Schnez2009,Gutinger2008,Guttinger2009,Cho2012,Hong2014,Li2013}. Although the high mobility of GaAs has enabled the rapid development of spin qubits \cite{Hanson2007}, the heavy atomic weights of gallium and arsenide atoms limit the spin relaxation time, making this material less ideal for upscaling. Silicon QDs do not suffer from a heavy atomic weight and have shown a sufficiently long spin relaxation time \cite{Zwanenburg2013}, but their mobility is limited by their doping mechanism. While research on these materials is ongoing, 2D materials such as graphene, transition metal dichalcogenides (TMDs) and topological insulators (TIs) have attracted considerable attention over the past few years because of their novel electronic properties \cite{Geim2013}. Graphene is a robust material for spintronics owing to its weak spin-orbit and hyperfine interactions. Over the last decade, attempts to confine and manipulate single charges in graphene quantum dots (GQDs) have been widely studied and reported, as noted in several review articles \cite{Stampfer2011,Molitor2011,Guttinger2012,Neumann2013}. However, early studies of GQDs on SiO$_{2}$ have indicated an absence of spin-related phenomena, such as spin blockade and the Kondo effect. In order to reduce the substrate disorder, which is one of the major sources of fast spin relaxation, recent efforts have been focused on GQDs on atomically flat substrates (e.g., hexagonal boron nitride [hBN]). Other 2D materials, such as TMDs, exhibit direct band gap in monolayer form and are promising for switch applications due to the high current on/off rates in their transistors. More importantly, the absence of inversion symmetry and the existence of strong spin-orbit coupling in TMDs allow the charge carriers to be simultaneously valley- and spin-polarized, providing more degrees of freedom that can be controlled as qubits. In addition, 3D topological insulators exhibit surface states whose excitations share similarities with Dirac fermions in graphene but with real - rather than pseudo - spin locked to the quasimomentum. These insulators hold promise for dissipationless spintronics and for the operation of quantum devices in the absence of an external magnetic field \cite{Kong2011,Cha2013,Hasan2010}. In this review, we aim to present the theoretical background on this field and provide an overview of experimental studies that are relevant to the development of spin qubits in 2D materials. We supply the entry-level knowledge for this field by first introducing the fundamental properties of 2D bulk materials followed by the theoretical background relevant to the physics of nanostructures. In the first part of the experimental review, we discuss the transport properties of graphene nanodevices fabricated on both SiO$_{2}$ and hBN at low temperatures and under high magnetic fields. Our primary focus is the single-electron tunneling regime in transport. In the second part, we extend our discussion to TMD and TI nanostructures. We review recent developments in the fabrication and understanding of the electronic properties of these 2D nanostructures, including MoS$_{2}$ nanoribbons, WSe$_2$ quantum dots, and Bi$_2$Se$_3$ nanowires and quantum dots. Finally, we outline the current challenges in manipulating the spin degree of freedom in graphene-based SETs and suggest how future work should pursue the development of spin qubits in 2D materials.

\subsection{1.1. Graphene and hBN}

\begin{figure}
\begin{center}
		\includegraphics[scale=0.72]{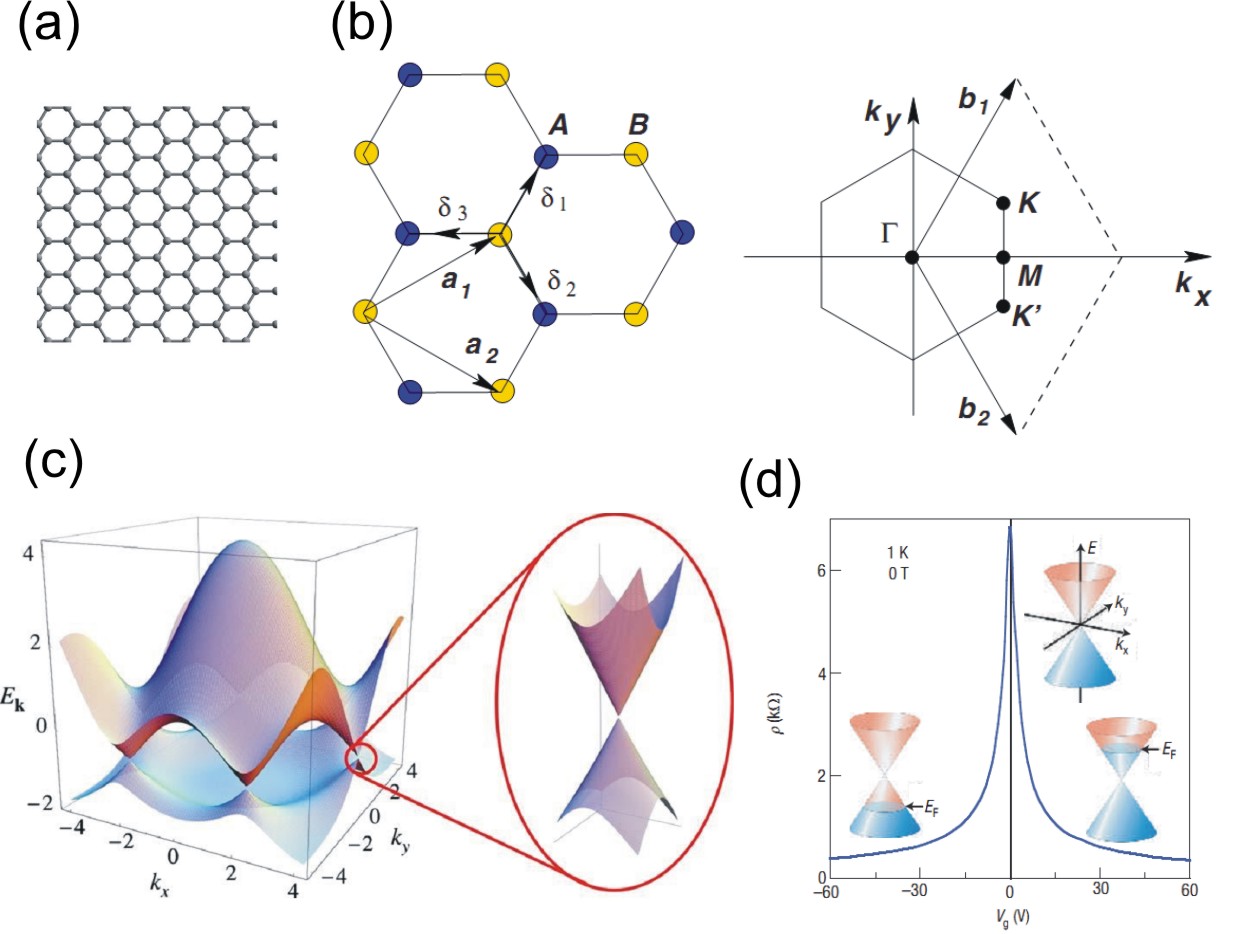}
\caption{(a) Graphene is a honeycomb lattice of carbon atoms. (b) Honeycomb lattice and its Brillouin zone. Left: lattice structure of graphene, made out of two interpenetrating triangular lattices ($\textbf{a}_{1}$ and $\textbf{a}_{2}$ are the lattice unit vectors, and $\vec{\delta_{i}}$, i =1, 2, 3 are the nearest-neighbor vectors). Right: corresponding Brillouin zone. The Dirac cones are located at the K and K' points. (c) The band structure of graphene calculated using a tight-binding model. The zoom in shows the conical dispersion relation around the Dirac point. (d) The ambi-polar electric field effect of graphene. V$_{g}$ is the back gate voltage and $\rho$ is resistivity. By varying V$_{g}$ one can shift the Fermi energy level and therefore determine the type of carriers (either electrons or holes) in graphene. (b, c) adapted with permission from ref. \cite{CastroNeto2009}. Copyright 2009 American Physical Society. (d) adapted with permission from ref. \cite{Geim2007}. Copyright 2007 Nature Publishing Group.}
\label{Figure 1.1}
\end{center}	
\end{figure}

Graphene is a single layer of carbon atoms packed tightly in a honeycomb lattice as shown in Fig. \ref{Figure 1.1}(a). An early study on few-layer graphene can be tracked back to 1948 by G. Ruess and F. Vogt, in which they occasionally observed extremely thin graphitic flakes in transmission electron microscope (TEM) images. However, no one isolated single layer graphene until 2004 when the physicists at the University of Manchester first isolated and spotted graphene on a chosen SiO$_{2}$ substrate \cite{Novoselov2004}. The first line of enquiry stems from graphene's unique gapless bandstructure. The unit cell of graphene consists of two carbon atoms, labeled as A and B sub-lattices, and can be described by the two lattice vectors $\textbf{a}_{1}$ and $\textbf{a}_{2}$, as shown in Fig. \ref{Figure 1.1}(b) (left panel). They include an angle of 60$^{\circ}$ and have a length of $\left|\textbf{a}_{1}\right|$ = $\left|\textbf{a}_{2}\right|$ = $\sqrt{3}$a$_{0}$ $-$ 2.461 $\dot{A}$, where a$_{0}$ is the carbon-carbon bond length (a$_{0}$ = 1.42 $\dot{A}$). The lattice vectors can be determined as $\textbf{a}_{1}$ = $\frac{a_{0}}{2}$ ($3$, $\sqrt{3}$) and $\textbf{a}_{2}$ = $\frac{a_{0}}{2}$ ($3$, -$\sqrt{3}$) and the reciprocal lattice is described by $\textbf{b}_{1}$ = $\frac{2\pi}{3a_{0}}$ ($1$, $\sqrt{3}$) and $\textbf{b}_{2}$ = $\frac{2\pi}{3a_{0}}$ ($1$, -$\sqrt{3}$), as shown in the right panel of Fig. \ref{Figure 1.1}(b). The lattice has high-symmetry points $\Gamma$, $\textit{K}$ and $\textit{M}$, where $\textit{K}$ = $(\frac{2\pi}{3a_{0}}, \frac{2\pi}{3\sqrt{3}a_{0}})$ and $K'$ = $(\frac{2\pi}{3a_{0}}, -\frac{2\pi}{3\sqrt{3}a_{0}})$ are two points at the corners of the hexagonal Brillouin zone \cite{CastroNeto2009}. Around the $\textit{K}$ point, a tight-binding calculation for the bandstructure of this lattice yields a 2D Dirac-like Hamiltonian $\hat{H}_K$ for massless fermions (and around the $K'$ point the Hamiltonian is simply $\hat{H}_{K'}$ $=$ $\hat{H}^{T}_K$): 

\begin{eqnarray}
&& \hat{H}_K \psi (\textbf{r}) = \hbar v_{F}\left( \begin{array}{cc} \ 0 & k_x - ik_y \\ \ k_x + ik_y & 0  \\\end{array} \right)\ \psi (\textbf{r}) = \nonumber -i \hbar v_{F} \left( \begin{array}{cc} \ 0 & \frac{\partial}{\partial{x}} - i\frac{\partial}{\partial{y}} \\ \ \frac{\partial}{\partial{x}} + i\frac{\partial}{\partial{y}} & 0  \\\end{array} \right)\ \psi (\textbf{r}) \\
&&=  -i \hbar v_{F} \vec{\sigma} \nabla \psi (\textbf{r}) = v_{F} \vec{\sigma} \cdot \vec{p} \psi (\textbf{r}) = \textit{E} \psi (\textbf{r})
\label{eqn 1}
\end{eqnarray} where $v_{F}$ is the Fermi velocity, $\vec{\sigma}$ $=$ ($\sigma_{x}$, $\sigma_{y}$) is the 2-D Pauli matrix and $\psi$($\textbf{r}$) is the two-component electron wavefunction. This Hamiltonian gives rise to the most important aspect of graphene's energy dispersion, $E=\hbar$$v_{F}k$, which is a linear energy-momentum relationship at the edge of the Brillouin zone as shown in Fig. \ref{Figure 1.1}(c). The two-component vector part of the wavefunction, which corresponds to the A or B sub-lattices, is the so-called pseudospin degree of freedom, since it resembles the two-component real spin vector. The Pauli matrices $\sigma_x$ and $\sigma_y$ combined with the direction of the momentum leads to the definition of a chirality in graphene ($h = \vec{\sigma} \cdot \vec{p}/2\left|\vec{p}\right|$), meaning the wavefunction component of A or B sub-lattice is polarized with regard to the direction of motion of electrons \cite{CastroNeto2009}. The existence of the $K$ and $K'$ points (as a result of graphene's hexagonal structure), where the Dirac cones for electrons and holes touch each other in momentum space [Fig. \ref{Figure 1.1}(b) and Fig. \ref{Figure 1.1}(c)], is sometimes referred as isospin, and gives rise to a valley degeneracy g$_{\nu}$ $=$ 2 for graphene. The linear dispersion along with the presence of potential disorder leads to a maximum resistivity in the limit of vanishing carrier density (or the so-called Dirac point), as shown in Fig. \ref{Figure 1.1}(d). To change the Fermi level, and hence the charge carrier density, voltage needs to be applied to a nearby gate capacitively coupled to the graphene, which in the case of Fig. \ref{Figure 1.1}(d) is a backgate - a doped Si substrate that is isolated from the graphene by a SiO$_{2}$ insulator layer.

\begin{figure}
\begin{center}
		\includegraphics[scale=0.72]{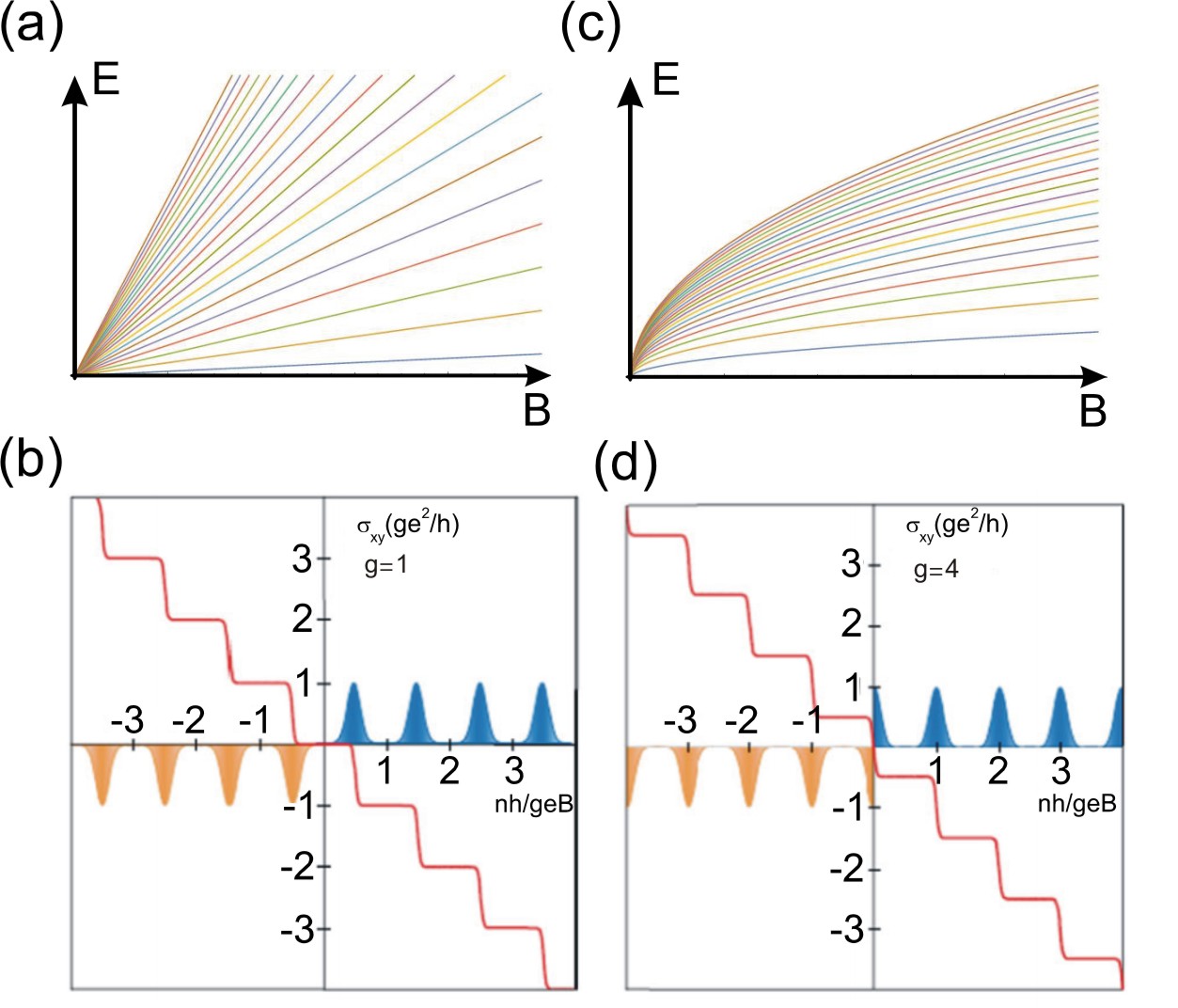}
\caption{(a) Schematic illustration of Landau levels for a standard 2DEG system. (b) Illustration of the integer Quantum Hall Effect (QHE) for a 2DEG system. (c) Same as (a) but for graphene. (d) Same as (b) but for graphene. The QHE plateau $\sigma_{xy}$ lie at half integers of 4e$^{2}$/h. (b, d) adapted with permission from ref. \cite{Novoselov2006}. Copyright 2006 Nature Publishing Group.}
\label{Figure 1.2}
\end{center}	
\end{figure}

There are rich physics originated from the Dirac nature of the fermions in graphene, such as its electronic, optical and mechanical properties \cite{CastroNeto2009,Basov2014,Amorim2016}. Here, we introduce an important phenomenon in graphene transport, which is relevant to the subjects to be discussed in this review: the extreme quantum Hall effect (QHE) that can be observed even at room temperature \cite{Geim2007}. Because the low-energy fermions in graphene are massless, it is obvious that for graphene we cannot apply the results valid for standard semiconductor two-dimensional electron gas (2DEG) systems. Charge carriers in a standard 2DEG have an effective mass, which is related with the parabolic dispersion relation of conduction (valence) band $via$ $E = E_{c} + \frac{\hbar^2k^2}{2m_e^\ast}$ ($E = E_{v} - \frac{\hbar^2k^2}{2m_h^\ast}$), where $E_{c(v)}$ is the conduction (valence) band minimum (maximum) and $m_{e(h)}^\ast$ is the effective mass for the electrons (holes). The band dispersion leads to a constant density of state (DOS) of $\frac{m_{e(h)}^\ast}{\pi\hbar^2}$ for the conduction (valence) band region. In a perpendicular magnetic field, the DOS of electrons in a 2DEG system is quantized at discrete energies given by:

\begin{eqnarray}
E_n &=& \pm\hbar\omega_c(n+1/2), \label{eqn 2}
\end{eqnarray} which is the so-called Landau Level (LL) energy, with $n$ the integer number and $\omega_c = eB/m_e^\ast$ the cyclotron frequency, as sketched in Fig. \ref{Figure 1.2}(a). The resulting Hall plateaus of a 2DEG lie at the conductivity values as follows: 

\begin{eqnarray}
\sigma_{xy} = \nu e^2/h, \label{eqn 3}
\end{eqnarray} where $\nu$ is the filling factor and takes only integer values, as illustrated in Fig. \ref{Figure 1.2}(b). For the QHE in graphene, the 2D massless Dirac equation must be solved in the presence of a perpendicular magnetic field $B$ to find the Landau Level energy $E_{n}$ \cite{Peres2006a,CastroNeto2009}. Thus, the Hamiltonian for graphene now reads:
\begin{equation}
v_{F} \vec{\sigma}\cdot(\vec{p}+\frac{e\vec{A}}{c}) \psi (\textbf{r}) = E \psi (\textbf{r}),
\label{eqn 50}
\end{equation} where momentum $\vec{p}$ in Eq. (\ref{eqn 1}) has been replaced by $\vec{p}+\frac{e\vec{A}}{c}$ and  $\vec{A}$ is the in-plane vector potential generating the perpendicular magnetic field $\vec{B}=B\hat{z}$. The solution of this equation gives rise to the eigenenergy of each Landau level for monolayer graphene: 

\begin{equation}
\ E_{n} = sgn(n) v_{F} \sqrt{2e\hbar B\left|n\right|}
\label{eqn 4}
\end{equation} with the Landau level index $n$ = 0, $\pm$1, $\pm$2, etc, and $sgn(n)$ stands for the sign of $n$. Unlike 2DEG, there will be a Landau level at zero energy ($n$ = 0) separating the positive and negative LLs, and their energies are proportional to $\sqrt{B}$ (instead of $B$ in 2DEG), as sketched in Fig. \ref{Figure 1.2}(c). In addition, the resulting Hall conductivity for monolayer graphene is given by:

\begin{equation}
\sigma_{xy} = 4e^{2}/h (n+\frac{1}{2}) = \nu e^2/h,
\label{eqn 5}
\end{equation}
where $n$ is an integer and the factor 4 is due to the double valley and double spin degeneracy \cite{Geim2007,DasSarma2011,CastroNeto2009}. Note the filling factor now reads: $\nu\equiv 4N/N_{\phi}$ = 4(n+1/2) = $\pm$2, $\pm$6, $\pm$10 etc, where $N$ is the total electron occupancy and $N_{\phi}$ is the magnetic flux divided by the flux quantum $h/e$. This result differs from the conventional QHE found in GaAs heterostructure 2DEGs [Fig. \ref{Figure 1.2}(b)] and is a hallmark of Dirac fermion in monolayer graphene. The quantization of $\sigma_{xy}$ has been observed experimentally \cite{Geim2007}, a sketch of the data is illustrated in Fig. \ref{Figure 1.2}(d). The lowest LL in the conduction band and the highest LL in the valence band merge and contribute equally to the joint level at $E = 0$, resulting in the half-odd-integer QHE. The factor 1/2 in Eq. (\ref{eqn 5}) is due to the additional Berry phase $\pi$ that the electrons, due to their chiral nature, acquire when completing a cyclotron trajectory (see section 1.3 for more details) \cite{Lukanchuk2004,Xue2013}. The observation of the QHE at room temperature is also a consequence of the Dirac nature of the fermions in graphene. Because in graphene $E_{n}$ is proportion to $v_{F}$$\sqrt{B\left|n\right|}$ (where $v_{F}$=10$^{6}$ m/s is the Fermi velocity), at low energy the energy spacing $\Delta E_{n}$ $\equiv$ $E_{n+1}$ $-$ $E_{n}$ between Landau Levels can be rather large. For example, for fields of the order of $B$ = 10 T, the cyclotron energy in a GaAs 2DEG system is of the order of 10 K, however, the same field in graphene gives rise to the cyclotron energy of the order of 1000 K, that is, two orders of magnitude larger. 

\begin{figure}
\includegraphics[scale=0.7]{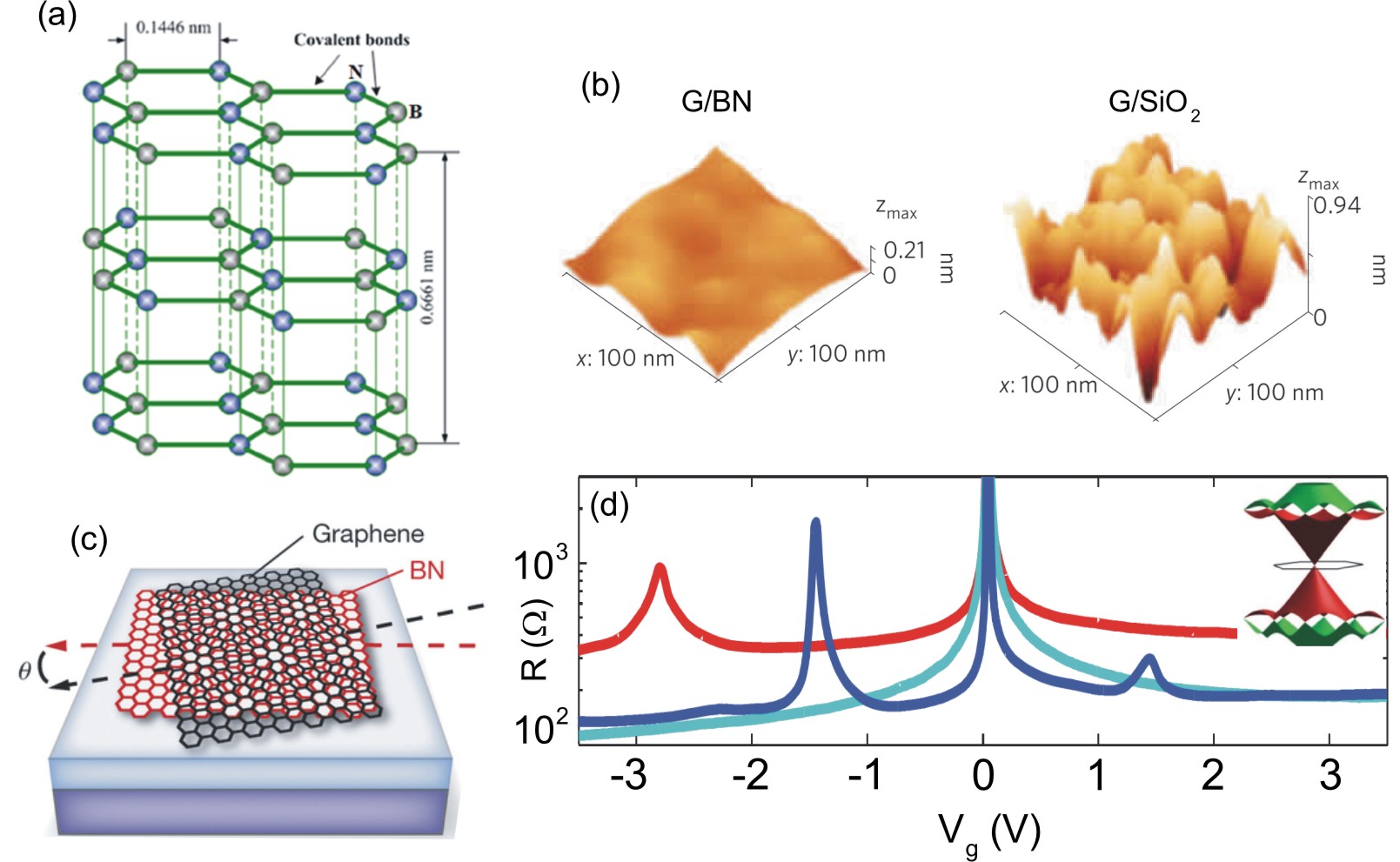}
\caption{(a) Structure of the multilayered Hexagonal Boron Nitride (hBN). (b) STM topographic images of monolayer graphene on hBN (left) and SiO$_{2}$(right) showing the underlying surface corrugations. (c) Schematic of the moir\'e pattern formed from a graphene/hBN stack. The moir\'e wavelength varies with the twist angle $\theta$. (d) Resistance as a function of gate voltage measured from three graphene/hBN stacks (with different moir\'e wavelengths), showing two extra Dirac peaks as a result of the superlattice minibands. Inset shows the band diagram of graphene on hBN. (a) adapted with permission from ref. \cite{Xu2013}. Copyright 2013 American Chemical Society. (b) adapted with permission from ref. \cite{Xue2011}. Copyright 2011 Nature Publishing Group. (c) adapted with permission from ref. \cite{Dean2013}. Copyright 2013 Nature Publishing Group. (d) adapted with permission from ref. \cite{Hunt2013}. Copyright 2013 American Association for the Advancement of Science.}  
\label{Fig9}
\end{figure}

Having briefly introduced graphene, we extend our discussion to Hexagonal Boron Nitride (hBN), which is isostructural to graphene but has boron and nitrogen atoms on the A and B sub-lattices, as shown in Fig. \ref{Fig9}(a). Due to the different onsite energy of A and B sub-lattices, the tight-binding calculation shows that hBN is an insulator with a large band gap of around 6 eV \cite{Golberg2010,Geim2013,Xu2013,Wang2014,Ferrari2015}. Traditionally, hBN has been used as a lubricant or a charge leakage barrier layer in electronic equipments \cite{Xu2013}. More importantly, recent studies have shown the use of hBN thin films as a dielectric layer for gating or as a flat substrate for graphene transistors can improve the electronic transport quality of devices by a factor of ten (or more), compared with the case of graphene on SiO$_{2}$ substrates \cite{R.2010,Dean2013,Ponomarenko2013,Hunt2013}. The high quality of graphene/hBN heterostructures originates from the atomic-level smooth surface of hBN that can suppress surface ripples in graphene. STM topographic images [Fig. \ref{Fig9}(b)] show that the surface roughness of graphene on hBN is greatly decreased compared with that of graphene on SiO$_{2}$ substrates. While graphene on SiO$_{2}$ exhibits charge puddles with diameters of 10$\sim$30 nm, the sizes of charge puddles in graphene on hBN are roughly one order of magnitude larger. The enhanced high mobility of graphene on hBN (up to 10$^{6}$ cm$^{2}$V$^{-1}$s$^{-1}$ reported \cite{Amet2015}) has enabled the studies of many-body physics and phase coherent transport that cannot be accessed in low-mobility samples, such as the observation of the fractional QHE and supercurrent in graphene \cite{Kou2014,Amet2015a}. 

Due to the similarity in lattice structure, when graphene is stacked on hBN with a small twist angle ($\leq$ 5$^{\circ}$), it can form a superlattice [called the moir\'e pattern, as shown in Fig. \ref{Fig9}(c)] with a wavelength ranging from a few to 14 nm \cite{Yankowitz2012,Hunt2013,Ponomarenko2013,Dean2013}. The superlattice with a relatively large wavelength compared to the bond length of carbon atom introduces additional minibands in graphene's band structure \cite{Hunt2013}. Fig. \ref{Fig9}(d) shows typical transfer curves for three graphene/hBN stacks with different moir\'e wavelengths, in which two extra Dirac peaks, situated symmetrically about the charge neutrality point ($V_{g}$ $=$ 0 V), are observed in all devices. These newly appeared Dirac peaks result from the superlattice minibands, which are away from the original Dirac point of graphene, as shown in the inset of Fig. \ref{Fig9}(d). Such hybrid band structures lend novel transport features to graphene; for example, the observation of the Hofstadter Butterfly spectrum in high magnetic fields \cite{Hunt2013,Ponomarenko2013,Dean2013}.

\subsection{1.2. Transition Metal Dichalcogenides}
2D materials with a hexagonal lattice structure (such as graphene or TMDs) possess valley of energy-momentum dispersion at the corner of the hexagonal Brillouin zone. In graphene, this dispersion at the $K$ and $-K$ points gives rise to a valley degeneracy (note that in this and the subsequent sections we use the notation $-K$ to replace $K'$ for simplicity). The situation is different in TMDs because of the absence of inversion symmetry, which allows the valley degree of freedom to be accessed independently (valleytronics), although it is still degenerate in energy. Semiconducting transition metal dichalcogenides (with a 2H-structure) have hexagonal lattices of MX$_2$, where M is a transition metal element from group VI (Mo or W) and X is a chalcogen atom (S, Se or Te), as illustrated in Fig. \ref{Figure 1.3}(a). Unlike graphene and hBN, the lattice structure of such a TMD consists of hexagons of M and X, with the M atom being coordinated by the six neighboring X atoms in a trigonal prismatic geometry, as shown in Fig. \ref{Figure 1.3}(b). A key aspect of semiconducting TMDs is the effect exerted by the number of layers on the electronic band structure. Fig. \ref{Figure 1.3}(c) shows the calculated band structure of a TMD (MoS$_2$), which exhibits a crossover from an indirect gap in the bulk form to a direct gap in the monolayer form as a result of a decreasing interlayer interaction. The photoluminescence (PL) from monolayer MoS$_2$ has shown the quantum yield to be two orders of magnitude larger than that from the multilayer material, providing evidence of such a crossover in the band gap \cite{Splendiani2010,Mak2010}. In the monolayer limit, the conduction and valence band edges are at the $\pm K$ points and are predominantly formed by the partially filled $d$-orbitals of the M atoms and have the following forms:

\begin{figure}
\begin{center}
		\includegraphics[scale=0.42]{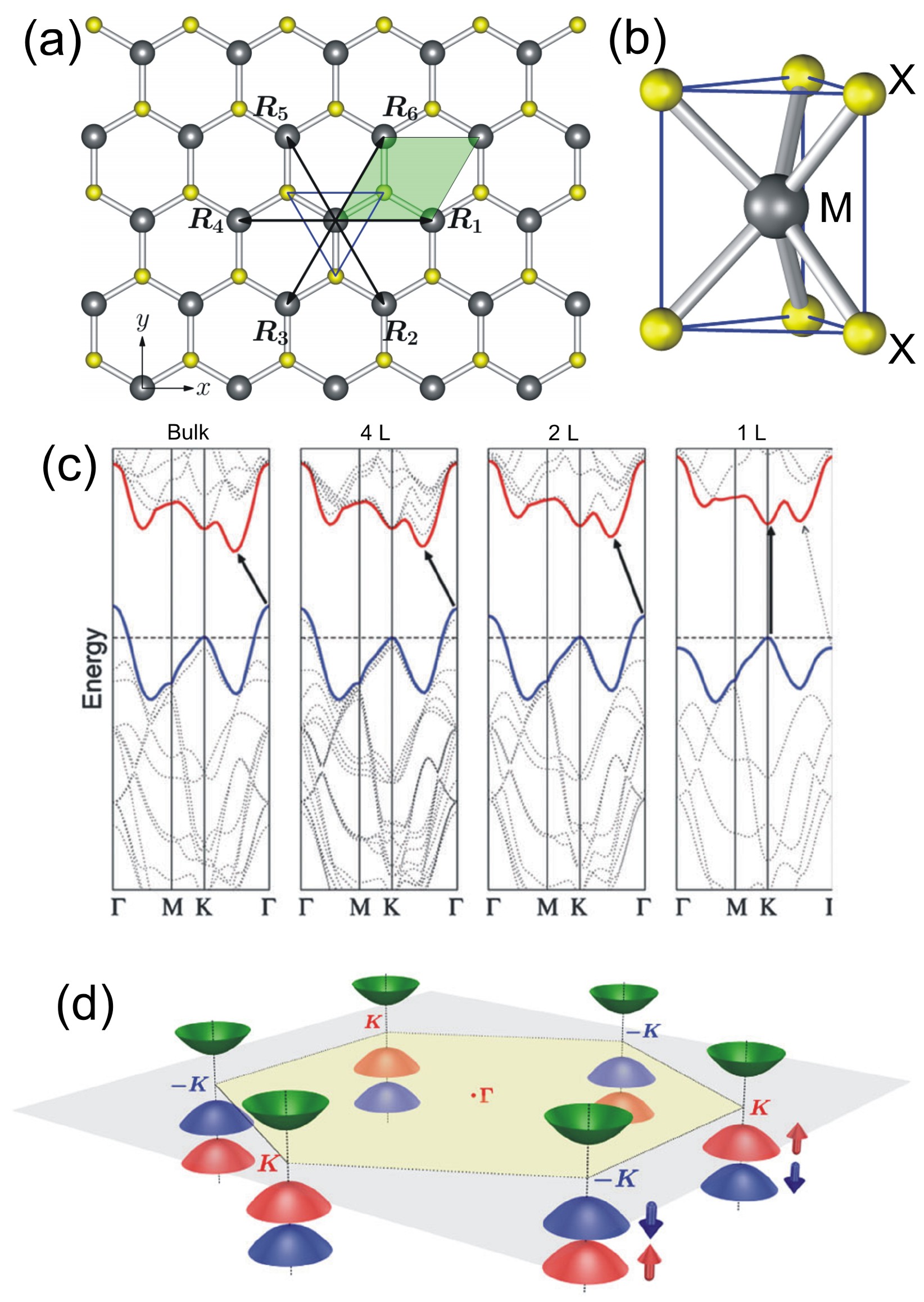}
\caption{(a) Top view of monolayer MX$_2$. Black balls represent M atoms, and yellow balls represent X atoms. The shadowed diamond region shows the 2D unit cell with lattice constant $a$. $\textbf{R}_1$ - $\textbf{R}_6$ denote the M-M nearest neighbors. (b) Schematic illustration for the structure of trigonal prismatic coordination, corresponding to a side view of the blue triangle in (a). (c) Energy dispersion in bulk, quadrilayer (4L), bilayer (2L) and monolayer (1L) MoS$_2$, from left to right, showing the transition from an indirect band gap to a direct band gap. (d) Schematic illustration of the band structure at the band edges located at the edges of the Brillouin zone. (a, b) adapted with permission from ref. \cite{Liu2013}. Copyright 2013 American Physical Society. (c) adapted with permission from ref. \cite{Duan2015}. Copyright 2015 Royal Society of Chemistry. (d) adapted with permission from ref. \cite{Xiao2012}. Copyright 2012 American Physical Society.}
\label{Figure 1.3}
\end{center}	
\end{figure}

\begin{eqnarray}
\left|\phi_{c}\right\rangle &=& \left|d_{z^2}\right\rangle \label{eqn 6.1} \\
\left|\phi_{v}^{\tau}\right\rangle &=& \frac{1}{\sqrt{2}}(\left|d_{x^2-y^2} \right\rangle + i\tau\left|d_{xy} \right\rangle), \label{eqn 6.2}
\end{eqnarray}
where $d_{z^2}$, $d_{x^2-y^2}$ and $d_{xy}$ are the $d$-orbitals of the M atom, the subscript $c$($v$) indicates the conduction (valence) band, and $\tau=\pm1$ is the valley index. At the valley points ($\pm K$), a two-band $k\cdot p$ Hamiltonian that takes the form of the massive Dirac fermion model is used to describe the dispersion at the conduction and valence band edges \cite{Xiao2012}:

\begin{equation}
H = at(\tau k_x \sigma_x + k_y \sigma_y) + \frac{\Delta}{2}\sigma_z - \lambda\tau \frac{\sigma_z -1}{2} \hat{S}_z, 
\label{eqn 7}
\end{equation}
where $\sigma$ denotes the Pauli matrices for the two basis functions given in Eq. (\ref{eqn 6.1}) and (\ref{eqn 6.2}), $a$ is the lattice constant, $t$ is the effective nearest neighbour hopping integral, and $\Delta$ is the band gap. The last term in Eq. (\ref{eqn 7}) represents the spin-orbit coupling (SOC), where 2$\lambda$ is the spin splitting at the top of the valence band and $\hat{S}_z$ is the Pauli matrix for spin. The spin splitting is due to the strong spin-orbit interaction arising from the $d$-orbitals of the heavy metal atoms. The conduction band-edge state consists of $d_{z^2}$ orbitals and remains almost spin-degenerate at the $\pm K$ points, whereas the valence-band-edge state shows a pronounced split. A schematic illustration of the band dispersion at the edges of the hexagonal Brillouin zone is shown in Fig. \ref{Figure 1.3}(d). Note that the spin splitting at the different valleys is opposite because the $K$ and $-K$ valleys are related to one another by time-reversal symmetry.

\begin{figure}
\begin{center}
		\includegraphics[scale=0.5]{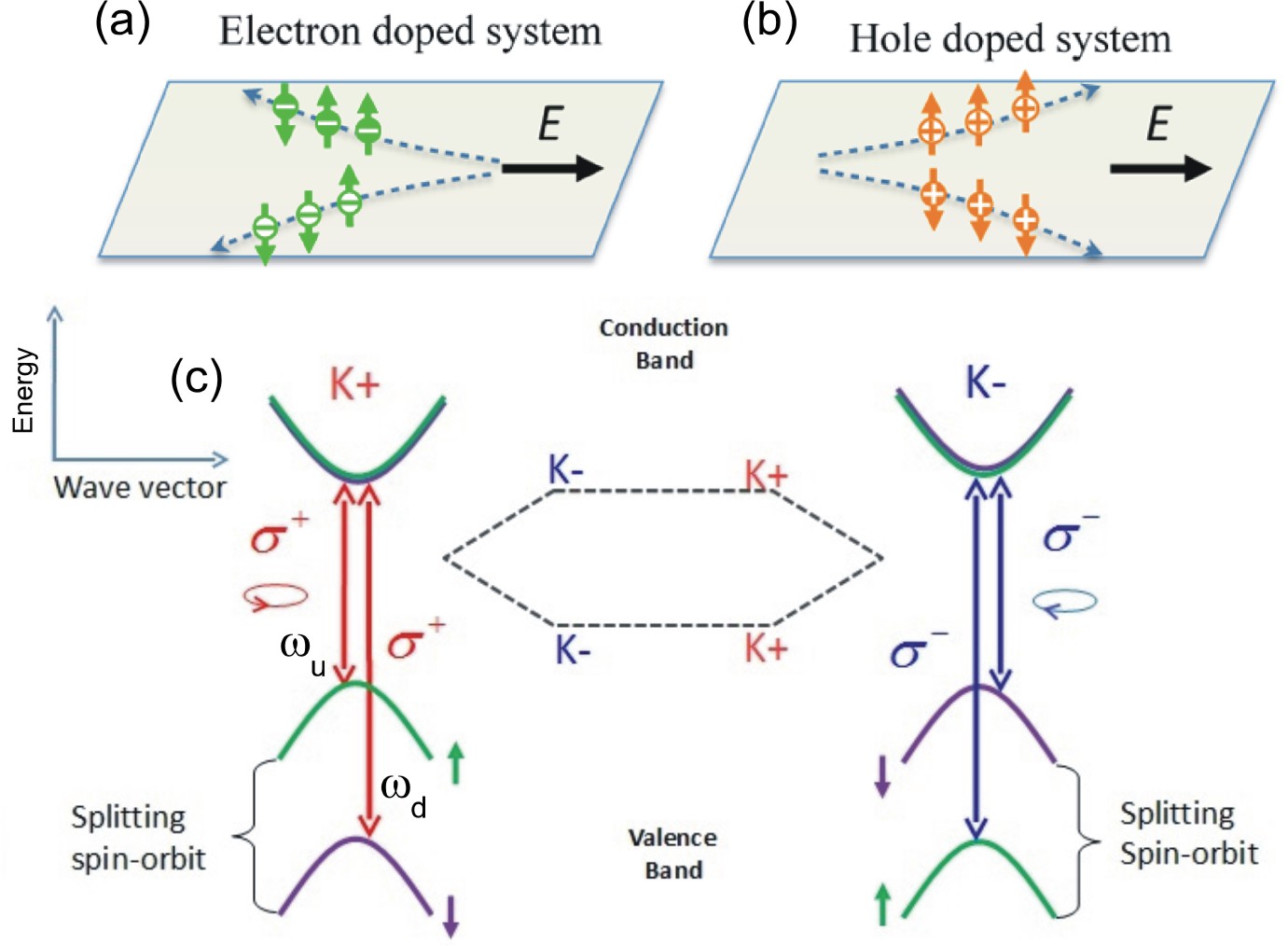}
\caption{Schematic illustration of the valley Hall effect in (a) the electron-doped regime and (b) the hole-doped regime. The electrons and holes in the $+K$ valley are denoted by white "$-$" and "$+$" symbols in dark circles and their counterparts in the $-K$ valley are denoted by an inverted color scheme. (c) Illustrations of the valley- and spin-dependent optical transition selection rules. (a, b) adapted with permission from ref. \cite{Xiao2012}. Copyright 2012 American Physical Society.}
\label{Figure 1.4}
\end{center}	
\end{figure}

Because of the large valley separation in momentum space, the valley index is expected to be robust against scattering by smooth deformations and long-wavelength phonons. To manipulate such a valley degree of freedom for valleytronic applications, measurable physical quantities that distinguish the $\pm K$ valleys are required. The Berry curvature ($\bf{\Omega}$) and the orbital magnetic moment (\textbf{m}) are two physical quantities for $\pm K$ valleys to have opposite values. The Berry curvature is defined as a gauge field tensor derived from the Berry vector potential $\textbf{A}_n$($\textbf{R}$) through the relation $\bf{\Omega}$$_n$($\textbf{R}$)=$\nabla_\textbf{R}$ $\times$ $\textbf{A}_n$($\textbf{R}$), where \textit{n} is the energy band index (in the case of TMDs and at the $\pm K$ points, $n$ is either the conduction or valence band) and $\textbf{R}$ is the parameter to be varied in a physical system (in the case below $\textbf{R}$ is the wavevector \textbf{k}) \cite{Xiao2010}. The Berry curvature can be written as a summation over the eigenstates as follows \cite{Xu2014}: 

\begin{eqnarray}
\bf{\Omega}_n(k) &=& i\frac{\hbar^2}{m^2}\sum_{i\neq n}\frac{\textbf{P}_{n,i}(\textbf{k}) \times \textbf{P}_{i,n}(\textbf{k})}{\left[E^{0}_{n}(\textbf{k}) - E^{0}_{i}(\textbf{k})\right]^2}
\label{eqn 8}
\end{eqnarray}
Here, $\textbf{P}$$_{n,i}$(\textbf{k}) $\equiv$ $\left\langle u_{n}(\bf{k})\right|$$\bf{\hat{p}}$$\left|u_{i}(\bf{k})\right\rangle$ is the interband matrix element of the canonical momentum operator $\bf{\hat{p}}$, where $u(\bf{k})$ is the periodic part of the Bloch wavefunction, and $E^{0}_{n(i)}(\bf{k})$ denotes the energy dispersion of the $n$($i$)-th band. Upon substituting the eigenfunctions of Eq. (\ref{eqn 7}) into Eq. (\ref{eqn 8}), the Berry curvature in the conduction band is given by:

\begin{eqnarray}
\bf{\Omega_c(k)} &=& -\tau \frac{2a^2 t^2 \Delta'}{(4a^2 t^2 k^2 + \Delta'^2)^{3/2}}
\label{eqn 9}
\end{eqnarray}
where $\tau$ is the valley index and $\Delta'$ $\equiv$ $\Delta - \tau S_z\lambda$ is the spin-dependent band gap. Note that the Berry curvature has opposite signs in opposite valleys, and this also occurs in the conduction and valence bands [$\bf{\Omega_v}$(\textbf{k}) = $-\bf{\Omega_c}$(\textbf{k})]. Here, we write the equations of motion for Bloch electrons under the influence of the Berry curvature and applied electric and magnetic fields \cite{Xiao2010}:
\begin{eqnarray}
\bf{\dot{r}} &=& \frac{1}{\hbar}\frac{\partial E_{n}(\textbf{k})}{\partial \textbf{k}}-\dot{\textbf{k}} \times \bf{\Omega_n(k)} \\ 
\hbar\dot{\textbf{k}} &=& -e\textbf{E} - e\dot{\textbf{r}} \times \bf{B}
\label{eqn 10}
\end{eqnarray}
It can be seen that in the presence of an in-plane electric field, carriers with different valley indices will acquire opposite velocities in the transverse direction because of the opposite signs of their Berry curvatures, leading to the so-called valley Hall effect, as illustrated in Fig. \ref{Figure 1.4}(a) and (b). Here, we note that this result is valid not only for monolayer TMDs but also for thin films with an odd number of layers because odd numbers of layers also exhibit inversion symmetry breaking, which is a necessary condition for the $\pm K$ valleys to exhibit valley contrast in the Berry curvature.

The valley contrast in TMDs can also reflect on the optical interband transitions from the top of the spin-split valence-band to the bottom of the conduction band at the $\pm K$ points. The coupling strength with optical fields of $\sigma_\pm$ circular polarization is given by \textbf{\textit{P}}$_{\pm}(\textbf{k})$ $\equiv$ \textbf{\textit{P}}$_{x}(\textbf{k})$ $\pm$ $i$\textbf{\textit{P}}$_{y}(\textbf{k})$, where \textbf{\textit{P}}$_{\alpha}$(\textbf{k}) $\equiv$ $m_0$$\left\langle u_{c}(\bf{k})\right|$$\frac{1}{\hbar}\frac{\partial\hat{H}}{\partial k_\alpha}$$\left|u_{v}(\bf{k})\right\rangle$ is the interband matrix element of the canonical momentum operator ($u_{c(v)}(\bf{k})$ is the Bloch function for the conduction (valence) band, and $m_0$ is the free electron mass). For transitions near the $\pm K$ points and for a reasonable approximation of $\Delta'$ $>>$ $atk$, this expression has the following form \cite{Xiao2012}:

\begin{eqnarray}
\left|\textit{\textbf{P}}_{\pm}(\textbf{k})\right|^2 &=& \frac{m_{0}^{2}a^{2}t^{2}}{\hbar^2}(1\pm\tau)^2 \label{eqn 11}
\end{eqnarray} It is evident that the coupling strength between circularly polarized light and the interband transitions is valley dependent; \textbf{\textit{P}}$_{+}(\textbf{k})$ has a non-zero value in the $+K$ valley, as does \textbf{\textit{P}}$_{-}(\textbf{k})$ in the $-K$ valley. This valley-dependent optical selection rule is illustrated in Fig. \ref{Figure 1.4}(c), where a $\sigma_{+(-)}$ circularly polarized optical field exclusively couples with the interband transitions at the $+(-)K$ valley. Note that spin is selectively excited through this valley-dependent optical selection rule, and consequently, the spin index becomes locked with the valley index at the band edges. For example, an optical field with $\sigma_{+}$ circular polarization and a frequency of $\omega_{d}$($\omega_{u}$) can generate spin-up (spin-down) electrons and spin-down (spin-up) holes in the $+K$ valley, whereas the excitation in the $-K$ valley is precisely the time-reversed counterpart of the above \cite{Xiao2012}.

\subsection{1.3. Topological Insulators}

Topology is a branch of mathematics concerned with the geometrical properties of objects that are insensitive to smooth deformations. For example, a sphere and a torus are topologically different because the former can not be smoothly deformed into the latter. These forms can be distinguished by an integer topological invariant called the genus, g, which is essentially equivalent to the number of holes in the object (for a sphere, g = 0; for a torus, g = 1). To mathematically characterize the topology of a given object, a theorem known as the Gauss-Bonnet theorem defines how an integer topological invariant called the Euler characteristic, $\chi$, is related to the genus g: 

\begin{eqnarray}
\chi = \frac{1}{2\pi} \int_{S}\textbf{\textit{K}}dA = 2-2g
\label{eqn 12}
\end{eqnarray}
Here, \textbf{\textit{K}} is the Gaussian curvature and the integral is performed over the surface of the given object. It can be easily checked that, for example, a sphere (g $=$ 0) of radius $R$ has \textbf{\textit{K}} = 1/$R^2$, which yields $\chi$ = 2, thus satisfying the relation. In analogy to geometric topology, materials can be identified as topologically equivalent if they can be transformed into one another by slowly changing their Hamiltonian, analogously to the smooth deformation of an object. To determine the topological classes of materials, a Gauss-Bonnet-like theorem is needed in quantum mechanics to topologically characterize their band structures. It turns out the Berry phase is a key conceptual tool for the analysis of topological phenomena in solid-state physics. The Berry phase describes the phase that a state acquires in an adiabatic cycle. In band theory, for a closed loop \textit{C} in \textbf{k} space, the Berry phase can be defined as follows:

\begin{figure}
\begin{center}
		\includegraphics[scale=0.5]{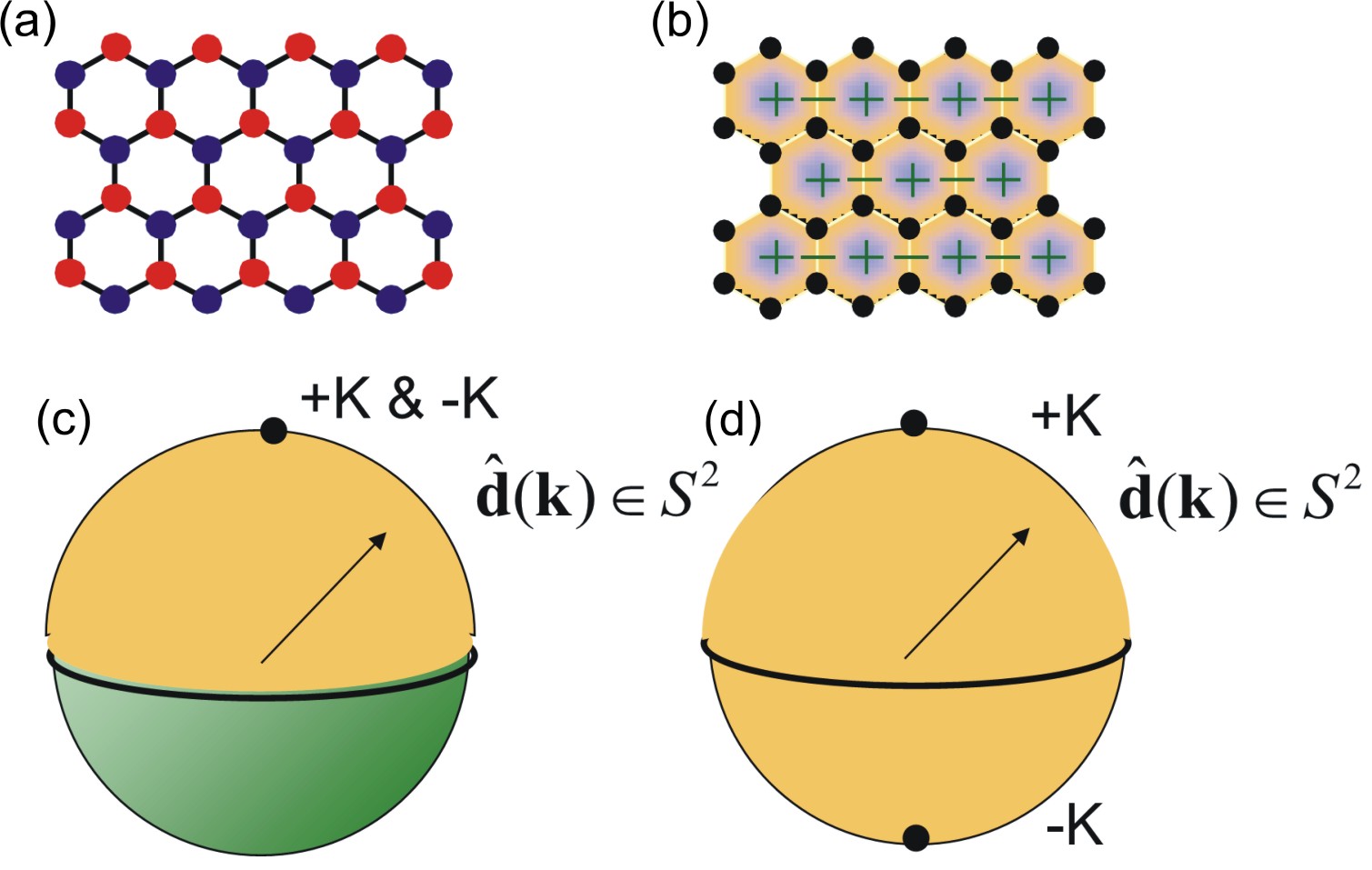}
\caption{(a) Schematic lattice structure of hBN (inversion symmetry breaking). (b) Schematic illustration of graphene in a periodic magnetic field, which is zero on the average (no net flux) but has all of the spatial symmetries of the honeycomb lattice. This condition is the so-called Haldane model (time reversal symmetry breaking). (c) Illustration of the solid angle swept out by \textbf{d}(\textbf{k}) for $m_+$ = $m_-$. (d) Illustration of the solid angle swept out by \textbf{\textit{d}}(\textbf{k}) for $m_+$ = $-m_-$. Adapted with permission from ref. \cite{Kane2013}. Copyright 2013 Elsevier B.V.}
\label{Figure 1.5}
\end{center}	
\end{figure}

\begin{eqnarray}
\gamma_C = \oint_{C}\textbf{A}\cdot d\textbf{K} = \int_{S} \Omega d^{2}\textbf{k}, \label{eqn 13}
\end{eqnarray} where \textbf{A} = -$i$ $\left\langle u_{n}(\bf{k})\right|$$\nabla_{\textbf{k}}$$\left|u_{i}(\bf{k})\right\rangle$ is the Berry vector potential and $\Omega$ = $\nabla \times$ \textbf{A} is the Berry curvature or Berry flux as noted in section 1.2 (\textbf{A} = $\sum_{n}$\textbf{A}$_n$ and $\Omega$ = $\sum_{n}$$\bf{\Omega}$$_n$). Note that \textit{S} denote the surface created by the closed loop $C$. In a two-level system whose Hamiltonian has the form $H$(\textbf{k}) = \textbf{\textit{d}}(\textbf{k}) $\cdot$ $\vec{\sigma}$ = $\left(\begin{array} {cc} \ d_{z} & d_{x} - id{_y} \ \\ d_{x} + id{_y} & -d_{z} \\ \end{array}\right)$, where \textbf{\textit{d}} is a physical variable and $\vec{\sigma}$ is the Pauli matrix, the Berry phase acquired by completing a loop $C$ is simply half of the solid angle swept out by \textbf{\textit{d}}(\textbf{k}). This relation can be seen from the example of graphene, in which \textbf{\textit{d}} is now the momentum \textbf{\textit{p}} and an electron acquires a Berry phase of $\pi$ when completing a closed trajectory (a 2$\pi$ rotation of \textbf{\textit{p}} in a plane), resulting in a factor of 1/2 in the quantum Hall conductivity [Eq. (\ref{eqn 5})]. The importance of the Berry phase lies in its close relation to an integer topological invariant called the Chern number, $n$, which is useful for characterizing the topology of a given Hamiltonian. The Chern number is defined as the Berry phase acquired after encircling the Brillouin zone boundary (or the total Berry flux in the Brillouin zone), namely,

\begin{eqnarray}
n = \frac{1}{2\pi}\int_{BZ}\Omega d^2\textbf{k} \label{eqn 14}
\end{eqnarray} Here, we take graphene as an example to consider two types of symmetry breaking: inversion symmetry breaking and time reversal symmetry breaking. We will discuss how the Chern number characterizes these two topological phases in a honeycomb lattice. Under symmetry breaking, a mass term $m$ will be introduced into the graphene Hamiltonian (the notation $\pm$ represent the valley index): 

\begin{eqnarray}
H(\textbf{q})= \hbar v_{F} \vec{\sigma}_{\pm} \cdot {\textbf{q}} + m_{\pm}\sigma_z, \label{eqn 15}
\end{eqnarray} where \textbf{q} $\equiv$ \textbf{k} $\mp$ \textbf{\textit{K}}, $\vec{\sigma}_{\pm}$ $=$ ($\sigma_{x}$, $\pm\sigma_{y}$) and m$_{\pm}$ denotes the mass term in the $\pm K$ valley. The energy dispersion associated with this Hamiltonian is $E(\textbf{q})$ = $\pm\sqrt{\hbar^2v_{F}^{2}\left|\textbf{q}\right|^2+m_{\pm}^{2}}$, which exhibits an energy gap of 2$\left|m\right|$ at \textbf{q} = 0. Depending on the symmetry that is lifted, $m_+$ and $m_-$ are related by different polarities. If it is the inversion symmetry that is broken, which means that the A and B atoms are not equivalent [ex: hBN; see Fig. \ref{Figure 1.5}(a)], then $m_+$ = $m_-$. If instead the broken symmetry is the time-reversal symmetry [ex: graphene in a periodic magnetic field; see Fig. \ref{Figure 1.5}(b)], then $m_+$ = $-m_-$. The Chern numbers for these two cases result in distinct integers, as illustrated in Fig. \ref{Figure 1.5}(c) and Fig. \ref{Figure 1.5}(d). In the case of inversion symmetry breaking, when cycling around the Brillouin zone [see Fig. \ref{Figure 1.1}(b), right panel], \textbf{\textit{d}} lies mostly on the equator except at the $\pm K$ points, where it visits the north pole (let $m_+$ be a positive value), as shown in Fig. \ref{Figure 1.5}(c). The net solid angle swept out by \textbf{\textit{d}} is zero, and thus, the Chern number is 0. By contrast, for time-reversal symmetry breaking, where $m_+$ = $-$$m_-$, \textbf{\textit{d}} visits both the north and south poles and loops around the sphere once, as shown in Fig. \ref{Figure 1.5}(d). Thus, the net solid angle swept out by \textbf{\textit{d}} is 2$\pi$ and the Chern number is 1. In an important 1982 paper \cite{Thouless1982}, Thouless, Kohmoto, Nightingale, and den Nijs (TKNN) developed the Kubo formula for $\sigma_{xy}$ and showed that the integer in the integer quantized Hall conductivity is precisely the Chern number. This results in values of $\sigma_{xy}$ = 0 for hBN, which is classified as an ordinary insulator, and $\sigma_{xy}$ = $e^2$/$h$ for the "graphene $+$ B-field" system (the so-called Haldane model), which is classified as a quantum Hall state. What is particularly interesting is the existence of gapless conducting states at interfaces where the topological invariant (the Chern number, in this case) changes if one places two topologically distinct systems together, as shown in Fig. \ref{Figure 1.6}(a). To describe such a system, let us consider $m$ as a function of $x$ [$m$ $\rightarrow$ $m(x)$], which changes sign across the interface at $x$ = 0, meaning that $m_{-}(x)$ $>$ 0 ($n$ = 0) for $x$ $>$ 0 and $m_{-}(x)$ $<$ 0 ($n$ = 1) for $x$ $<$ 0. Such a condition is met when a magnetic field is present in a stripe of graphene ($n$ = 1) that has an upper boundary with vacuum ($n$ = 0), as shown in Fig. \ref{Figure 1.6}(b). From Eq. (\ref{eqn 15}), it can be easily found that there is a zero-energy mode $\phi_0$ near the interface (where $m$ = 0) that satisfies a band dispersion of $E_{0}(q_{y})$ = $v_F$$q_{y}$. This band of states (the so-called chiral Dirac fermions) intersects the Fermi energy $E_F$ with a positive group velocity $dE/dq_y$ = $v_F$ and defines a rightward moving chiral edge mode, as shown in Fig. \ref{Figure 1.6}(c). These edge states are insensitive to disorder because there are no states available for backscattering $-$ a fact that underlies the perfectly quantized electronic transport in the quantum Hall effect. 

\begin{figure}
\begin{center}
		\includegraphics[scale=0.5]{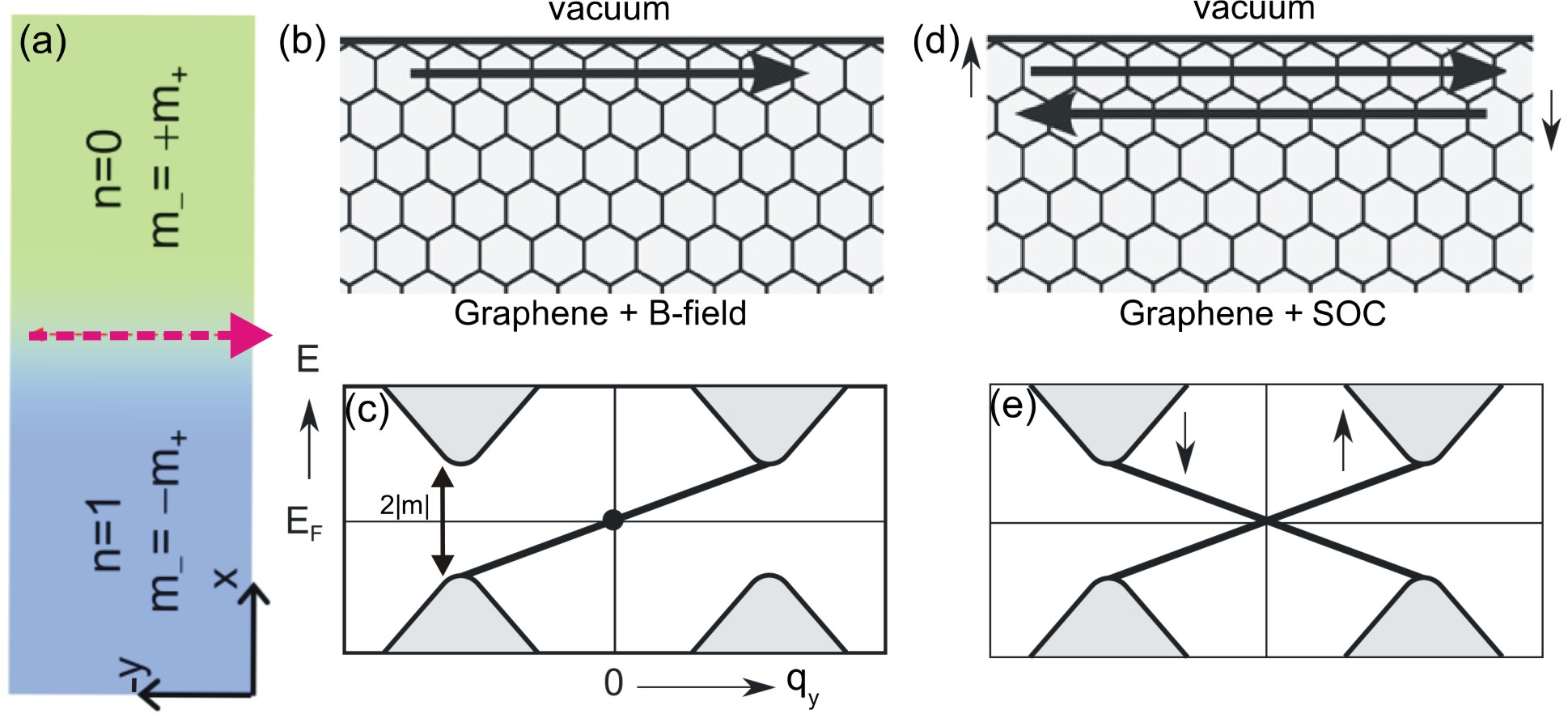}
\caption{(a) The interface between a quantum Hall state ($n$ = 1) and an ordinary insulator ($n$ = 0) has a chiral edge mode. (b) Illustration of the chiral edge state near the interface between a semi-infinite graphene strip and vacuum in a periodic magnetic field. (c) The electronic structure of (b). A single edge state connects the valence band to the conduction band. (d) Illustration of the spin Hall edge states near the interface between a semi-infinite graphene strip and vacuum. (e) The electronic structure of (d). The edge states for upward and downward spins propagate in opposite directions. Adapted with permission from ref. \cite{Kane2013}. Copyright 2013 Elsevier B.V.}
\label{Figure 1.6}
\end{center}	
\end{figure}

\begin{figure}
\begin{center}
		\includegraphics[scale=0.6]{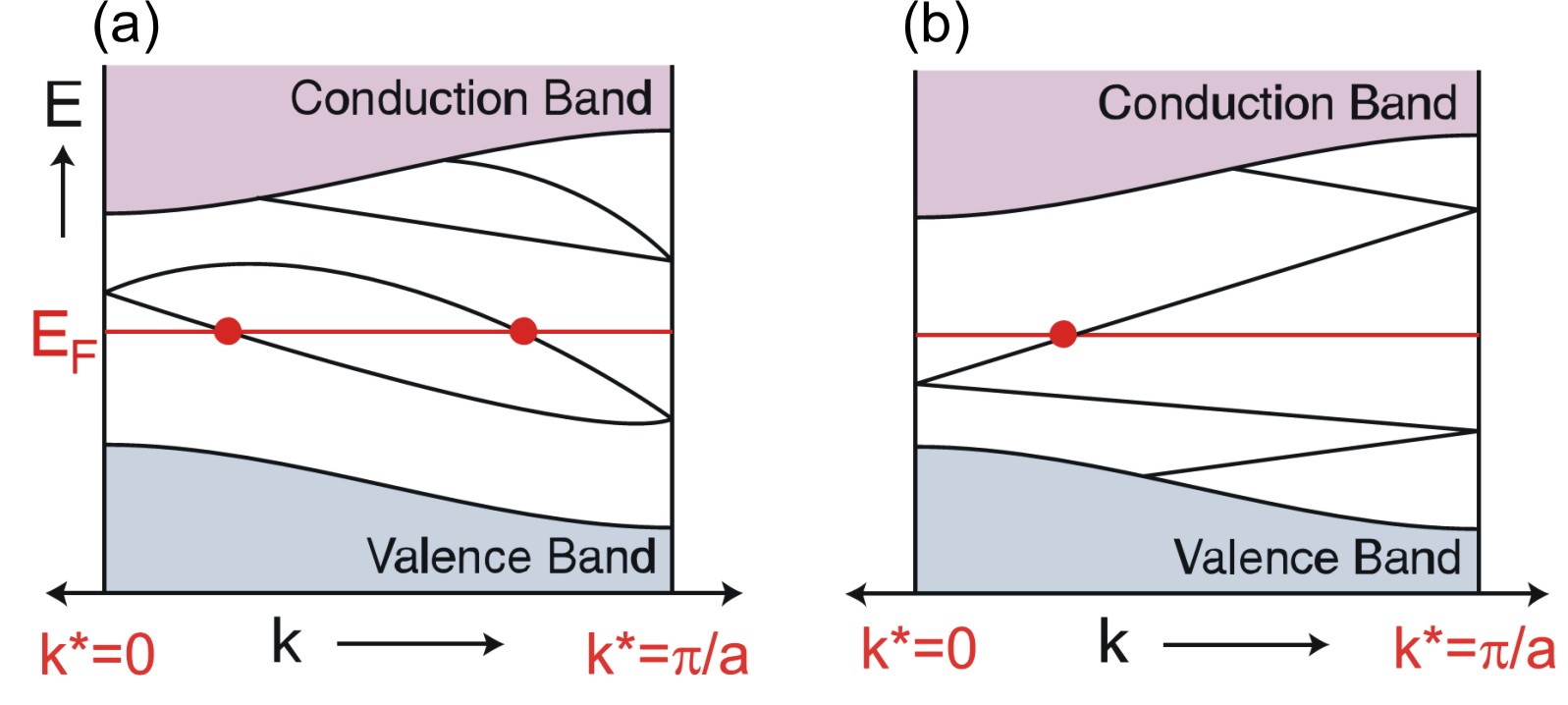}
\caption{Electronic dispersion between two Kramers-degenerate boundary points. In (a), the surface states crosses the Fermi energy $E_{F}$ an even number of times, whereas in (b) they crosses $E_{F}$ an odd number of times. An odd number of crossings leads to topologically protected metallic boundary states. Adapted with permission from ref. \cite{Kane2013}. Copyright 2013 Elsevier B.V.}
\label{Figure 1.7}
\end{center}	
\end{figure}

The chiral edge states discussed above can only arise when time-reversal symmetry (TRS) is broken either by an external magnetic field or by magnetic order. However, can such a dissipationless edge state exist in the absence of magnetic fields? The answer to this question leads to a so-called quantum spin Hall (QSH) effect in which time-reversal symmetry is unbroken. Here, we will show that the spin-orbit coupling allows a different topological class of insulating band structures, which is protected by a $Z_2$ topological invariant. The spin-orbit interaction induces a new mass term $\Delta_{SO}$ in the graphene Hamiltonian; thus, we rewrite Eq. (\ref{eqn 15}) as follows:
\begin{eqnarray}
H = \hbar v_{F} \vec{\sigma}_{\pm} \cdot {\textbf{q}} \pm \Delta_{SO}\sigma_zs^z, \label{eqn 16} \end{eqnarray} where $\pm$ is the valley index and $s^z = \pm 1$ is the spin index. This Hamiltonian is invariant under time reversal and thus satisfies $\Theta H(\textbf{q}) \Theta^{-1}$ = $H(-\textbf{q})$, where $\Theta$ = $e^{i\pi S_y/\hbar}K$ is the time-reversal operator ($S_y$ is the spin operator and $K$ denotes complex conjugation). For each spin component ($s^z = \pm 1$), the Hamiltonian has exactly the same form as that for the spinless electrons in a magnetic field [see Eq. (\ref{eqn 15})] but with upward spins and downward spins possessing opposite B-field signs, which gives rise to quantized Hall conductivity $\pm e^2/h$. Since it is equivalent to two copies of a quantum Hall state, the quantum spin Hall state must have gapless edge states (when the "graphene $+$ SOC" system is in contact with a vacuum). These edge states are protected by TRS and have the special "spin-filter" property in which upward and downward spins propagate in opposite directions, as shown in Fig. \ref{Figure 1.6}(d) and (e). Thus, an applied electric field will result in zero net current but a net spin current \textbf{J}$_s$ = ($\hbar/2e$)(\textbf{J}$_\uparrow$ $-$ \textbf{J}$_\downarrow$) flowing along the edge. Like the edge mode in a quantum Hall state, the quantum spin Hall edge states cannot be localized even in the case of strong disorder, meaning that an incident electron should be transmitted perfectly across the disordered region unless TRS is broken. An insulator that has quantum spin Hall edge states is called a 2D topological insulator. The fact that the edge states of a topological insulator are robust suggests that there must be a topological distinction between a topological insulator and an ordinary insulator. Note that the Chern number for a quantum spin Hall insulator is zero ($n_{\uparrow} + n_{\downarrow}$ = 0) because of TRS and thus cannot be used to distinguish the QSH states from ordinary insulators. To find out the correct topological invariant, we start with the Kramers' theorem. The Kramers' theorem requires that there must be a twofold degeneracy at the time-reversal invariant momenta $k_y$ = 0 and $k_y$ = $\pi/a$, whereas away from these points, the spin-orbit interaction allows the degeneracy to break. This permits two (and only two) ways for the states at $k_y$ = 0 and $k_y$ = $\pi/a$ to connect, as shown in Fig. \ref{Figure 1.7}. They can either connect in pairs, as shown in Fig. \ref{Figure 1.7}(a), in which case the edge states intersect with $E_F$ an even number of times, or they can switch partners and connect as shown in Fig. \ref{Figure 1.7}(b), in which case the edge states intersect with $E_F$ an odd number of times. Note that the edge states in Fig. \ref{Figure 1.7}(a) can be eliminated by pushing all of the bound states out of the gap, whereas those in Fig. \ref{Figure 1.7}(b) cannot be eliminated (topologically protected). This condition provides a distinction between an ordinary insulator and a topological insulator, in that the change in the $Z_2$ topological invariant $\Delta\nu$ across the interface is related to the number of Kramers pairs of edge modes intersecting $E_F$. If $\Delta\nu$ is even (including 0), the insulator is trivial, whereas if $\Delta\nu$ is odd, then the insulator is a topological insulator with protected edge states. Mathematical formulations for the $Z_2$ topological invariant $\nu$ have been developed by several groups; one approach that is applied for systems with inversion symmetry is to use a unitary matrix $w_{mn}(\textbf{k})$ $=$ $\left\langle u_{m}(\bf{k})\right|$$\Theta$$\left|u_{n}(-\bf{k})\right\rangle$ constructed from the occupied Bloch functions $\left|u_m(\textbf{k})\right\rangle$. Since $\Theta^2$ = $-$1, the matrix is antisymmetric $w(\textbf{k})$ $=$ $-w^T(-\textbf{k})$. There are four special points $\Lambda_a$ in the bulk 2D Brillouin zone where $\textbf{k} = \Lambda_a = -\Lambda_a$, and thus, $w(\Lambda_a)$ $=$ $-w^T(\Lambda_a)$. The determinant of an antisymmetric matrix is the square of its Pfaffian, based on which we can define $\delta_a$ $=$ PF[$w(\Lambda_a)$]/$\sqrt{Det[w(\Lambda_a)]}$ $=$ $\pm1$ and the $Z_2$ topological invariant $\nu$ is related to $\delta_a$ via:

\begin{eqnarray}
(-1)^{\nu} = \prod^{4}_{a=1}\delta_a \label{eqn 17} \end{eqnarray}This formulation can be generalized to 3D topological insulators and involves the eight special points in the 3D Brillouin zone.

\begin{figure*}[!t]	
\includegraphics[scale=0.7]{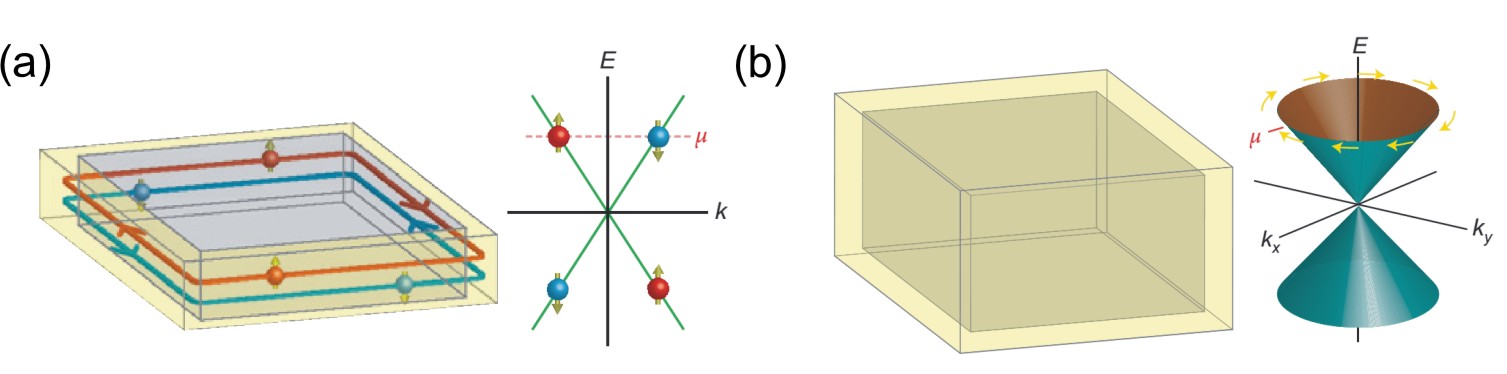}
\caption{(a) (Left): Schematic illustration of the metallic edge (shown in yellow) of a 2D topological insulator, in which spin-up and spin-down electrons counter-propagate. (Right): The corresponding idealized spin-resolved band structure of the edge states. $\mu$ denotes the Fermi level. (b) (Left): Schematic illustration of the metallic surfaces (shown in yellow) of a 3D topological insulator. (Right): The corresponding idealized band structure of the surface states, revealing chiral lefthanded spin-texture which is sometimes referred as spin-momentum locking. Adapted with permission from ref. \cite{Kong2011}. Copyright 2011 Nature Publishing Group.}  
\label{Figure 1.8}
\end{figure*}

The first experimentally demonstrated TI is a quantum well structure in which a HgTe thin film is sandwiched between two layers of CdTe \cite{Konig2007}. The opposite parities of the $p$ and $s$ levels in HgTe compared with those in CdTe give rise to a band crossing in HgTe, from which the edge states can be formed. The resulting 2D TI and the corresponding edge-state dispersion are illustrated in Fig. \ref{Figure 1.8}(a). It can be seen from the edge dispersion that an electron with a reversed wavevector has the opposite spin polarization, and therefore, the spin-up and spin-down electrons counter-propagate along the edge. 3D topological insulators, such as Bi$_{1-x}$Sb$_{x}$ and Bi$_{2}$Se$_{3}$, were predicted and identified experimentally more recently \cite{Fu2007,Moore2007,Roy2009,Hsieh2008,Xia2009,Zhang2009a}. Compared with 2D TIs, the edge states now form metallic surface states covering the entire material, as shown in Fig. \ref{Figure 1.8}(b). The band structure [right panel of Fig. \ref{Figure 1.8}(b)] shows that an electron with a positive surface wavevector $k$ has the opposite spin orientation to that of an electron with a negative surface wavevector -$k$; this phenomenon is generally referred to as spin-momentum locking. The surface states in 3D TIs can be viewed as Dirac fermions in graphene but without the twofold valley and spin degeneracies, and they are topologically protected from backscattering by time-reversal symmetry.

In this chapter, we introduced the fundamental physics relevant to various 2D materials that are to be discussed in this review. Combined with the quantum dot physics presented in the next chapter, this discussion will serve as a basis for our examination of the experimental results presented in subsequent chapters. 

\maketitle
\section{2. Theoretical background on quantum dots}

\subsection{2.1. Single quantum dot}
\begin{figure}
\begin{center}
		\includegraphics[scale=0.7]{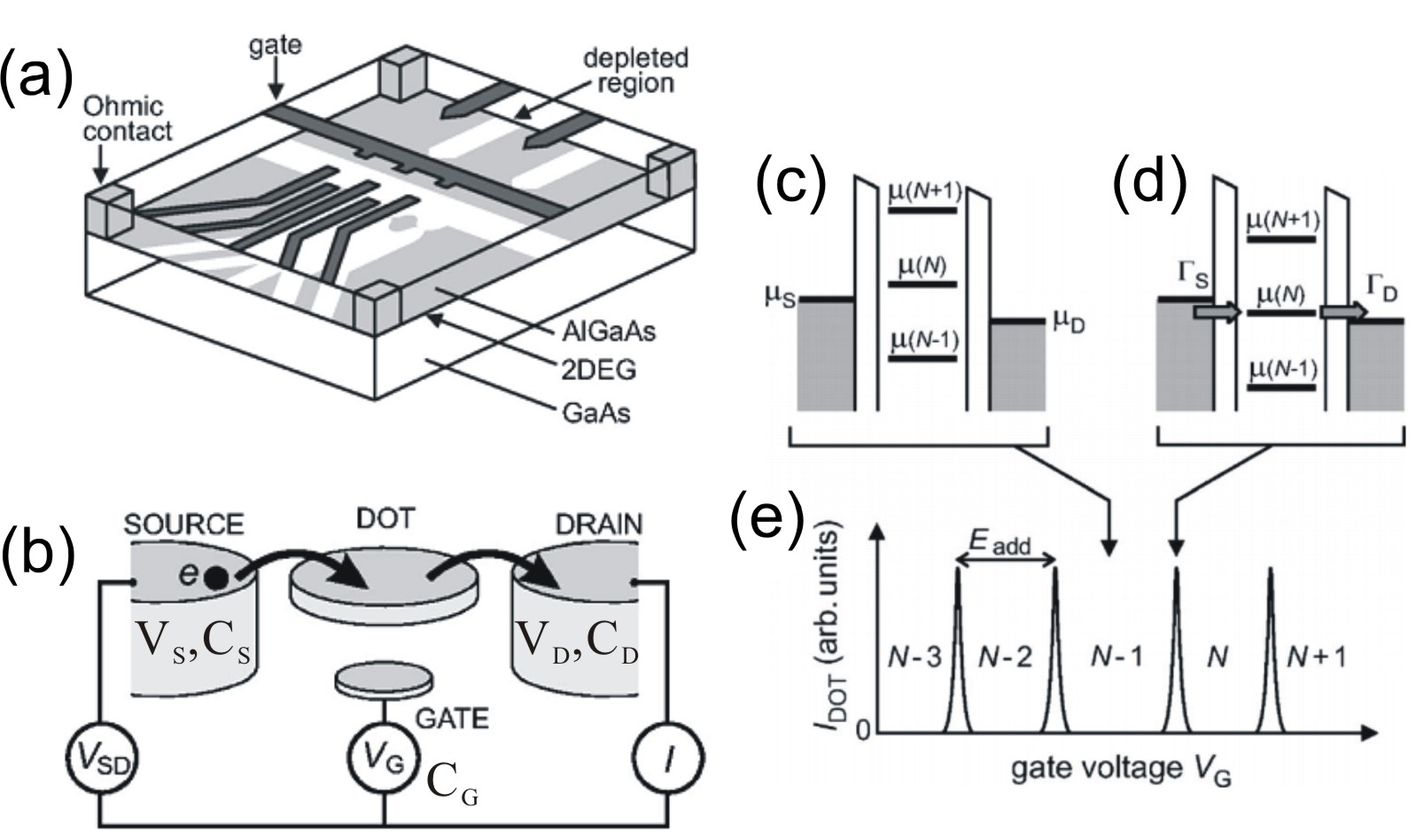}
\caption{(a) Schematic view of a lateral quantum dot device defined by metal surface electrodes on a GaAs/AlGaAs 2DEG system. (b) Electrical network diagram of a single quantum dot. (c) The electrochemical potential energies for a single dot with transport blocked due to Coulomb blockade. (d) With one of the electrochemical potentials lying within the bias window transport through the dot is then permitted. (e) Schematic plot of the current through a single dot against gate voltage showing periodic oscillations with the dot charge. Adapted with permission from ref. \cite{Hanson2007}. Copyright 2007 American Physical Society.}
\label{Figure 2.1}
\end{center}	
\end{figure}

A quantum dot is an artificially structured system that can be filled with only a few electrons or holes \cite{Hanson2007}. The charged carriers in such a system are generally confined in a submicron area, and the confinement potential in all directions is so strong that it gives rise to quantized energy levels that can be observed at low temperatures. The electronic properties of quantum dots are dominated by several effects \cite{Hanson2007}. First, the Coulomb repulsion between electrons on the dot leads to an energy cost called charging energy $E_C$ $=$ $e^{2}/C$, where $C$ is the total capacitance of the dot, for adding an extra electron to the dot. Because of this charging energy, the tunneling of electrons to or from the reservoirs can be suppressed at low temperatures (when $E_C$ $>$ $k_BT$), which leads to a phenomenon called Coulomb blockade. Second, the tunnel barrier resistance $R_{t}$, which describes the coupling of the dot to both the source and drain reservoirs, has to be sufficiently opaque such that the electrons are located either in the source, in the drain, or on the dot. 
The minimal $R_{t}$ can be estimated using the uncertainty principle, $\Delta E$$\cdot$$\Delta t$ $>$ $h$. From $\Delta E$ $=$ $e^{2}/C$ and $\Delta t$ $=$ $R_{t}C$, the condition $R_{t}$ $>$ $h/e^{2}$ for $R_{t}$ can be found. This means that the energy uncertainty corresponding to the tunneling time can not be greater than the charging energy; otherwise, it would lead to uncertainty in the number of carriers occupying the dot. Third, if the confinement in all three directions is strong enough for electrons residing on the dot to form quantized energy levels $E_{n}$ (often denoted as single-particle energy), the energy spacing $\Delta E$ $=$ $E_{n}$ $-$ $E_{n-1}$ can be observed on top of charging energy if $\Delta E$ $>$ $k_BT$. Because of this discrete energy spectrum $E_{n}$, quantum dots behave in many ways as artificial atoms. Fig. \ref{Figure 2.1}(a) shows an example of a quantum dot formed in a GaAs/AlGaAs 2DEG system, where the dot is defined by a gate-depleted area and is tunnel coupled to the reservoir on each side. Varying the voltages on the surface gates enables several important parameters, such as the number of electrons and the tunnel barrier resistance, to be finely tuned. To understand the dynamics of a single quantum dot, a constant interaction model has been proposed \cite{Hanson2007} and is illustrated in Fig. \ref{Figure 2.1}(b). The model is based on two assumptions. First, the Coulomb interactions among electrons in the dot, and between electrons in the dot and those in the environment, are parameterized by a single, constant capacitance $C$. This capacitance is the sum of the capacitance between the dot and the source $C_{S}$, the drain $C_{D}$ and the gate $C_{G}$: $C$=$C_{S}$+$C_{D}$+$C_{G}$. The second assumption is that the single-particle energy spectrum $E_{n}$ is independent of the Coulomb interaction, therefore of the number of electrons in the dot. Using this model, the total energy of a single dot with $N$ electrons in the ground state is given by \cite{Hanson2007}:
\begin{eqnarray}
U(N) &=&  \frac{\left(-\left|e\right|\left(N-N_0\right)+ C_S V_S + C_D V_D + C_G V_G\right)^2}{2C} + \sum_{n=1}^{N} E_n
\label{eqn 18}
\end{eqnarray}
where -$\left|e\right|$ is the electron charge, $N_0$ is the charge on the quantum dot due to the positive background charge of the donors and $V_S$, $V_D$ and $V_G$ are the voltages of the source, drain and gate respectively. The last term is a sum over the occupied single-particle energy levels $E_{n}$ which depend on the characteristics of the confinement potential.

The electrochemical potential of the dot $\mu \left(N\right)$ is defined as the energy needed to add the $N$-th electron to a dot with $N$-1 occupied electrons \cite{Hanson2007}:
\begin{eqnarray}
\mu \left(N\right) 	 &=& \nonumber U\left(N\right) - U\left(N-1\right) \\
&= & \left(N - N_0 - \frac{1}{2}\right)E_C - \frac{E_C}{e}\left(C_S V_S + C_D V_D + C_G V_G\right) + E_N
\label{eqn 19}
\end{eqnarray}
where 
\begin{eqnarray}
E_{C}=e^{2}/C
\label{eqn 48}
\end{eqnarray} is the charging energy. The addition energy is then given by the energy difference between two successive electrochemical potentials: 
\begin{equation}
E_{add}\left(N\right) = \mu \left(N+1\right) - \mu \left(N\right) \\
= E_C + \Delta E
\label{eqn 20}
\end{equation} 
where $\Delta E = E_{N+1} - E_N$ is the single-particle energy spacing, and is independent of the electron number on the dot (the second assumption).

When the temperature is low enough ($k_{B}$$T \ll \Delta E, E_C$), the transport through the quantum dot depends on whether the dot electrochemical potentials align with bias window, which is defined as the spacing between the electrochemical potentials of the source and drain, i.e., $-eV_{SD}$ $\equiv$ $\mu_S - \mu_D$ $=$ $-eV_{S}-(-eV_{D})$. In the low bias regime where $-eV_{SD} < E_C$, electron tunneling can only happen when the dot electrochemical potential lies in a small bias window, such that $\mu_D < \mu \left(N\right) < \mu_S$ as shown in Fig. \ref{Figure 2.1}(d). When the electrochemical potential is outside the bias window the transport is blocked and no current flows through the dot, which is the Coulomb blockade regime as shown in Fig. \ref{Figure 2.1}(c). When a gate $V_G$ constantly tunes the electrochemical potential of the quantum dot, an on-off current can be observed as peaks with constant spacing ($E_{add}$) between each other as shown in Fig. \ref{Figure 2.1}(e). Each current forbidden regime corresponds to a different electron number on the dot, so in this way the number of electrons on the dot can be varied.

\begin{figure}[!]
\begin{center}
		\includegraphics[scale=0.7]{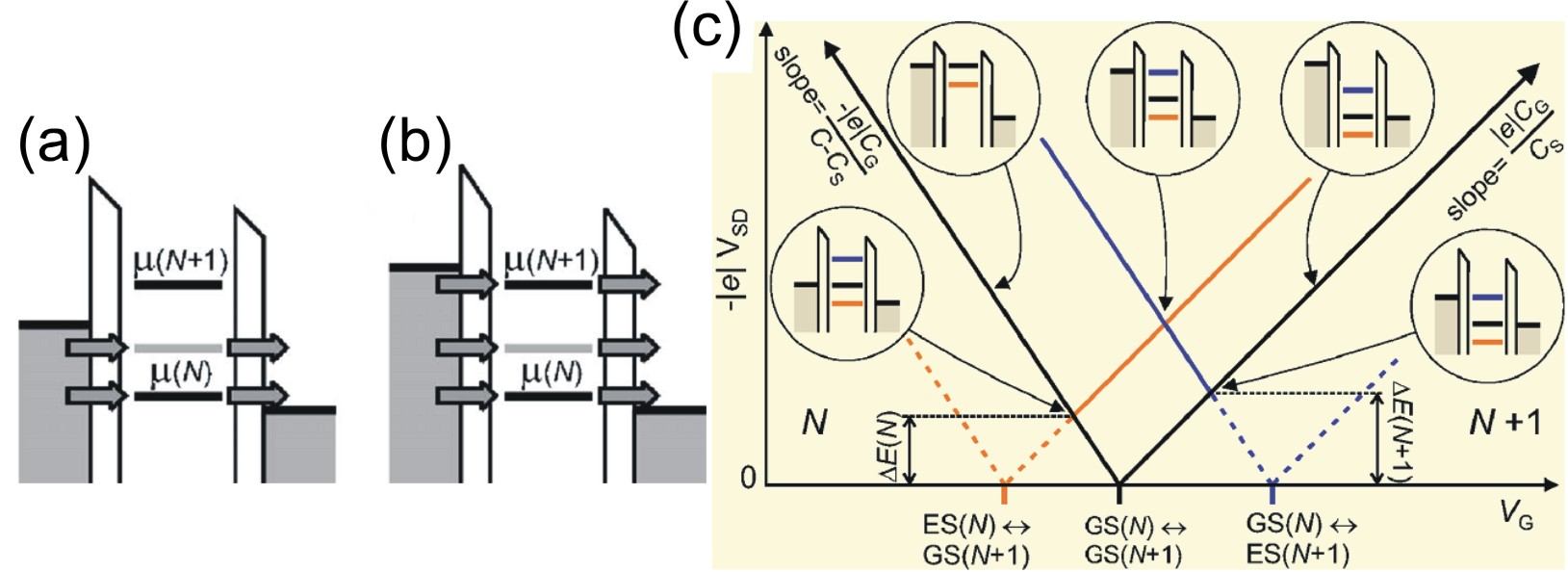}
\caption{ Schematic diagrams of the electrochemical potential levels of a quantum dot in the high bias regime. (a) $V_{SD}$ exceeds $\Delta$E so the electron transport couple to an excited state (the grey level). (b) $V_{SD}$ exceeds the addition energy so the electron transport couples to two successive ground state levels. (c) Differential conductance $dI_{DOT}/dV_{SD}$ through a quantum dot as a function of gate $V_{G}$ and bias voltage -$\left|e\right|V_{SD}$. The insets show different configurations of the dot level with respect to the lead potential in the $V_{SD}-V_{G}$ plane. Adapted with permission from ref. \cite{Hanson2007}. Copyright 2007 American Physical Society.}
\label{Figure 2.2}
\end{center}	
\end{figure}

In the high bias regime where $-eV_{SD} > \Delta E$ and/or $-eV_{SD} > E_{add}$, more dot levels are allowed to lie within the bias window and give rise to multiple tunneling paths as shown in Fig. \ref{Figure 2.2}(a) and (b). Depending on how wide the bias window is, the transition can involve a ground sate and its excited state as shown in Fig. \ref{Figure 2.2}(a), or in an even wider window ($-eV_{SD} >E_{add}$) it can couple to two successive ground states as shown in Fig. \ref{Figure 2.2}(b). From Eq. (\ref{eqn 19}) the electrochemical potential is a function of $V_S$, $V_D$ and $V_G$. Since $\mu_{S(D)}=-eV_{S(D)}$, if we measure the conductance of dot as a function of bias $eV_{SD}$ and gate voltage $V_G$ a spectrum called "Coulomb diamond" is formed as shown in Fig. \ref{Figure 2.2}(c). Since larger biases require a wider spacing in gate voltage for dot levels being pulled out of the window, the V-shape feature can be expected. In Fig. \ref{Figure 2.2}(c) along the left (right) edge of the black V-shape following the slope at $\frac{-\left|e\right|C_{G}}{C-C_{S}}$ ($\frac{\left|e\right|C_{G}}{C_{S}}$), the level of the $N$-electron ground state is aligned with the source (drain) level while the bias window is becoming wider. The black V-shape shows the transition between the $N$-electron ground state and $N+1$-electron ground state, and defines the regimes of blockade (outside the V-shape) and tunneling (within the V-shape). The orange and blue V-shapes shown in Fig. \ref{Figure 2.2}(c) correspond to two different transitions between the dot states which are, the $N$-electon excited state to $N+1$-electron ground state ($ES(N)$$\rightarrow$$GS(N+1)$) and the $N$-electon ground state to $N+1$-electron excited state ($GS(N)$$\rightarrow$$ES(N+1)$). Since the excited state energy $ES(N)$ and $ES(N+1)$ are separated from the gorund states $GS(N)$ and $GS(N+1)$ by $\Delta$$E(N)$ and $\Delta$$E(N+1)$ respectively [see Fig. \ref{Figure 2.2}(c)], Coulomb diamond measurements are very useful for studying the excited state spectroscopy in a quantum dot system. The insets shown in Fig. \ref{Figure 2.2}(c) represent a different configuration of dot levels with respect to the source-drain level. Note that the $ES(N)$$\rightarrow$$GS(N+1)$ and $GS(N)$$\rightarrow$$ES(N+1)$ transitions are forbidden outside the black V-shape as $ES(N)$ and $ES(N+1)$ states only exist when the $GS(N)$$\rightarrow$$GS(N+1)$ transition is within the bias window. Finally, the dimension of the Coulomb diamond (current-suppressed region) in the bias direction is a direct measure of $E_{add}$ or the charging energy $E_C$, because beyond the edge of the diamond the bias window is greater than $E_{add}$ and transport is no longer blocked.

\begin{figure}
\begin{center}
		\includegraphics[scale=0.74]{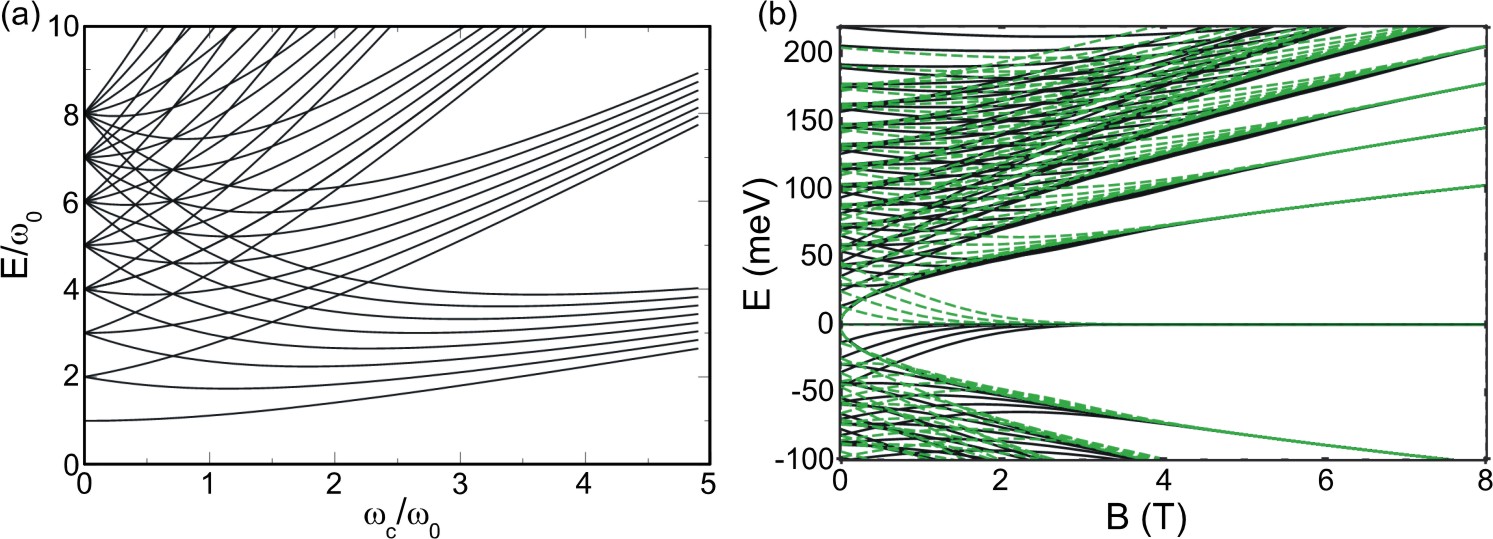}
\caption{(a) Fock-Darwin spectrum of a 2DEG symmetric quantum dot up to a quantum number of $n$ $=$ 7. (b) Energy spectrum of a graphene quantum dot with $R$ $=$ 70 nm for $m$ $=$ -4,...,4 and $n$ $=$ 1,...,6. The energy states for $\tau = +$1 are drawn as solid black lines, and those for $\tau = -$1 are drawn as dashed green lines. (b) adapted with permission from ref. \cite{Schnez2008}. Copyright 2008 American Physical Society.}
\label{Figure 36}
\end{center}	
\end{figure}

Here, we discuss the effect exerted by a magnetic field on the single-particle energy of QDs. The energy spectrum of a 2DEG quantum dot in the presence of a magnetic field is typically solved using a single-particle approximation with a parabolic confinement potential \cite{Fock1928,Kouwenhoven2001}. Such a spectrum is called the Fock-Darwin diagram which describes how 0D levels evolve with respect to an applied perpendicular magnetic field. The symmetric parabolic potential can be approximated as $U(x,y) = \frac{m^\ast}{2}\omega^{2}_{0}(x^2+y^2)$, where $m^\ast$ is the effective mass and $\omega^{2}_{0}$ denotes the strength of the confinement potential. Thus, the Hamiltonian of an electron in the dot can be written as follows: 

\begin{eqnarray}
H = \frac{1}{2m^\ast}(\textbf{p}+e\textbf{A})^2 + \frac{m^\ast}{2}\omega^{2}_{0}(x^2+y^2)
\label{eqn 46}
\end{eqnarray} If we choose the symmetric gauge for the vector potential $\textbf{A}$ $=$ ($-By/2$, $Bx/2$, 0), then the energy spectrum of the Hamiltonian can be solved as follows: 
\begin{eqnarray}
E_{n_+,n_-} = (n_{+} + 1)\hbar\Omega + \frac{1}{2}\hbar\omega_cn_-
\label{eqn 47}
\end{eqnarray}with $\Omega^2$ $\equiv$ $\omega^2_0 + \frac{\omega^2_c}{4}$ where $\omega_c$ $=$ $\frac{\left|eB\right|}{m^\ast}$ is the cyclotron frequency, and with quantum numbers $n_{\pm}$ $=$ $n_x$ $\pm$ $n_y$ where $n_x$, $n_y$ $=$ 0, 1, 2, ..., etc. This spectrum is plotted in Fig. \ref{Figure 36}(a). For $B = 0$ the spectrum has a constant level spacing and is simply the spectrum of the two-dimensional harmonic oscillator. In the high-field limit, the spectrum goes over into that of the Landau levels [see Fig. \ref{Figure 1.2}(a)], with the confinement effects of the dot playing an ever-decreasing role. 

In a graphene quantum dot, the Fock-Darwin spectrum is notably different from that in the 2DEG case owing to the existence of a Landau level (LL) at zero energy, which does not shift in energy with increasing magnetic field \cite{Schnez2008,Recher2009}. Together with quantum confinement, the unique linear band dispersion of graphene results in an electron-hole crossover in GQD's magneto-transport \cite{Guttinger2009,Chiu2012}. To solve the Fock-Darwin spectrum for a graphene quantum dot, we start from a free Dirac equation with a circular confinement potential $V(r)$ and include a perpendicular magnetic field, where the symmetric gauge $A$ $=$ $\frac{B}{2}(-y, x, 0)$ $=$ $\frac{B}{2}$($-r$sin$\phi$, $r$cos$\phi$, 0) for the vector potential is used ($\phi$ is the polar angle). Thus, the Hamiltonian now reads (ignoring spin) \cite{Schnez2008}:  

\begin{eqnarray}
H = \nu_F(\textbf{p} + e\textbf{A})\cdot\vec{\sigma} + \tau V(r)\sigma_z,
\label{eqn 48}
\end{eqnarray} where $\vec{\sigma} = (\sigma_x, \sigma_y)$ represents Pauli's matrices and $\tau$ $=$ $\pm$1 is the valley index for $\pm K$. Note that the quantum confinement effect is introduced in the Hamiltonian $via$ a mass-related potential $V(r)$ coupling to the $\sigma_z$ Pauli matrix. We let the mass in the dot to be zero, i.e., $V(r) = 0$ for $r < R$, but let it tend toward infinity at the edge of the dot, i.e., $V(r) = \infty$ for $r > R$. In this way, charge carriers are confined inside the quantum dot which has a radius of $R$. This leads to a boundary condition, which yields the simple relation that $\frac{\psi_2}{\psi_1}$ $=$ $\tau i$ exp[$i\phi$] for circular confinement \cite{Schnez2008}, where $\psi$ $=$ ($\psi_1, \psi_2$) is the eigenfunction of the Hamiltonian. Hence, in the following, we can set $V(r) = 0$; thus, the energy $E$ is related to the wavevector $k$ $via$ $E = \hbar \nu_F k$, and we can determine $k$ using the boundary condition. Following ref. \cite{Schnez2008}, the implicit equation for determining the wavevector $k$ (and therefore the energy $E$) that satisfies the boundary condition is given by:

\begin{eqnarray}
\left(1-\tau\frac{kl_B}{R/l_B}\right)L\left(\frac{k^2l^2_B}{2}-(m+1), m, \frac{R^2}{2l^2_B}\right)+L\left(\frac{k^2l^2_B}{2}-(m+2), m+1, \frac{R^2}{2l^2_B}\right) = 0,
\label{eqn 49}
\end{eqnarray} where $l_B$ $=$ $\sqrt{\hbar/eB}$ is the magnetic length and $m$ is the angular momentum quantum number. The functions $L(a, b, x)$ are generalized Laguerre polynomials, which are oscillatory functions. Hence, there are an infinite number of wavevectors $k_n$ for a given $B$, $m$, and $\tau$ that fulfill Eq. (\ref{eqn 49}). This condition defines the radial quantum number $n$, from which the energy spectrum $E (n, m, \tau)$ $=$ $\hbar \nu_F k_n$ can be plotted, as shown in Fig. \ref{Figure 36}(b) for a QD of radius R $=$ 70 nm. Note that $-E (n, m, \tau)$ $=$ $E (n, m, -\tau)$, which gives rise to the electron-hole symmetry in the spectrum. We discuss Eq. (\ref{eqn 49}) under two particular limits. For $B$ $\rightarrow$ 0, Eq. (\ref{eqn 49}) can be written as follows: 

\begin{eqnarray}
\tau J_m(kR) = J_{m+1}(kR),
\label{eqn 50}
\end{eqnarray} where $J_m$ is Bessel function. This relation yields the single-particle energy spectrum and can be used to estimate the energy of the excited states on a graphene dot with $N$ confined charge carriers [$\Delta (N)$ = $\hbar \nu_F$/($d \sqrt{N}$), where $d$ is an effective dot diameter; see ref. \cite{Ponomarenko2008,Schnez2009}]. In addition, there is no state at zero energy under zero magnetic field, which leads to an energy gap separating the states of negative and positive energies. By contrast, at high field, where $R/l_B$ $\rightarrow$ $\infty$, Eq. (\ref{eqn 49}) gives rise to the following:
\begin{eqnarray}
E_m = \hbar \nu_F k_m = \pm\nu_F\sqrt{2e\hbar B(m+1)}
\label{eqn 51}
\end{eqnarray} which are the Landau levels for graphene. Therefore, as the B-field increases, there will be a transition governed by the parameter $R/l_B$, from a regime in which the confinement play an important role ($R \leq l_B$) to the Landau-level regime ($R \geq l_B$). Note that the resonances on both sides of the electron-hole crossover have opposite slopes and merge into the zeroth Landau level. An experimental observation of this effect would constitute clear identification of this crossover, as will be presented in section 3.2.2.

\subsection{2.2. Double quantum dot}
When two single dots are placed in series and seperately connected to a source and drain reservoir, a double quantum dot (DQD) with a network of source-dot-dot-drain is formed. To apply the constant interaction model in such a system \cite{Wiel2002}, a schematic diagram of its equivalent electrical network is shown in Fig. \ref{Figure 2.3}(a). In this model, the dots QD1 (QD2) are capacitatively coupled to their nearest plunger gate PG1 (PG2) via a capacitance $C_{g1}$ ($C_{g2}$), however, they are also coupled to the further gate PG2 (PG1) through the cross capacitance $C_{g21}$ ($C_{g12}$). The dots themselves also couple to each other through an interdot capacitance $C_{m}$ and to the source and drain reservoir through $C_{S}$ and $C_{D}$ individually. The voltages applied to plunger gate 1, plunger gate 2, source and drain are denoted by $V_{PG1}$, $V_{PG2}$, $V_{S}$ and $V_{D}$ respectively as shown in Fig. \ref{Figure 2.3}(a). The charge and its equivalent voltage on QD1 (QD2) are denoted by $Q_{1(2)}$ and $V_{1(2)}$, also shown in Fig. \ref{Figure 2.3}(a). Based on this model the charge at each dot is given by the vector $\bf{Q} = \bf{CV}$ where $\bf{C}$ is the capacitance matrix, $\bf{Q}$=($Q_{1}$, $Q_{2}$) is the vector of charges and $\bf{V}$=($V_{1}$, $V_{2}$) is the vector of electrostatic potentials. Therefore the components of $\bf{Q}$ are given by \cite{Wiel2002}

\begin{equation}
\left( \begin{array}{cc}
\ Q_{1}+C_{S}V_{S}+C_{g1}V_{g1}+C_{g21}V_{g2} \\
\ Q_{2}+C_{D}V_{D}+C_{g2}V_{g2}+C_{g12}V_{g1}  \\ \end{array} \right)\ =
\left( \begin{array}{cc}
\ C_{1} & -C_{m} \\
\ -C_{m} & C_{2}  \\ \end{array} \right)\
\left( \begin{array}{cc}
\ V_1 \\
\ V_2  \\ \end{array} \right)\
\label{eqn 20}
\end{equation}, where $C_{1(2)}$ = $C_{S(D)}+C_{g1(2)}+C_{g21(12)}+C_{m}$ is the total capacitance of dot 1(2). Making the substitution $Q_{1(2)}=-N_{1(2)}e$, and taking $V_{S}=V_{D}=0$ [in the low bias regime and $N_{1(2)}$ is the electron number in dot 1(2)], Eq. (\ref{eqn 20}) then reads \cite{Wiel2002}:

\begin{figure}[!]
\begin{center}
		\includegraphics[scale=0.60]{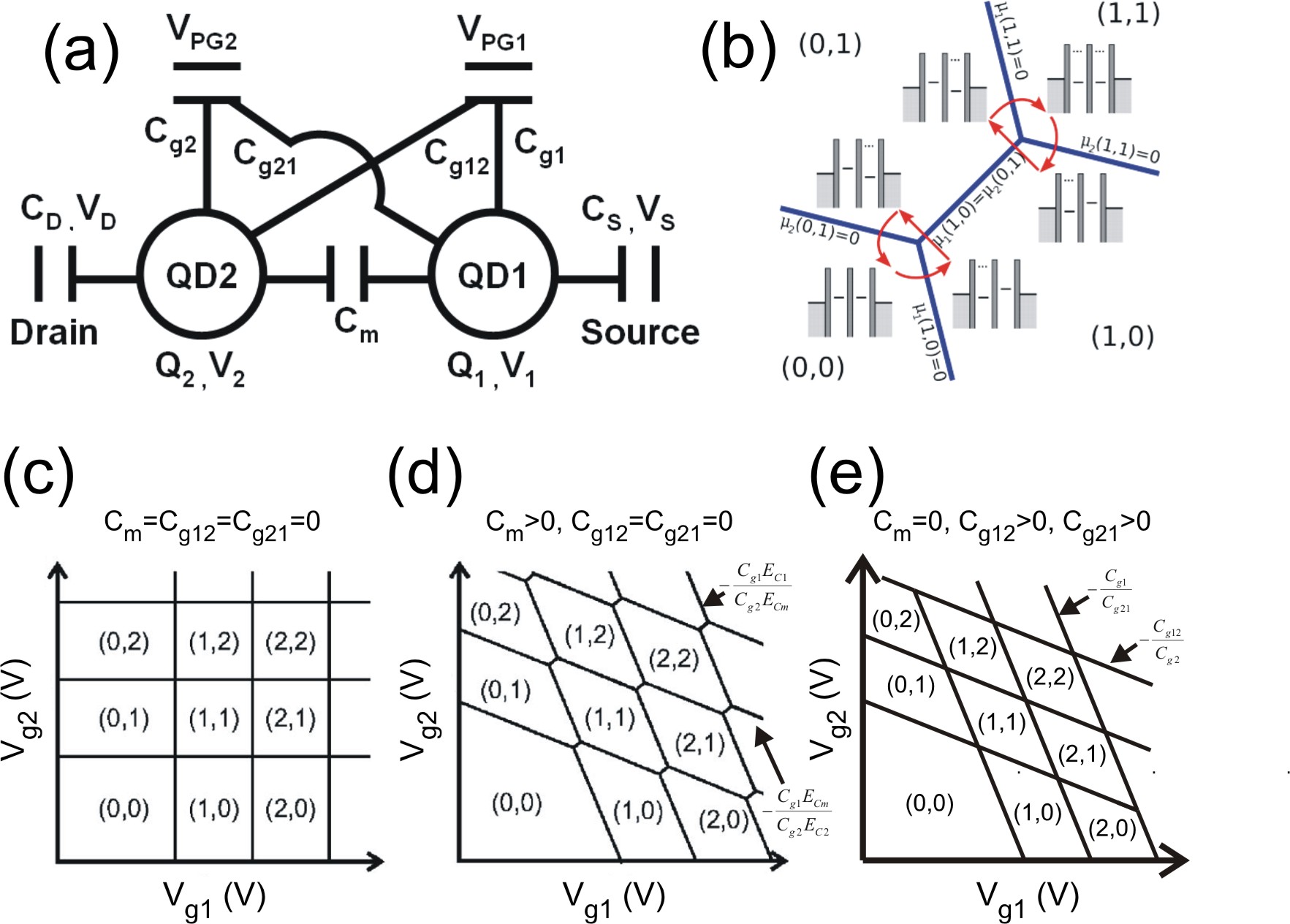}
		\end{center}		
\caption{(a) Electrostatic network model for a double quantum dot considering the cross-capacitance coupling. (b) Charge evolution during conduction at the triple points. The outer numbers in brackets give the stable electron number for that region. The inserts shown are the energy levels of the stable regions at points close to the triple point. (c)-(e) Charge-stability diagrams for a double quantum dot with (c) no interdot and cross-capacitance coupling ($C_{g21}=C_{g21}=E_{Cm}=0$), (d) intermediate interdot coupling but no cross-capacitance coupling ($C_{g21}=C_{g12}=0$, $E_{Cm}\neq0$) and (e) no interdot coupling but intermediate cross-capacitance coupling ($C_{g21}\neq0,C_{g12}\neq0$, $E_{Cm}=0$). (c, d) adapted with permission from ref. \cite{Wiel2002}. Copyright 2003 American Physical Society.}
\label{Figure 2.3}
\end{figure}

\begin{equation}
\left( \begin{array}{cc}
\ V_1 \\
\ V_2  \\ \end{array} \right)\ =\frac{1}{C_{1}C_{2}-C^{2}_{m}} \left( \begin{array}{cc}
\ C_{2} & C_{m} \\
\ C_{m} & C_{1}  \\\end{array} \right)\ \left( \begin{array}{cc}
\ -N_{1}e+C_{g1}V_{g1}+C_{g21}V_{g2} \\
\ -N_{2}e+C_{g2}V_{g2}+C_{g12}V_{g1}  \\ \end{array} \right)\
\label{eqn 21}
\end{equation}

So the total electrostatic energy of such a system is given by \cite{Hanson2007},
\begin{eqnarray}
U\left(N_1,N_2\right) &=& \nonumber \frac{1}{2} \bf{Q}\cdot\bf{V}\\ 
&=& \nonumber \frac{1}{2}N^{2}_{1}E_{C1}+N_{1}N_{2}E_{Cm}+\frac{1}{2}N^{2}_{2}E_{C2}-\frac{1}{2e}C_{g1}V_{g1}N_{1}E_{C1}- \frac{1}{2e}C_{g21}V_{g2}N_{1}E_{C1} \\ \nonumber
&& -\frac{1}{2e}C_{g2}V_{g2}N_{1}E_{Cm}-\frac{1}{2e}C_{g12}V_{g1}N_{1}E_{Cm}-\frac{1}{2e}C_{g1}V_{g1}N_{2}E_{Cm} \\
&&  -\frac{1}{2e}C_{g21}V_{g2}N_{2}E_{Cm}-
\frac{1}{2e}C_{g2}V_{g2}N_{2}E_{C2}-\frac{1}{2e}C_{g12}V_{g1}N_{2}E_{C2}
\label{eqn 22}
\end{eqnarray}
where the charging energies for the dots $E_{C1}$ and $E_{C2}$ and the coupling energy $E_{Cm}$ are given by
\begin{eqnarray}
E_{C1} &=& \frac{e^2}{C_1} \frac{1}{1-\frac{C_{m}^2}{C_1 C_2}} \label{eqn 23}\\
E_{C2} &=& \frac{e^2}{C_2} \frac{1}{1-\frac{C_{m}^2}{C_1 C_2}} \label{eqn 24}\\
E_{Cm} &=& \frac{e^2}{C_m} \frac{1}{\frac{C_1 C_2}{C_{m}^2}-1} \label{eqn 25}
\end{eqnarray}

The electrostatic potentials for the dots are then given by \cite{Hanson2007},

\begin{eqnarray}
\mu_{1}\left(N_1,N_2\right) &=& \nonumber U\left(N_1,N_2\right) - U\left(N_1-1,N_2\right)\\
 &=& \nonumber \left(N_1 - 1\right)E_{C1}  + N_2 E_{Cm} \\ 
&&  - \frac{1}{2e}\left(C_{g1}V_{g1}E_{C1}  +  C_{g2}V_{g2}E_{Cm}+C_{g21}V_{g2}E_{C1}+ C_{g12}V_{g1}E_{Cm}\right) \label{eqn 26}\\
\mu_{2}\left(N_1,N_2\right) &=& \nonumber U\left(N_1,N_2\right) - U\left(N_1,N_2-1\right)\\
 &=& \nonumber \left(N_2 - 1\right)E_{C2}  + N_1 E_{Cm} \\ 
&&  - \frac{1}{2e}\left(C_{g1}V_{g1}E_{Cm}  +  C_{g2}V_{g2}E_{C2}+C_{g12}V_{g1}E_{C2}+ C_{g21}V_{g2}E_{Cm}\right) \label{eqn 27}
\end{eqnarray}

The physical meaning of each term, for example, $(N_{1}-1)E_{C1}$ and $N_{2}E_{Cm}$, stand for the Coulomb statistic energy increases on dot 1 and dot 2 when the $N_{1}$-th electron is added to dot 1. The term $C_{g1}V_{g1}E_{C1}$ in Eq. (\ref{eqn 26}) is the direct coupling energy between PG1 and QD1, while $C_{g2}V_{g2}E_{Cm}$ is the indirect coupling energy between PG2 and QD1 in which PG2 couples to QD2 first then QD2 influences QD1 through the interdot coupling. The last two terms in Eq. (\ref{eqn 26}) shows the cross-coupling effect where $C_{g21}V_{g2}E_{C1}$ is the cross-coupling energy between PG2 and QD1 and $C_{g12}V_{g1}E_{Cm}$ is the indirect cross-coupling energy that PG1 couples to QD2 first and QD2 influences QD1 through the interdot coupling. 
At low temperature ($k_{B}T$ $<$ $e^{2}/C$), the electrical transport through the DQD is only possible in the case where the energy levels in both dots are aligned with the source-drain bias window and this gives rise to the charge-stability diagram as shown in Fig. \ref{Figure 2.3}(b). The outer numbers in brackets ($N_{1}$, $N_{2}$) are the stable electron numbers residing in the dot for that region and the condition for electron transport is met whenever three charge states meet in one point (the so-called triple point). The arrows in Fig. \ref{Figure 2.3}(b) circling each triple point mark the route around the stability diagram that the system takes as electrons shuttle through. The counterclockwise path follows the sequence of charge state $(N_{1}, N_{2})\rightarrow(N_{1}+1, N_{2})\rightarrow(N_{1}, N_{2}+1)\rightarrow(N_{1}, N_{2})$, corresponding to moving an electron to the right. The clockwise path follows the sequence of charge state $(N_{1}+1, N_{2}+1)\rightarrow(N_{1}+1, N_{2})\rightarrow(N_{1}, N_{2}+1)\rightarrow(N_{1}+1, N_{2}+1)$, corresponding to moving a hole to the left. We here try to find a specific slope for $\mu_{1(2)}$ in the $V_{g1}$-$V_{g2}$ plane, along which $\mu_{1(2)}$ will remain constant for a given ($N_{1}$, $N_{2}$). We make the second row of Eq. (\ref{eqn 26}) and Eq. (\ref{eqn 27}) $=$ 0 which gives:

\begin{eqnarray}
V_{g1}\left(C_{g1}E_{C1}+C_{g12}E_{Cm}\right)=-V_{g2}\left(C_{g2}E_{Cm}+C_{g21}E_{C1}\right) \nonumber \\ \Rightarrow \frac{V_{g2}}{V_{g1}}=-\left(\frac{C_{g1}E_{C1}+C_{g12}E_{Cm}}{C_{g2}E_{Cm}+C_{g21}E_{C1}}\right),  (for \  \mu_{1}) \label{eqn 28} \\
V_{g2}\left(C_{g2}E_{C2}+C_{g21}E_{Cm}\right)=-V_{g1}\left(C_{g1}E_{Cm}+C_{g12}E_{C2}\right) \nonumber \\ \Rightarrow \frac{V_{g2}}{V_{g1}}=-\left(\frac{C_{g1}E_{Cm}+C_{g12}E_{C2}}{C_{g2}E_{C2}+C_{g21}E_{Cm}}\right),  (for \  \mu_{2}) \label{eqn 29}
\end{eqnarray}

We discuss the stability diagram for a double quantum dot with three different coupling regimes.
\begin{enumerate}
\item \textbf{No interdot and cross-capacitance coupling}\\
If we do not consider the cross-capacitance and interdot coupling (i.e., $C_{g12}=C_{g21}=E_{Cm}=0$), so that PG1 only influences QD1 and PG2 only influences QD2, Eq. (\ref{eqn 28}) and Eq. (\ref{eqn 29}) now read:
\begin{eqnarray} 
\frac{V_{g2}}{V_{g1}} &=& -\infty,  (for \  \mu_{1}) \label{eqn 30} \\
\frac{V_{g2}}{V_{g1}} &=& 0,  (for \  \mu_{2}) \label{eqn 31}
\end{eqnarray}
The resulting stability diagram is shown as Fig. \ref{Figure 2.3}(c) where the lines for $\mu_{1(2)}$ to stay constant appear as vertical (horizontal) lines.
\item \textbf{Finite interdot but no cross-capacitance coupling}\\
As the interdot coupling or the cross-capacitance coupling opens, the gate PG1(2) has the ability to influence QD2(1). We first consider the case that interdot coupling is finite but the cross-capacitance coupling is weak, so $C_{g21}=C_{g12}=0$, $E_{Cm}\neq0$. In such a case the only way that PG1(2) influences dot 2(1) is to influence dot 1(2) first and through interdot capacitance to tune the other dot indirectly. So now Eq. (\ref{eqn 28}) and Eq. (\ref{eqn 29}) read,
\begin{eqnarray} 
\frac{V_{g2}}{V_{g1}}=-\left(\frac{C_{g1}E_{C1}}{C_{g2}E_{Cm}}\right),  (for \  \mu_{1}) \label{eqn 32} \\
\frac{V_{g2}}{V_{g1}}=-\left(\frac{C_{g1}E_{Cm}}{C_{g2}E_{C2}}\right),  (for \  \mu_{2}) \label{eqn 33}
\end{eqnarray}
The resulting stability diagram is shown as Fig. \ref{Figure 2.3}(d). Instead of appearing as vertical (horizontal) lines, now $\mu_{1(2)}$ has a slope which is determined by the strength of the interdot coupling $E_{Cm}$. The larger $E_{Cm}$ is, the more $\mu_{1(2)}$ deviates from a vertical(horizontal) line. 
\item \textbf{No interdot but finite cross-capacitance coupling}\\
Finally, in the case of no interdot coupling but with cross-capacitance coupling, i.e., $C_{g12}\neq0$, $C_{g21}\neq0$, $E_{Cm}=0$, Eq. (\ref{eqn 28}) and Eq. (\ref{eqn 29}) read:
\begin{eqnarray} 
\frac{V_{g2}}{V_{g1}}=-\frac{C_{g1}}{C_{g21}},  (for \  \mu_{1}) \label{eqn 34} \\ 
\frac{V_{g2}}{V_{g1}}=-\frac{C_{g12}}{C_{g2}},  (for \  \mu_{2}) \label{eqn 35}
\end{eqnarray}
and the resulting stability diagram is shown in Fig. \ref{Figure 2.3}(e) where the slopes are now determined by the ratio between the direct capacitance $C_{g1(2)}$ and the cross-capacitance $C_{g21(12)}$.
\end{enumerate}

\begin{figure}[!]
\begin{center}
		\includegraphics[scale=0.68]{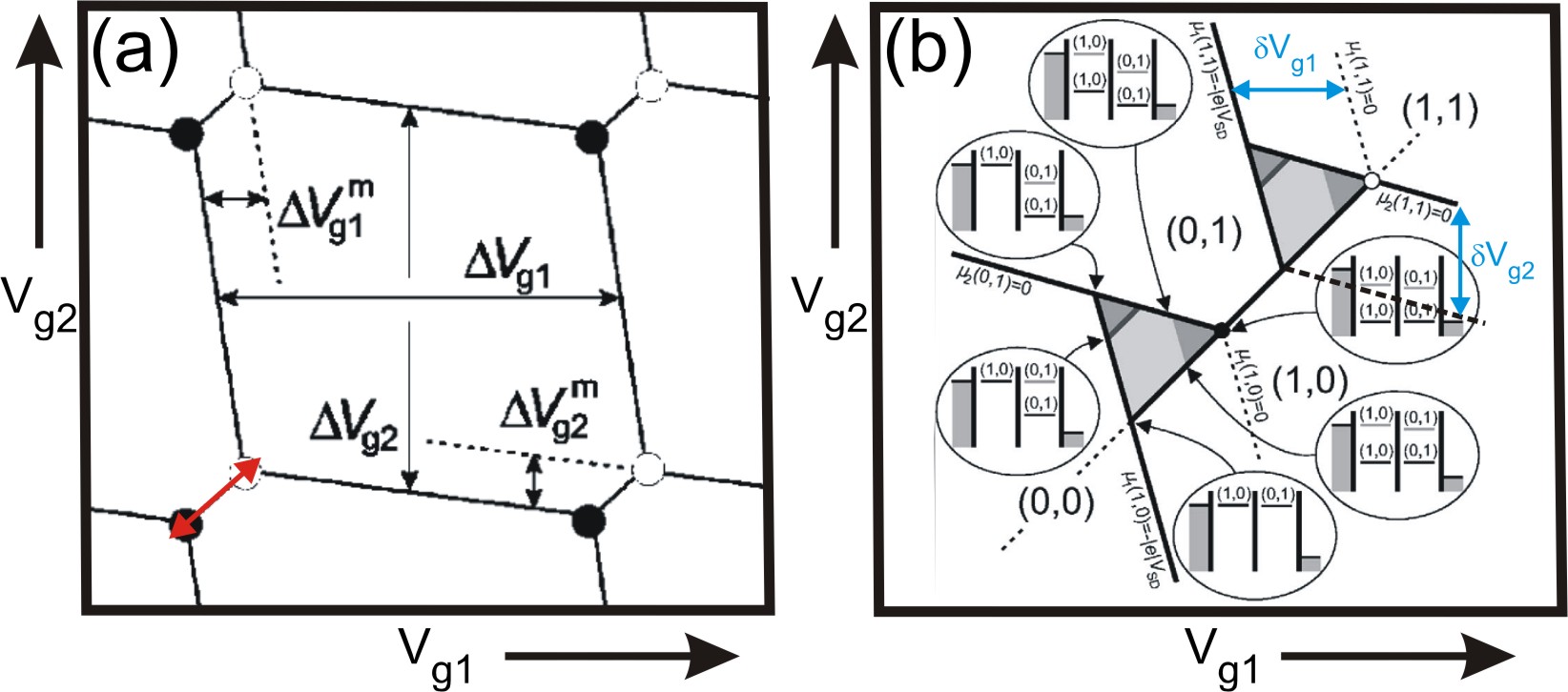}
\caption{(a) An enlarged stability diagram of Fig. \ref{Figure 2.3}(d), with the dimensions for one hexagonal region. (b) Schematic for transport through the double dot at the triple points in the high bias regime. The electrochemical potentials at different points of the triangular regions are also indicated. (a) adapted with permission from ref. \cite{Wiel2002}. Copyright 2003 American Physical Society. (b) adapted with permission from ref. \cite{Hanson2007}. Copyright 2007 American Physical Society.}
\label{Figure 2.4}
\end{center}	
\end{figure}

Usually a double-dot system has a finite interdot and weak cross-capacitance coupling strength so the charge-stability diagram is made up of hexagonal regions of a fixed charge, as shown in Fig. \ref{Figure 2.3}(d) [also an enlarged illustration in Fig. \ref{Figure 2.4}(a)]. The dimensions of the hexagonal regions as indicated in Fig. \ref{Figure 2.4}(a) are given by \cite{Wiel2002}:
\begin{eqnarray}
\Delta V_{g1} &=& e/C_{g1} \label{eqn 36} \\
\Delta V_{g2} &=& e/C_{g2} \label{eqn 37} \\
\Delta V_{g1}^m &=& \Delta V_{g1}\frac{C_m}{C_{2}} \label{eqn 38} \\ 
\Delta V_{g2}^m &=& \Delta V_{g2}\frac{C_m}{C_{1}} \label{eqn 39} 
\end{eqnarray} Note that the separation between two nearby triple points is determined by the interdot capacitance $C_{m}$; the larger the interdot coupling strength the greater the separation of the two triple points. We can also work out the slope of the line connecting two triple points which is highlighted by a red double-arrow in Fig. \ref{Figure 2.4}(a). Letting Eq. (\ref{eqn 28}) $=$ Eq. (\ref{eqn 29}) which means making $\mu_{1}=\mu_{2}$ in the $V_{g1}$-$V_{g2}$ plane, we have 

\begin{equation}
\frac{V_{g2}}{V_{g1}}=\frac{C_{g1}\left(E_{C1}-E_{Cm}\right)+C_{g12}\left(E_{Cm}-E_{C2}\right)}{C_{g2}\left(E_{C2}-E_{Cm}\right)+C_{g21}\left(E_{Cm}-E_{C1}\right)}
\label{eqn 40} 
\end{equation}

In the high bias regime, the triple points evolve into bias-dependent triangular regions where the two dot levels lie within the bias window as shown in Fig. \ref{Figure 2.4}(b). The dimensions of the triangles are related with the applied bias via \cite{Wiel2002}:
\begin{eqnarray}
\alpha_1 \delta V_{g1} & = & \frac{C_{g1}}{C_1}e \delta V_{g1} = |eV_{SD}| \label{eqn 42} \\
\alpha_2 \delta V_{g2} & = & \frac{C_{g2}}{C_2}e \delta V_{g2} = |eV_{SD}| \label{eqn 43}
\end{eqnarray}
Here $\alpha_{1(2)}$ is the conversion factor between gate voltage and energy which could be extracted from the dimension of the bias triangle. Therefore, the charging energy of the dots and the interdot coupling energy can be found:

\begin{eqnarray}
E_{C1} &=& e^{2}/C_{1} = \frac{\alpha_1e}{C_{g1}} = \alpha_1\Delta V_{g1} \label{eqn 44} \\
E_{C2} &=& e^{2}/C_{2} = \frac{\alpha_1e}{C_{g2}} = \alpha_2\Delta V_{g2} \label{eqn 45} \\
E_{Cm} &=& \alpha_1\Delta V^{m}_{g1} = \alpha_2\Delta V^{m}_{g2} \label{eqn 46}
\end{eqnarray}

\begin{figure}
\begin{center}
		\includegraphics[scale=0.67]{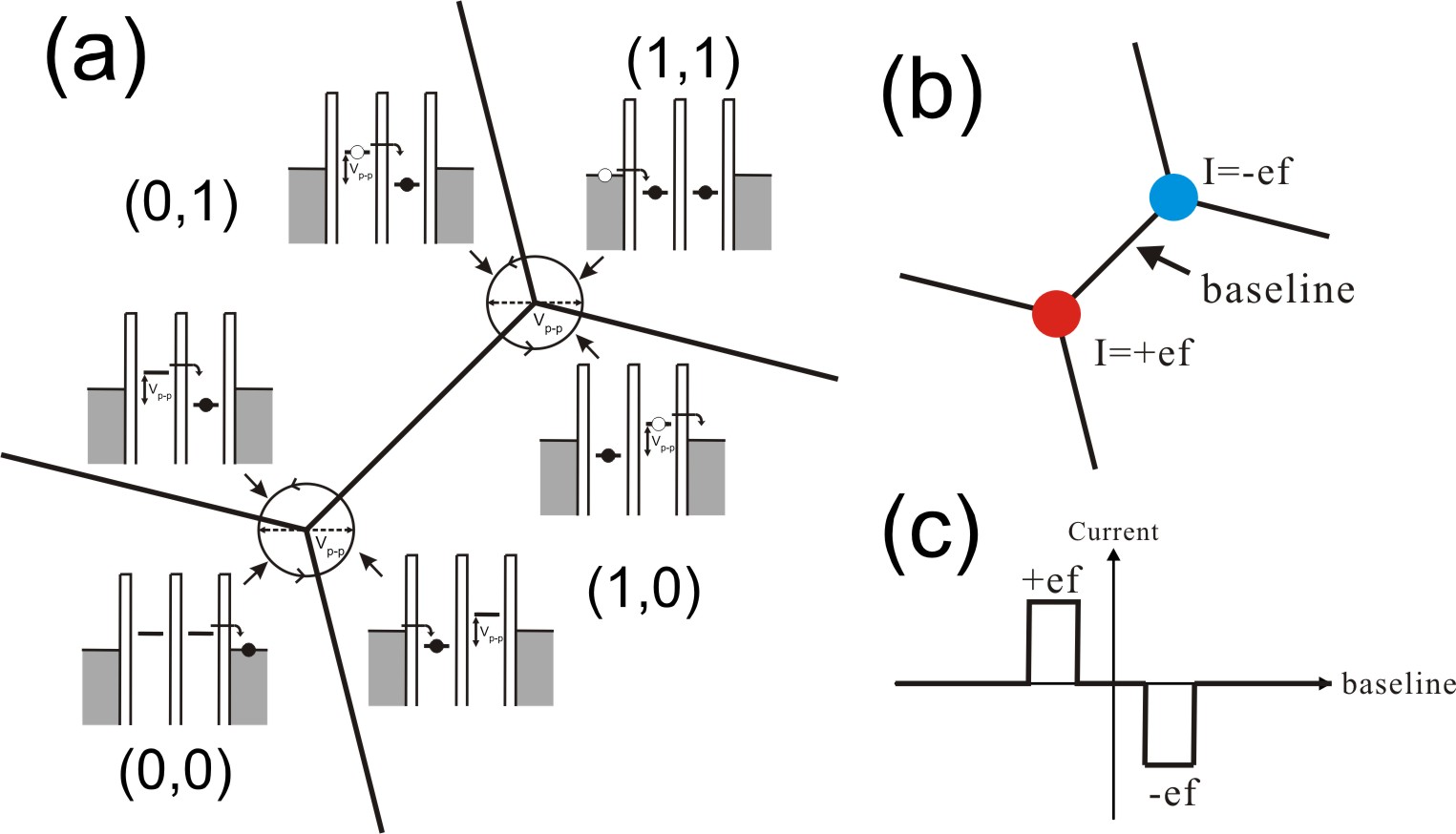}
\caption{(a) The pump loop (counterclockwise close trajectory) and its corresponding charge configuration obtained by modulating the gate voltages by two out-of-phase RF signals, induces one electron to go around the circuit. V$_{p-p}$/2 is the RF amplitude. (b) Schematic for the resulting current map from (a). If the pump is working, one should observe a quantized current $I$ = $\pm ef$ at the location of two nearby triple points. (c) Schematic for the linecut along the baseline in (b). Reproduced from K. L. Chiu, PhD Thesis, 2012.}
\label{Figure 2.5}
\end{center}	
\end{figure}

A phenomenon closely related to the manipulation of the dot levels in double quantum dots is charge pumping. Charge pumping refers to a quantized number $n$ of electrons transferred from the source to the drain at a driven frequency $f$, leading to a total current $I$ $\equiv$ $nef$ even if zero bias is applied ($e$ is the elementary charge). Such quantized charge transport was first demonstrated in single-electron turnstile devices in which an external radio-frequency (RF) signal was applied to linear arrays of tunnel junctions. By doing so electrons could be clocked through each tunnel junction one at a time by exploiting the Coulomb blockade effect \cite{Geerligs1990,Kouwenhoven1991}. When an RF signal is applied to the plunger gates (instead of barriers) of a double quantum dot device, it is also possible to generate an accurate and frequency-dependent quantized current through the device. The AC voltages on both plunger gates with a phase difference between them drives the DQD into different charge states around the triple point. The schematic diagram to illustrate such a pumping mechanism is shown in Fig. \ref{Figure 2.5}(a). Assuming the voltages applied on the plunger gates are AC sinusoidal waves with a phase difference of 90 degrees, it effectively forms a circular pump loop in the stability diagram. The radius of the circle is determined by the amplitude of the sinusoidal wave ($V_{p-p}$/2). When the circular route passes through three charge states around the triple point, it corresponds to shuttling a charge carrier from source reservoir to drain reservoir and generating a current. If the AC amplitude is small enough for the pump loop to just enclose a triple point and the frequency is large enough to produce a measurable pumping current, a current $I=ef$ will follow even when zero source-drain voltage is applied. Depending on the type of triple point that the pumping circle encloses, it generates a different direction of current; i.e., positive current for the electron-transport-type triple point and negative current for the hole-transport-type triple point. So if the pumping is successful the current recorded around two nearby triple points will present a circular shape with equal values but different signs as shown in Fig. \ref{Figure 2.5}(b). If we do a linecut along the baseline of triple points, the current appears as two plateaus as shown in Fig. \ref{Figure 2.5}(c).

In summary, we have introduced the quantum dot physics relevant to the transport properties of nanostructures. In the following chapters, we will begin to review the experiments that have been performed on nanostructures made of various 2D materials.   

\maketitle
\section{3. Graphene nanostructures on SiO$_{2}$/Si substrates}
In this chapter, we will review the early development of graphene nanostructures fabricated on SiO$_{2}$/Si substrates. After briefly introducing graphene nanoribbons and their function as tunnel barriers, we will focus mainly on graphene quantum dots and their transport properties.  
\subsection{3.1. Graphene Nanoribbons}

\begin{figure}
\begin{center}
		\includegraphics[scale=0.7]{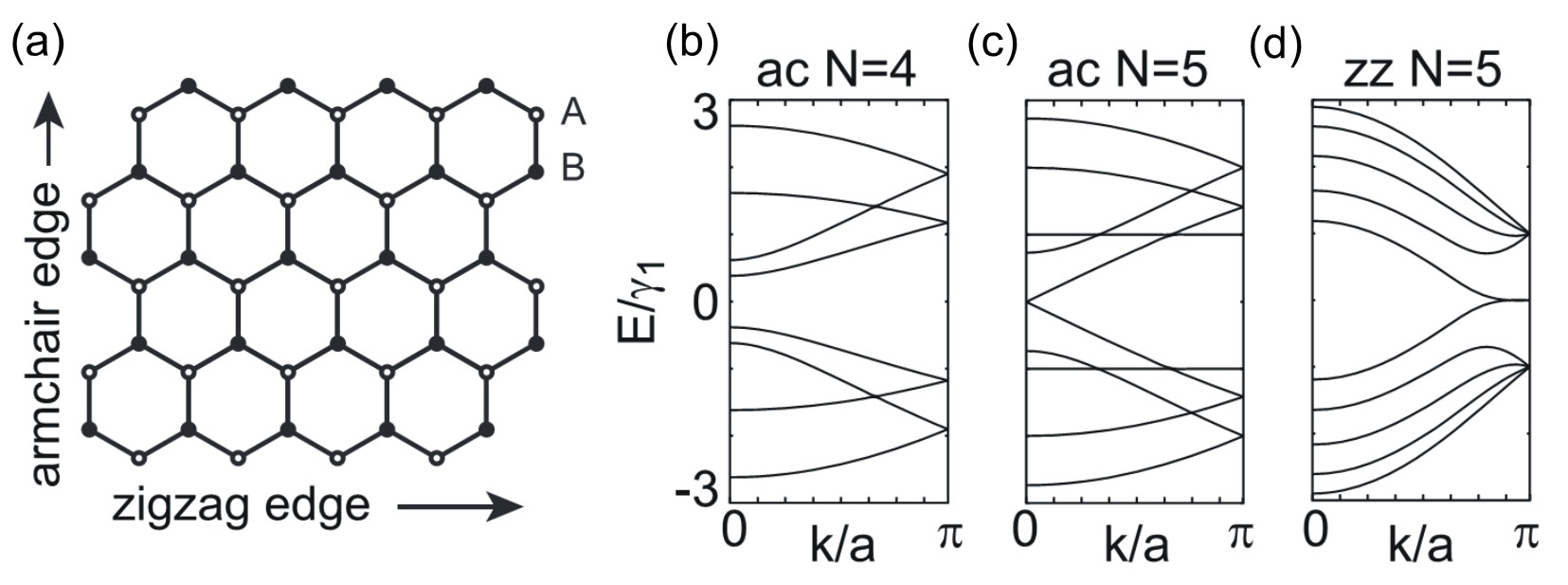}
\caption{(a) Lattice of a zigzag (armchair) graphene nanoribbon by extension in the x-(y-) direction. (b-d) Tight-binding calculations of the nanoribbon subbands for (b) an $N$ = 4 semiconducting armchair (ac) nanoribbon, (c) an $N$ = 5 metallic armchair (ac) nanoribbon and (d) an $N$ = 5 metallic zigzag (zz) nanoribbon. $N$ denotes the number of the dimer (carbon-site pair) lines for the armchair ribbon and the number of the zigzag lines for the zigzag ribbon, respectively. Adapted with permission from ref. \cite{Guttinger2012}. Copyright 2012 IOP Publishing Ltd.}
\label{Figure 37}
\end{center}	
\end{figure}

Although graphene is a superb conductor which offers advantages in terms of sensing and analog electronics, its gapless bandstructure hinders its use in logic circuit applications. Owing to the absence of a band gap, the current in graphene cannot be completely turned off, leading to low on/off ratios that are insufficient for switches \cite{Geim2007}. Engineering band gaps in graphene is thus a major challenge that must be addressed to enable the use of graphene-based transistors in digital electronics. First-principle calculations predict that cutting graphene into one-dimensional nanoribbons can open up a scalable band gap $E_{g}$ = $\alpha$/$w$, where $w$ is the nanoribbon width and $\alpha$ is in the range of 0.2 eV$\cdot$nm to 1.5 eV$\cdot$nm, depending on the model and the crystallographic orientation of the edges \cite{Lin2008a,Stampfer2011}. Similar results are also obtained from tight-binding calculations \cite{Nakada1996,Wakabayashi1999}. A GNR can have two possible types of edge terminations, namely, armchair and zigzag edges, as shown in Fig. \ref{Figure 37}(a). These two edge types correspond to different boundary conditions, from which the energy band dispersion can be found. The tight-binding calculated energy band structures for armchair GNRs (of two different ribbon widths) and zigzag GNRs are shown in Fig. \ref{Figure 37}(b) - (d), where $N$ denotes the number of dimer (carbon-site pair) lines (for the armchair ribbons) or the number of zigzag lines (for the zigzag ribbons). The band dispersion for an armchair nanoribbon with $N$ = 3$m$ $-$ 2 dimers exhibits a band gap (semiconducting), whereas for an armchair nanoribbon with $N$ = 3$m$ $-$ 1 dimers, the dispersion is metallic ($m$ is an integer). For semiconducting ribbons, the direct gap decreases with increasing ribbon width and tends toward zero in the limit of very large $N$. Zigzag nanoribbons always exhibit metallic behavior [Fig. \ref{Figure 37}(d)] regardless of how the width ($N$) is varied. The predicted existence of band gaps in GNRs has motivated an experimental effort to establish whether nanostructuring graphene is a feasible route for preparing graphene-based switches \cite{Han2007,Todd2008,Molitor2009a,Bai2010,Connolly2011,Wang2011,Jiao2010,Wei2013b}. GNRs can be fabricated by means of O$_{2}$ plasma etching using physical masks \cite{Han2007,Todd2008,Molitor2009a,Bai2010,Connolly2011}, unzipping carbon nanotubes \cite{Wang2011,Jiao2010,Wei2013b}, gas phase etching \cite{Wang2010a} or functionalization \cite{Withers2011,Lee2011}. Such devices have been tested for their transport properties at various temperatures, and the general results will be discussed below.

\begin{figure*} 
\includegraphics[scale=0.9]{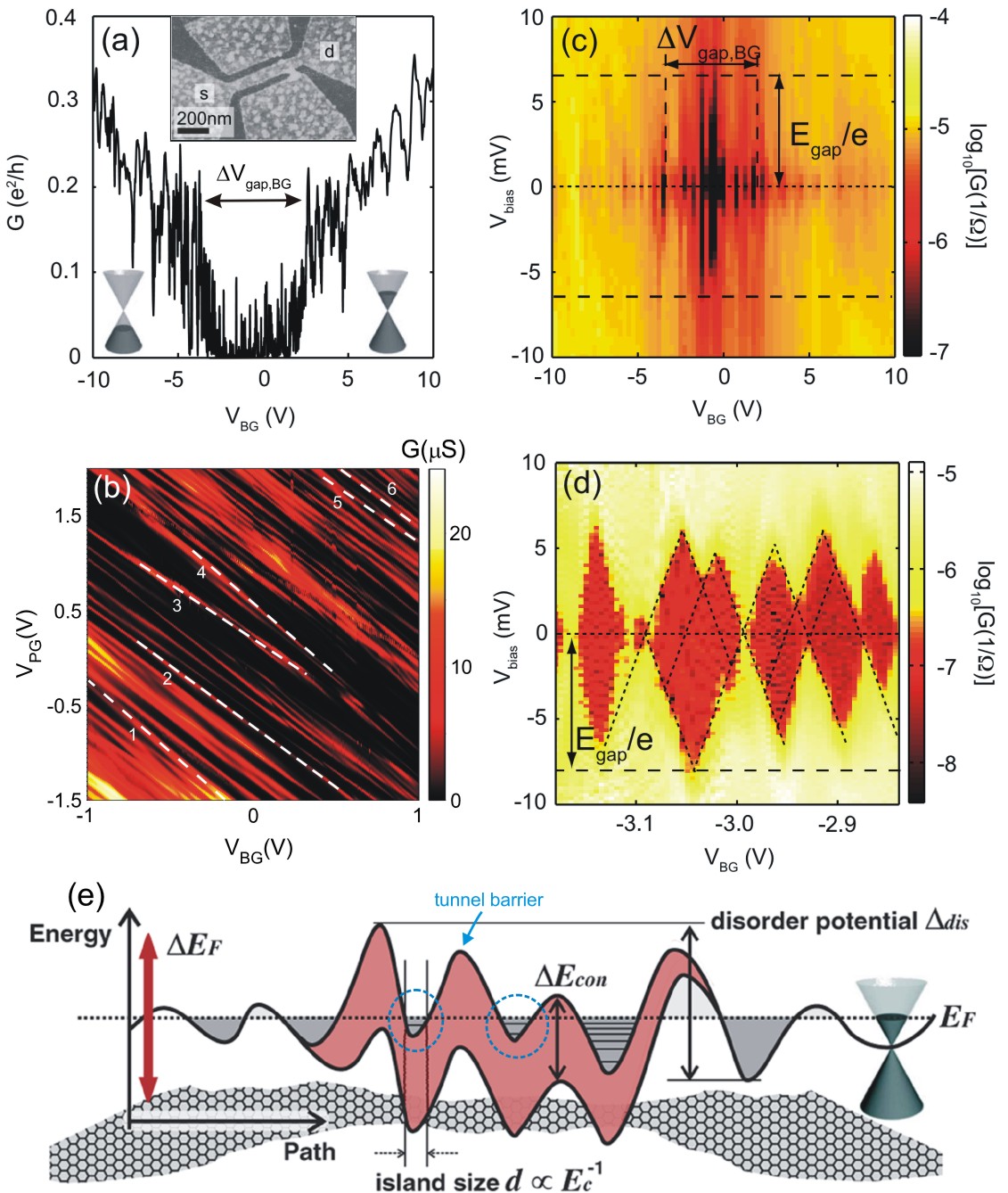}
\caption{(a) Conductance through a nanoribbon (shown in the insert) as a function of the back-gate voltage $V_{BG}$, recorded under an applied bias voltage of $V_{bias}$ = 300 $\mu$V and at a temperature of $T$= 2 K. Insert: Atomic force microscopy image of a graphene nanoribbon ($w$ = 85 nm, $l$ = 500 nm) etched using O$_{2}$ plasma. (b) Conductance as a function of $V_{BG}$ and the plunger-gate voltage $V_{PG}$ of another GNR device, showing a variation in the slopes of the Coulomb resonances (indicated by dashed lines and numbers). (c) Logarithmic conductance as a function of $V_{bias}$ and $V_{BG}$ at $T$= 2 K for the device shown in (a), with indication of the extent of the transport gap $\Delta V_{gap,BG}$ in the back-gate direction and bias-gap $E_{gap}$ in the bias direction. (d) Zoom-in of the region of suppressed conductance depicted in (c). (e) Schematic illustration of the formation of localized states induced by disorders. $\Delta_{dis}$ characterizes the strength of the charge-neutrality point fluctuation and $\Delta E_{con}$ is a confinement gap induced by local constriction. $\Delta E_{F}$ denotes the Fermi energy spacing that the transport gap has to overcome. (a, c, d) adapted with permission from ref. \cite{Molitor2009a}. (b) reproduced from K. L. Chiu, PhD Thesis, 2012. (e) adapted with permission from ref. \cite{Stampfer2009}. Copyright 2009 American Physical Society.}  
\label{Fig1}
\end{figure*}

Fig. \ref{Fig1}(a) shows the conductance of an O$_{2}$ plasma etched GNR [inset of Fig. \ref{Fig1}(a)] as a function of the voltage applied to the back-gate. This back-gate sweep shows a typical V-shape, with a region around 0 V separating the hole- from electron-transport regime where the conductance is strongly suppressed. In contrast to the prediction of energy gaps in clean GNRs (i.e., without considering bulk disorder and edge roughness), where transport should be completely pinched-off, this gap exhibits a large number of conductance peaks reminiscent of Coulomb blockade resonances in quantum dots. The nature of these resonances can be interrogated by varying the potential of the GNR. Fig. \ref{Fig1}(b) shows the conductance as a function of both back-gate and plunger-gate (an in-plane gate close to the GNR) voltages within the transport gap. The conductance resonances exhibiting a range of relative lever arms indicated by dashed lines are present over a wide range of $V_{BG}$ and $V_{PG}$ voltages. One explanation for this behavior draws on its similarity to a series of charge islands (or QDs), each coupled to the plunger-gate through different capacitive coupling strength, assuming the lever arm of the back-gate to the charge islands is nearly constant all over the GNR. More information about such localized states in the GNR can be gleaned by the Coulomb diamond measurements (see section 2.1), in which the differential conductance as a function of back-gate voltage and source-drain bias is recorded, as shown in Fig. \ref{Fig1}(c). Within this picture, the extent in bias voltage of the diamond-shaped regions of suppressed current [see $E_{gap}$/$e$ in Fig. \ref{Fig1}(c) and its zoom-in in Fig. \ref{Fig1}(d)] is a direct indication of the charging energy of the dots (see section 2.1), which fluctuates strongly with $V_{BG}$ and extends to $\approx$ 8.5 meV. The overlapping diamonds in Fig. \ref{Fig1}(d) resembles the behavior of a QD network \cite{Dorn2004}, supporting the notion that multiple QDs form along the GNR. In addition, the gap in Fermi energy $\Delta E_{F}$ corresponding to the transport gap $\Delta V_{gap,BG}$ can be estimated using $\Delta E_{F}$ $\approx$ $\hbar \nu_{F}$$\sqrt{2\pi C_{g}\Delta V_{gap,BG}/\left|e\right|}$, where $C_{g}$ is the back-gate capacitance per area and $\nu_{F}$ is the Fermi velocity in graphene \cite{Stampfer2009,Guttinger2012}. This leads to an energy gap $\Delta E_{F}$ $\approx$ 110 - 340 meV which is significantly larger than the observed $E_{gap}$ (8.5 meV) and the band gaps $\Delta E_{con}$ ($\leq$ 50 meV) estimated from calculations of a GNR with width $W$ = 45 nm \cite{Guttinger2012}.

A schematic model shown in Fig. \ref{Fig1}(e) is able to qualitatively explain the findings described above \cite{Stampfer2009}. This model consists of a combination of quantum confinement energy gap $\Delta E_{con}$ (the intrinsic band-gap of a clean GNR) and strong bulk and edge-induced disorder potential fluctuation $\Delta _{dis}$. The confinement energy $\Delta E_{con}$ alone can neither explain the observed energy scale $\Delta E_{F}$, nor the dots formation in the GNR. However, superimposing a fluctuation in the disorder potential ($\Delta _{dis}$) can result in tunnel barriers separating different localized states (i.e., puddles or QDs), as shown in in Fig. \ref{Fig1}(e). Therefore, transport in such a system is described by a percolation between the puddles [in Fig. \ref{Fig1}(e) the dashed circles indicate the puddles, whereas the blue arrow indicates the tunnel barrier]. Within this model, $\Delta E_{F}$ depends on both the confinement energy gap and the disorder potential fluctuation, and can be approximated using the relation $\Delta E_{F}$ = $\Delta _{dis}$ $+$ $\Delta E_{con}$. $\Delta _{dis}$ can be estimated from the bulk carrier density fluctuations $\Delta n$ (due to substrate disorder) using $\Delta _{dis}$ = $\hbar \nu_{F}$$\sqrt{4\pi \Delta n}$, where $\Delta n$ $\approx$ $\pm$2 $\times$ 10$^{11}$ is extracted from ref. \cite{Martin2008}. This in turns gives $\Delta E_{F}$ = $\hbar \nu_{F}$$\sqrt{4\pi \Delta n}$ $+$ $\Delta E_{con}$ $\approx$ 126 meV \cite{Guttinger2012}, which is comparable to the experimental value (110 - 340 meV). The energy gap in the bias direction ($E_{gap}$) is not directly related with the magnitude of the disorder potential but rather with its spatial variation. When the Fermi energy (or said $V_{BG}$) lies in the center of the transport gap, the smaller localized states are more likely to form, giving rise to the larger charging energies (larger Coulomb diamonds). By contrast, when the Fermi energy is tuned away from the charge-neutrality point, the size of the relevant diamonds gets generally smaller due to the merging of individual puddles. 

Although the localized states in GNRs pose additional complications, their tunability in resistances still allows them to be used as tunnel barriers for transport in GQDs. While a large number of studies on GNRs have been reported in the field; however, in this review, we will focus primarily on GQDs in which GNRs are used as tunnel barriers. Further discussion of the transport properties of GNRs can be found in ref. \cite{Bischoff2015b}.

\subsection{3.2. Graphene single quantum dots}

\begin{figure}
\includegraphics[scale=1.15]{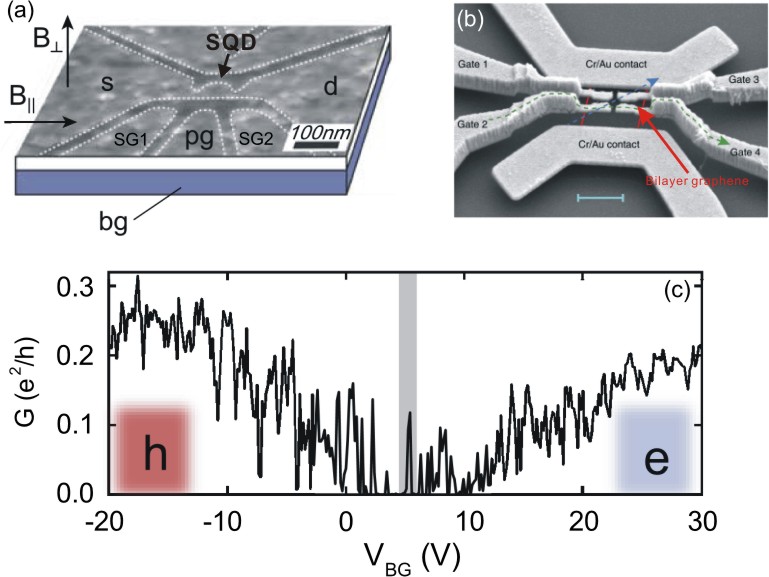}
\caption{(a) Atomic Force Microscope image of a graphene single quantum dot ($\approx$ 50 nm wide and $\approx$ 80 nm long) etched by O$_{2}$ plasma. (b) Scanning Electron Micrograph of a suspended bilayer GQD device. Bilayer graphene (highlighted by red line) is suspended between two electrodes below local top gates. Scale bar, 1 $\mu$m. (c) Source-drain conductance as a function of back-gate voltage $V_{BG}$ at bias $V_{b}$ = 4 mV measured from the device shown in (a). (a, c) adapted with permission from ref. \cite{Guttinger2009}. Copyright 2009 American Physical Society. (b) adapted with permission from ref. \cite{Allen2012}. Copyright 2012 Nature Publishing Group.}  
\label{Fig2}
\end{figure}

Owing to the expected long spin relaxation time, graphene quantum dots (GQDs) are considered to be a viable candidate for preparing spin qubits and spintronic devices \cite{Trauzettel2007}. Over the past decade, GQDs have proven to be a useful platform for confining and manipulating single electrons \cite{Gutinger2008,Volk2013,Guttinger2009,Chiu2012,Guttinger2010,Liu2010,Volk2011,Connolly.2013a}. In this section, we will review a few relevant transport experiments performed on graphene single quantum dots (GSQDs) fabricated on SiO$_{2}$/Si substrates. These include the Coulomb blockade at zero field, Fock-Darwin spectrum, spin states and charge relaxation dynamics, as will be discussed in the subsequent sections.

\subsection{3.2.1. Coulomb blockade at zero field}

GQDs can be formed by etching isolated islands connected to source and drain graphene reservoirs via nanoconstrictions that are resistive enough to act as tunnel barriers \cite{Gutinger2008,Volk2013,Guttinger2009}. An example of such a device is shown in Fig. \ref{Fig2}(a), in which in-plane graphene side and plunger gates (SG1, SG2, PG) are used to locally tune the potential of the tunnel barriers and the 50 nm diameter dot, while the doped-silicon back-gate (BG) is used to adjust the overall Fermi level. Another way to define a GQD is to induce a band-gap in bilayer graphene by applying an electric field perpendicular to the layers; in this way, charges are confined in an island defined by top gate geometry \cite{Goossens2012,Allen2012}. Such a structure can be seen in Fig. \ref{Fig2}(b), where a bilayer graphene is suspended between two Cr/Au electrodes and sits below suspended local top gates that are used to break interlayer symmetry. Graphene quantum dots can also be formed from the disorder potential \cite{Zhang2009,Amet2012}, strain engineering \cite{Klimov2012} and gated GNRs \cite{Liu2010}, in all of which Coulomb blockade can be observed. 

\begin{figure}[!t]	
\includegraphics[scale=1.2]{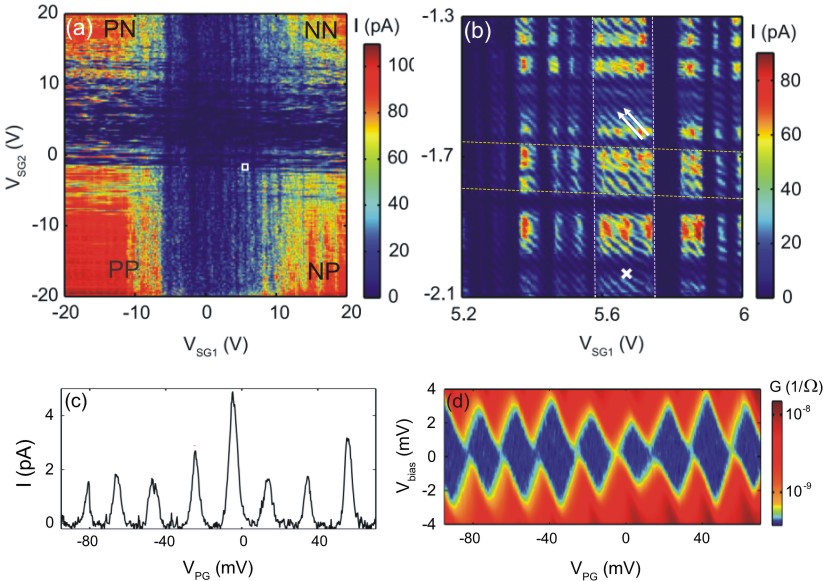}
\caption{(a) Current through a GQD (diameter $\approx$ 180 nm) as a function of two side-gate voltages $V_{SG1}$ and $V_{SG2}$. (b) Current as a function of $V_{SG1}$ and $V_{SG2}$ in the range indicated by the white square in (a). (c) Current as a function of plunger gate voltage $V_{PG}$ at $V_{SG1}$ = 5.67 V and $V_{SG2}$ = -2.033 V [the white cross in (b)]. (d) Coulomb diamonds associated with the Coulomb resonances in (c). Adapted with permission from ref. \cite{Molitor2011}.}  
\label{Fig2_B}
\end{figure}

Fig. \ref{Fig2}(c) shows the back-gate sweep (conductance as a function of back-gate voltage) of the device shown in Fig. \ref{Fig2}(a). The measurement shows a transport gap ranging from 0 $\leq$ $V_{BG}$ $\leq$ 10 V, in which current is suppressed except for multiple sharp Coulomb resonances, separating hole- from electron-transport regime. The transport gap resulting from the GNR tunnel barriers can be lifted using the side-gate voltage. Fig. \ref{Fig2_B}(a) shows the current measurements of another GQD (diameter $\approx$ 180 nm) as a function of its side-gate voltages $V_{SG1}$ and $V_{SG2}$ at a fixed back-gate voltage within the transport gap. There is a cross-like region of suppressed current separating four large conductance regions, which correspond to different doping configurations of the constrictions, labeled as NN, NP, PP and PN at the corners of the diagram, respectively. For example, keeping $V_{SG1}$= -20 V constant and sweeping $V_{SG2}$ from -20 V to $+$20 V keeps constriction 1 in the $p$-doped regime whereas constriction 2 is tuned from $p$-doped to $n$-doped (PP to PN transition). In order to observe single electron transport, it is necessary to operate in a region of gate space where both tunnel barriers are resistive (i.e., within the center of the cross-like current suppressed regime). Fig. \ref{Fig2_B}(b) shows the case with the Fermi energy located at the edge of the transport gap for both constrictions [marked by the white square in Fig. \ref{Fig2_B}(a)]. The measurement shows broaden vertical and horizontal resonances [white and yellow dashed lines in Fig. \ref{Fig2_B}(b)], which correspond to resonant transmission through the localized states in the left and right constrictions, tuned with the respective side-gate. The fact that those lines are almost perfectly vertical and horizontal indicates that the side-gate only influences its adjacent constriction. A closer inspection of Fig. \ref{Fig2_B}(b) shows a series of diagonal lines (indicated by arrows), which correspond to the Coulomb blockade resonances from the central quantum dot, where both side gates are expected to have a similar lever arm. These 0D Coulomb resonances can be unambiguously resolved as a series of well-defined and regular peaks, as shown in Fig. \ref{Fig2_B}(c), by sweeping a plunger gate voltage $V_{PG}$ with sides gates fixed at $V_{SG1}$ = 5.67 V and $V_{SG2}$ = −2.03 V [the white cross in Fig. \ref{Fig2_B}(b)]. A Coulomb diamond measurement of these resonances further confirms their origin. A charging energy $E_{C}$ $\approx$ 3.2 meV is extracted from the vertical extent of the Coulomb diamonds shown in Fig. \ref{Fig2_B}(d), in reasonable agreement with the dot diameter if the Disc plate capacitance model $E_{C}$=$e^{2}/8\epsilon\epsilon_{0}r$, where $r$ is the radius of the quantum dot, is used \cite{Molitor2011}. In the following sections, we discuss how these Coulomb blockade peaks evolve with the applied perpendicular and in-plane magnetic fields.

\subsection{3.2.2 Electron-hole crossover in perpendicular magnetic field}

\begin{figure*}[!t]	
\includegraphics[scale=0.7]{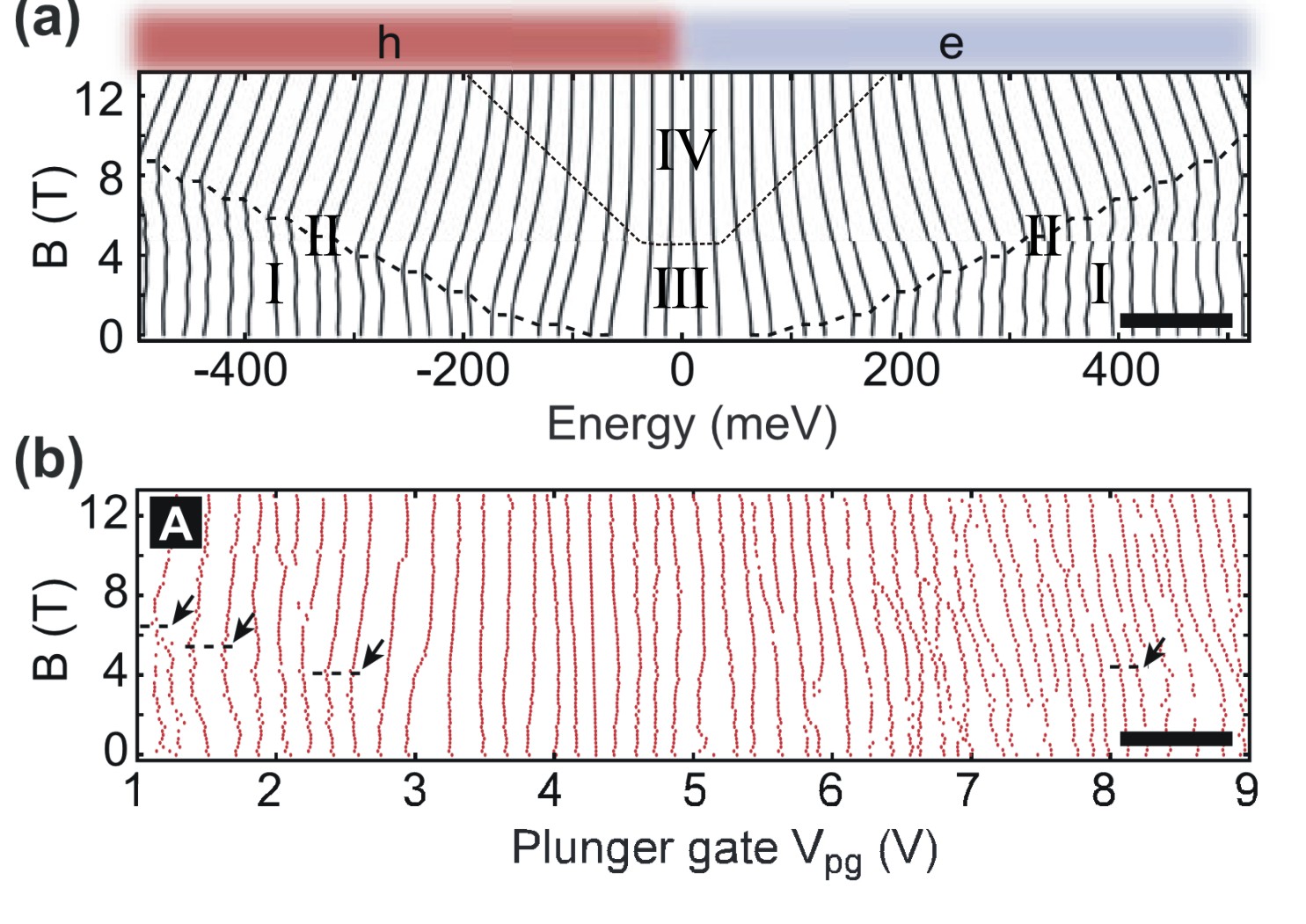}
\caption{(a) The Fock-Darwin spectrum of a 50 $\times$ 80 nm GQD calculated by assuming a constant charging energy and spin degenerate states. The dashed line (regime II) indicates filling factor $\nu$=2 above which all eigenstates continuously evolve into the zero-energy Landau level. The captions I-IV denote different regimes mentioned in the content. (b) Coulomb peak position as a function of perpendicular magnetic field, measured from the device shown in Fig. \ref{Fig2}(a). The arrows indicate the filling factor $\nu$=2 kinks. Adapted with permission from ref. \cite{Guttinger2009}. Copyright 2009 American Physical Society.}  
\label{Fig3}
\end{figure*}

In section 2.1, we have shown the calculated Fock-Darwin spectrum of a graphene quantum dot. Here, we consider a more practical case where a charging energy is included in the spectrum. Fig. \ref{Fig3}(a) shows a tight-binding simulated Fock-Darwin spectrum of a 50 $\times$ 80 nm GQD, where a constant charging energy $E_{C}$=18 meV have been added to each single-particle level spacing ($\approx$4 meV in average). Several key features seen from the spectrum are summarized in the following. At low $B$-field, the 0D levels fluctuate but stay at roughly the same energy, as can be seen in the regime I of Fig. \ref{Fig3}(a). This fluctuation of the Coulomb blockade resonances at low $B$ is due to the continuously crossing of different unfilled states at low energy, as seen in Fig. \ref{Figure 36}(b). This situation changes when the second lowest LL (LL$_1$) is full, at which point the levels show a kink (regime II) indicating that the electrons (or holes) start to condense into the lowest Landau level (i.e., LL$_0$ at energy $E_{0}$), and the $B$-field onset of this kink increases with increasing number of particles in the quantum dot. Beyond this $B$-field, the levels tend to move towards the charge-neutrality point (regime III), meaning the hole levels move to higher energies while the electron levels move to lower energies. At large enough $B$-field, eventually the levels stop moving and stay roughly at the same energy again (regime IV), indicating the full condensation of electrons/holes into the lowest LL. The Fock-Darwin spectrum of the GQD in Fig. \ref{Fig2}(a) has been studied experimentally by tracking the position of Coulomb peaks under the influence of perpendicular magnetic fields, as shown in Fig. \ref{Fig3}(b). Comparing the numerical simulation and the experimental data [Fig. \ref{Fig3}(a) and (b)], one can find the same qualitative trend of states running toward the center ($E_{0}$). The arrows in Fig. \ref{Fig3}(b) indicate the kinks beyond which all the levels start to fall into the lowest Landau level. These kinks in the magnetic-field dependence of Coulomb resonances can be used to identify the few-carrier regime in graphene quantum dots. The opposite energy shift for electrons and holes in the Fock-Darwin spectrum also provides a method to estimate the charge neutrality point in GQDs \cite{Chiu2012}, but the precise first electron to hole transition is difficult to identify. This can be attributed to the formation of localized states near the Dirac point, which exhibit a weak magnetic-field dependence that alters the spectrum. It is also worth noting that the parasitic magnetic resonances in the tunnel barrier GNRs can also alter the magnetotransport in the GQD \cite{Chiu2012}, which complicates a direct comparison with the simulated Fock-Darwin spectrum.

\subsection{3.2.3. Spin states in in-plane magnetic field}

\begin{figure}	
\includegraphics[scale=0.8]{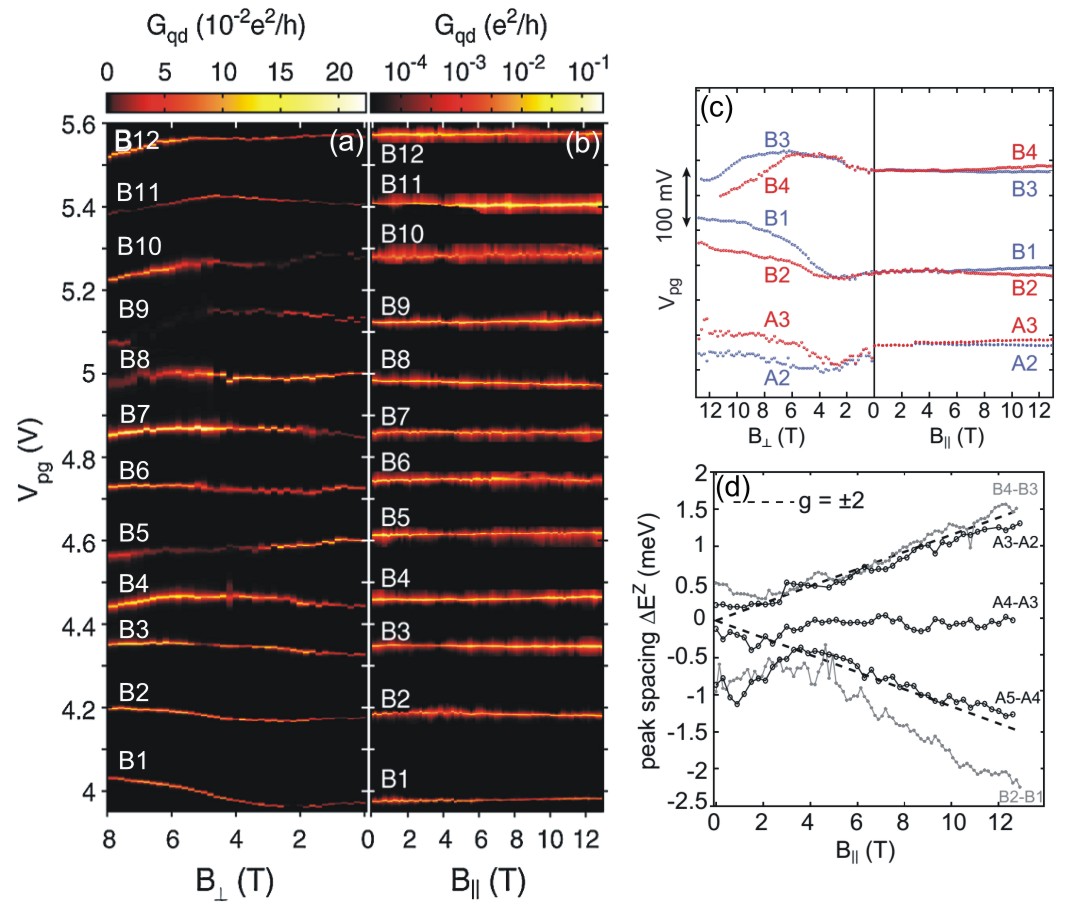}
\caption{(a) Coulomb peaks as a function of perpendicular magnetic field recorded at $V_{b}$=100 $\mu$V, measured from the device shown in Fig. \ref{Fig2}(a). (b) The same Coulomb peaks in (a) but measured in parallel (in-plane) magnetic field. (c) Comparing the evolution of three peak pairs in perpendicular (left) and parallel (right) magnetic field. The peak positions are extracted by fitting the data in (a) and (b), and are offset in $V_{PG}$ voltage such that the pairs coincide at $B$=0 T. (d) Peak spacing as a function of in-plane magnetic field for the three pairs in (c). The dashed lines represent the Zeeman splitting $\Delta E$$^{Z}$=$\pm$$\left|g\right|$$\mu_{B}B$ for a $g$-factor $\left|g\right|$=2. Adapted with permission from ref. \cite{Guttinger2010}. Copyright 2010 American Physical Society.}  
\label{Fig4}
\end{figure}
Perpendicular magnetic fields strongly affect the component of the electron wavefunctions in a QD, resulting in the Fock-Darwin spectrum. In-plane magnetic fields, on the other hand, leave the orbital component unaffected, making it possible to explore Zeeman splitting of QD states \cite{Folk2001,Lindemann2002,Guttinger2010}. It is critical to perfectly align the sample plane to the magnetic field to reduce the perpendicular components, which can be technically difficult. However, this problem can be minimized if one can analyze spin pairs, i.e., two subsequently filled electrons occupying the same orbital state with opposite spin orientation. In this case, the orbital contributions can be significantly reduced by subtracting the positions of individual peaks sharing the same orbital shift in perpendicular magnetic field. Potential spin pairs can be identified by tracking the evolution of two subsequent Coulomb peaks with increasing perpendicular magnetic field, as shown in Fig. \ref{Fig4}(a). For example, the lowest two peaks (B1 and B2) and the following two (B3 and B4) are identified as potential spin pairs due to their similar peak evolution. Fig. \ref{Fig4}(b) shows a measurement of the same peaks in Fig. \ref{Fig4}(a) but with increasing in-plane magnetic fields after the sample is carefully rotated into an orientation parallel to the applied $B$-field. The peaks show a small energy shift with in-plane $B$-field, indicating the orbital effect is negligible. In order to analyze the movement of the peaks in more details, Fig. \ref{Fig4}(c) show the fit of the data selected from Fig. \ref{Fig4}(a) and (b), in which two adjacent peaks (a spin pair) are plotted with suitable offsets in $V_{pg}$ such that pairs coincide at \textit{B}=0 T. As can be seen from the left panel of Fig. \ref{Fig4}(c), the orbital states of each pair have approximately the same $B_{\bot}$ dependence, hence spurious orbital contributions (from slight misalignment) to the peak spacing in $B_{\left|\right|}$ are limited, resulting in a resolvable Zeeman splitting [the right panel of Fig. \ref{Fig4}(c)]. The energy scale of the Zeeman splitting for the spin pairs in Fig. \ref{Fig4}(c) and for two additional peak spacings [A3-A4 and A5-A4, not shown in Fig. \ref{Fig4}(c)] are plotted in Fig. \ref{Fig4}(d). The spin differences between three successive spin ground states take the integer
values $\Delta^{(2)}$ = 0, $\pm$1, ....[e.g., for two successive states, the spin difference can be 1/2 (-1/2) for adding a spin-up (spin-down) electron or 3/2 (-3/2) for adding a spin-up (spin-down) electron while flipping another spin from down (up) to up (down)]. Therefore, apart from the slight deviation of B2-B1, all spin pairs in Fig. \ref{Fig4}(d) follow the relation $\Delta E$$^{Z}$ = $\Delta^{(2)}$$g$$\mu_{B}B$ and a $g$-factor value of approximately 2 can be extracted. The study of Zeeman splitting on spin pairs enables the extraction of the spin-filling sequence in a GQD, which follows an order of $\downarrow\uparrow\uparrow\downarrow\downarrow\uparrow\uparrow\downarrow$ (data not shown) \cite{Guttinger2010}. It is deviated from a sequence of $\uparrow\downarrow\uparrow\downarrow$ observed in the low carrier regime of carbon nanotube quantum dots \cite{Buitelaar2002,Cobden2002}. This phenomenon has been attributed to the exchange interaction between the charge carriers in graphene, which is comparable to the single-particle energy spacing in GQDs and can therefore lead to a ground-state spin polarization \cite{Guttinger2010}.

The spin states in GQDs can in principle be considered as a candidate of spin qubits. However, the spin related transport in graphene has shown to suffer from the extrinsic perturbations \cite{Tombros2007,Han2010a,Han2011}. We will address this issue again in section 3.3 and chapter 7. 

\subsection{3.2.4. Charge relaxation time}

\begin{figure*}[!t]	
\includegraphics[scale=0.72]{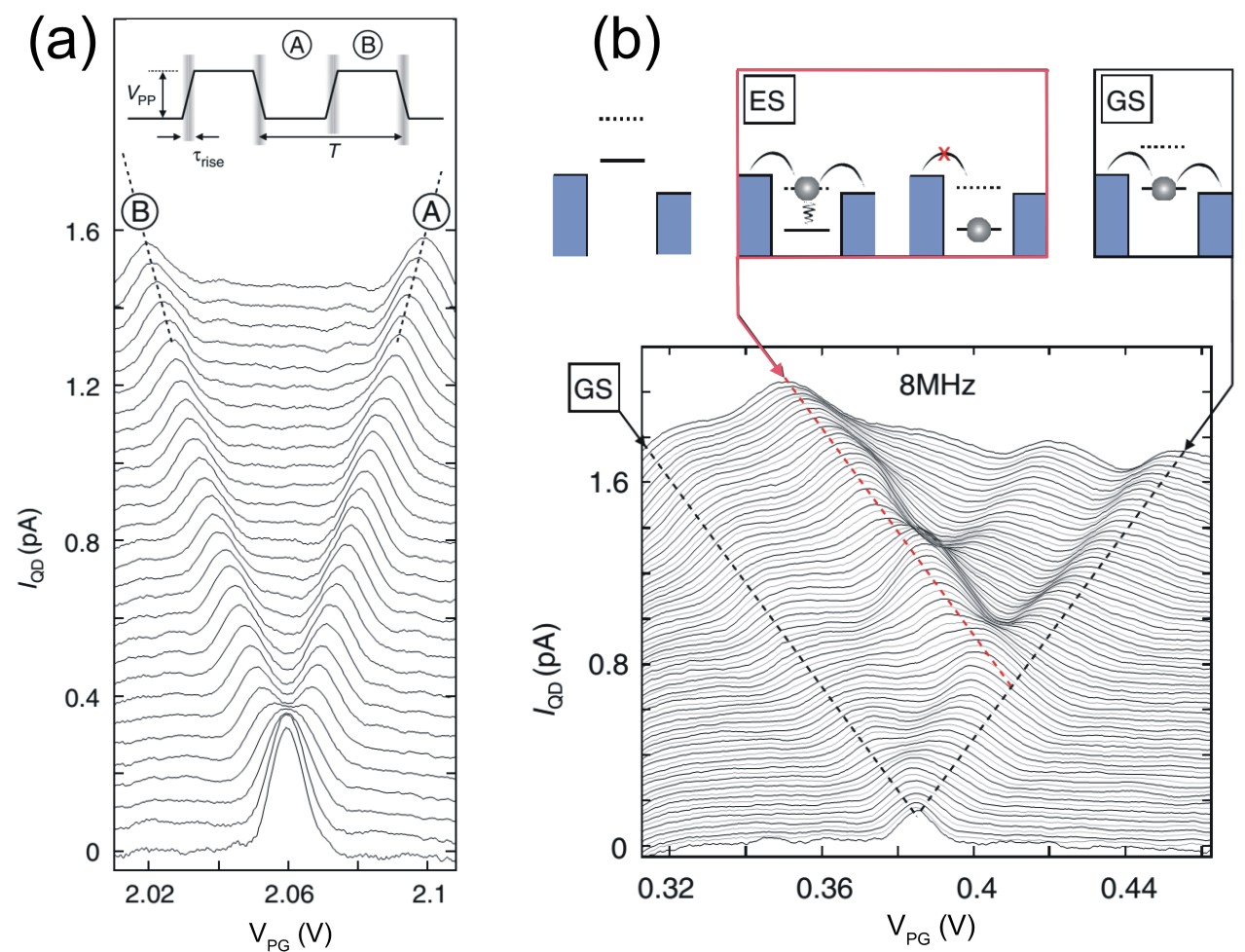}
\caption{(a) Current through the dot at V$_{SD}$=1.5 mV while applying a 100-kHz pulse. Different lines (from bottom to top) correspond to $V_{pp}$ being varied from 0 to 1.4 V in steps of 50 mV. Inset: Sketch of the pulse scheme employed in the measurements presented in this figure. Low and high pulse-level are labeled A and B, and T is the period of the pulse. (b) Top panel: Schematic of transport via GS, ES and, on the left, of a possible initialization stage. Bottom panel: Measurement similar to the ones shown in (a), but with a higher frequency of 8 MHz. $V_{pp}$ is varied from 0 to 2 V in steps of 25 mV (from bottom to top). Adapted with permission from ref. \cite{Volk2013}. Copyright 2013 Nature Publishing Group.}  
\label{Fig5}
\end{figure*}

Pulsed gating, in which a radio-frequency (RF) voltage is applied to the gates, is a powerful tool to manipulate electron spin and to study the spin relaxation time in 2DEG quantum dot systems \cite{Hanson2007}. In this section, we will describe how pulse gating can be used to investigate the charge relaxation dynamics of excited states (ESs) in GSQDs \cite{Droscher2012,Volk2013}. In these measurements, a rectangular pulse $V_{pp}$ with a duration $T$ [inset in Fig. \ref{Fig5}(a)] is applied on top of a DC voltage ($V_{PG}$) to the plunger gate located in the vicinity of the GQD. If the frequency of the pulse is low (2/\textit{T} $\leq$ $\Gamma_{R}$, $\Gamma_{L}$, where $\Gamma_{R(L)}$ is the tunneling rates of the right (left) barrier), the square-wave modulation of the gate voltage results simply in the splitting of the Coulomb resonance into two peaks. Fig. \ref{Fig5}(a) shows such a behavior when pulses with increasing amplitude (from bottom to top) are applied to the plunger gate. These peaks [labeled A and B in Fig. \ref{Fig5}(a)] result from the QD ground state (GS) entering the bias-window at two different values of $V_{PG}$, one for the lower pulse-level (A) and one for the upper one (B). This situation changes dramatically at higher frequencies (2/\textit{T} $\geq$ $\Gamma_{R}$, $\Gamma_{L}$), as shown in the bottom panel of Fig. \ref{Fig5}(b), where the splitting is broadened due to the reduced electron tunneling probability set by $T$ (black dashed line), and a number of additional peaks appear due to transient transport through the excited states of GQD (red dashed line). Each of these additional resonances corresponds to a situation in which the QD levels are pushed well outside the bias-window in the first half of the pulse [Fig. \ref{Fig5}(b), top left panel], and then brought into a position where transport can occur only through the ESs in the second one [Fig. \ref{Fig5}(b), top middle panel]. When the ES lie within the bias-window, an electron occupying the GS, either because of tunneling from the leads or relaxation from the ES, will block the current. Therefore, the additional resonances can be resolved in the DC-current measurements only if the frequency of the pulse is higher than the characteristic rate $\gamma$ of the blocking processes. As both tunneling and ES relaxation lead to the occupation of the GS, $\gamma$ is approximately given by $\gamma$ $\approx$ $\Gamma$$ + $1/$\tau$, where $\Gamma$ is the tunneling rate from lead to dot and $\tau$ is the intrinsic relaxation time of the ES. Since the lowest frequency at which signatures of transport through ESs emerge provides an upper bound for $\gamma$, and $\Gamma$ can be determined by the fitting of peak current through the dot. This in turn gives a lower bound $\tau$ $\geq$ 78 ns for the charge relaxation time of the GQD ESs \cite{Volk2013}.

The ES relaxation timescale is related to the lifetime of charge excitations, which is limited by electron-phonon interactions. The main potential source that induces the charge relaxation in supported graphene is through coupling to the longitudinal-acoustic (LA) phonon $via$ deformation potential (due to an area change of the unit cell) \cite{Ando2005,Kaasbjerg2012,Volk2013}. However, the fact that the observed timescale is a factor 5-10 larger than what has been reported in III-V QDs \cite{Fujisawa2001,Fujisawa2002,Fujisawa2002a} indicates that the electron-phonon interaction in sp$^{2}$-bound carbons is relatively weak, which is likely due to the absence of piezoelectric phonons in graphene \cite{Volk2013}.

\subsection{3.3. Graphene double quantum dots}
Graphene double quantum dots (GDQDs) are formed when two graphene islands are located close enough such that they are capacitively coupled to each other and individually coupled to the adjacent gates. Double quantum dots (DQDs) in a wide range of semiconductors are a model system for investigating the spin dynamics of electrons \cite{Hanson2007,Wiel2002,Pfund2006,Schroer2011,Pecker2013}. For example, spin-to-charge conversion using the Pauli spin blockade phenomenon and measurements of spin decoherence time were pioneered in GaAs and later realized in carbon nanotube and silicon DQDs \cite{Hanson2007,Ono2002,Johnson2005,Johnson2005a,Petersson2010,Chorley2011,Zwanenburg2013}. Graphene has been predicted to be particularly suitable for preparing spin-based qubits because of its weak spin-orbit interaction and hyper-fine effect \cite{Trauzettel2007}, which should lead to a long spin relaxation time. However, although the 0D states in a GQD have shown the ability to store spin (see section 3.2.3), spin-related transport phenomena such as the Kondo-effect \cite{Goldhaber-Gordon1998} and spin blockade have thus far not been observed in GQDs \cite{Moriyama2009,Molitor2010,Liu2010,Volk2011,Wang2012,Wei2013,Chiu2015,Deng2015}. It has been reported that the spin relaxation time in monolayer graphene ranges from 100 ps to 2 ns, significantly shorter than theoretically predicted \cite{Tombros2007,Han2010a,Han2011,Maassen2012,Drogeler2014}. Two mechanisms have been proposed to explain this observation. One involves local magnetic moments, which enhance spin relaxation through the resonant scattering of electrons off magnetic moments. Adatoms, organic molecules, vacancies, or spin-active edges are the possible sources of such local magnetic moments \cite{Kochan2014}. The other mechanism is related to the interplay between the spin and pseudospin quantum degrees of freedom when disorder does not induce valley mixing \cite{Tuan2014}. Due to these possibilities, the spins of electrons in GDQDs can be rapidly flipped, thus lifting the spin blockade. Although attempts to probe the spin dynamics in such a system have failed, the control of confined charges in GDQDs can still be achieved. These include gate-tunable interdot coupling \cite{Molitor2009,Liu2010,Volk2011,Wei2013} and charge pumping \cite{Connolly.2013a}, which are discussed in the following sections.

\subsection{3.3.1. Coulomb blockade and magneto-transport}

\begin{figure}[!t]	
\includegraphics[scale=1.33]{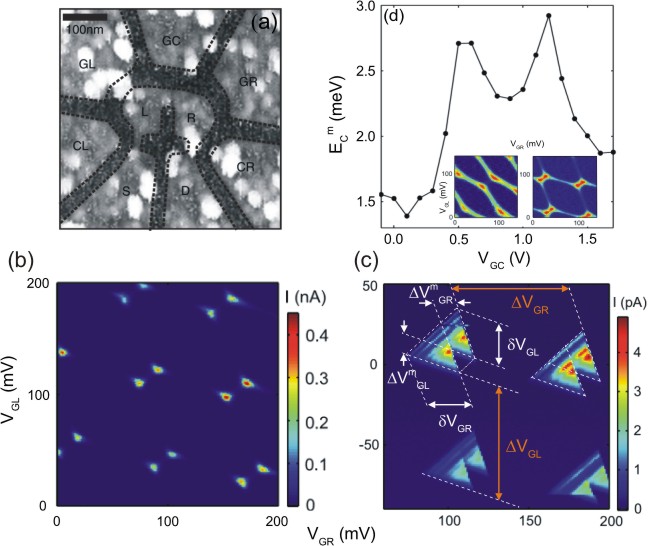}
\caption{(a) Atomic Force Microscope image of a GDQD etched by O$_{2}$ plasma. (b) Current through the GDQD in (a) as a function of $V_{GR}$ and $V_{GL}$ measured at a low bias voltage $V_{b}$ = 500 $\mu$V showing the triple points. (c) The same as (b) but at a higher bias voltage $V_{b}$ = 6 mV shows the bias triangles. (d) Mutual capacitive coupling between the two dots as a function of central plunger gate $V_{GC}$. All the data points correspond to the same triple point. Inset: Current as a function of $V_{GR}$ and $V_{GL}$ for two different central plunger gate voltages $V_{GC}$ = −1.9 V (left) and $V_{GC}$ = 0 V (right). Adapted with permission from ref. \cite{Molitor2010}. Copyright 2010 European Physical Society.}  
\label{Fig6}
\end{figure}

GDQDs can be fabricated lithographically by O$_{2}$ plasma etching out of a graphene flake or by defining the potential landscape using top gates on an etched GNR \cite{Moriyama2009,Molitor2010,Liu2010,Volk2011,Wang2012,Wei2013}. Fig. \ref{Fig6}(a) shows an AFM image of an etched GDQD device on SiO$_{2}$/Si substrate. Two plunger gates $V_{GR(GL)}$ are used to tune the energy levels in QD$_{R(L)}$ while three side gates ($V_{CL,GC,CR}$) are used to tune the tunnel barriers. Fig. \ref{Fig6}(b) shows the current through the device as a function of $V_{GR}$ and $V_{GL}$ at $V_{b}$= 500 $\mu$V, in which a honeycomb-like charge stability pattern typical for a double quantum dot device can be seen. In this low bias regime, transmission is only possible within small areas (known as triple points) in the stability diagram where the levels of two dots are aligned with a small bias window. When the applied bias is large, the current flow is possible over a wider range in gate space, resulting in current measured in the bias-dependent triangle-shaped regions (known as bias triangles), as shown in Fig. \ref{Fig6}(c). The dimensions of bias triangle allow the determination of the conversion factors between gate voltage and energy. The charging energies for the left dot $E^{L}_{C}$=$\alpha_{L}$$\cdot$$\Delta$V$_{GL}$=13.2 meV and for the right dot $E^{R}_{C}$=$\alpha_{R}$$\cdot$$\Delta$V$_{GR}$=13.6 meV are obtained using the voltage-energy conversion factor $\alpha_{L(R)}$=$eV_{b}$/$\delta$V$_{GL(GR)}$, which can be extracted from the bias triangles shown in Fig. \ref{Fig6}(c) (see section 2.2). The interdot coupling energy can also be determined from the splitting of the triangles [Fig. \ref{Fig6}(c)]: $E^{m}_{C}$=$\alpha_{L}$$\cdot$$\Delta V$$^{m}_{GL}$=$\alpha_{R}$$\cdot$$\Delta V$$^{m}_{GR}$=2.2 meV (see section 2.2). It is possible to modulate the interdot coupling strength by changing the voltage applied to the central gate, i.e., $V_{GC}$. The inset in Fig. \ref{Fig6}(d) shows examples of two charge stability diagrams recorded with exactly the same parameters, except for the voltage applied to the central plunger gate. This is also shown in Fig. \ref{Fig6}(d) where the interdot coupling energy $E^{m}_{C}$ extracted from the data is plotted as a function of $V_{GC}$. The oscillating behavior has been also reported in three different GDQD devices and was attributed to resonances induced by disorder states either in the middle GNR (connecting two dots) or in the graphene gate itself \cite{Molitor2009,Liu2010,Wei2013}. Since large gate-voltage ranges are used, the capacitive coupling of the gates to the disorder states can add or subtract charges discretely to these localized states, thus altering the entire environment abruptly and unpredictably. Consequently, the wavefunction in DQD needs to reconstruct itself, leading to the non-monotonic changes in the inter-dot coupling strength with gate voltage. 

\begin{figure}
\begin{center}
		\includegraphics[scale=0.53]{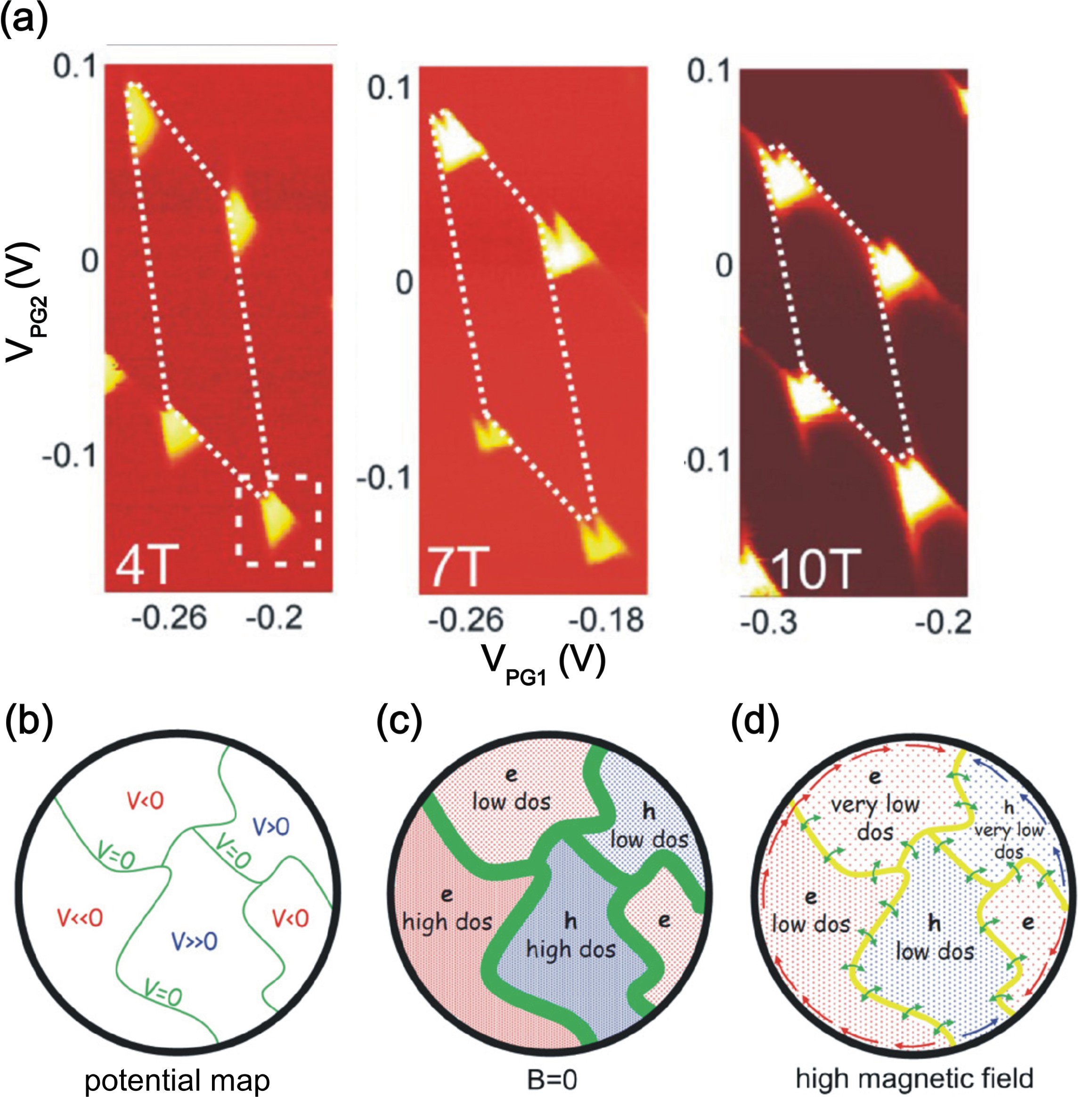}
\caption{(a) The evolution of the charge stability diagram of a large GDQD under the influence of perpendicular magnetic field from 4 to 10 Tesla, measured at $V_{b}$ = -1 mV. (b) Example of a potential distribution in a large disordered quantum dot. (c, d) Expected DOS distributions in the dot at zero magnetic field and high magnetic field, respectively. Adapted with permission from ref. \cite{Chiu2015a}. Copyright 2015 American Physical Society.}
\label{Figure 38}
\end{center}	
\end{figure}

When a perpendicular magnetic field is applied to a large graphene dot in which substrate disorder plays an important role [meaning that the size of the QD is greater than the size of the disorder-induced charge puddles; see Fig. \ref{Fig9}(b)], it is possible to induce charge redistribution due to the merging of charge puddles in the dot. Charge stability diagrams of DQDs reveal a wealth of information about their charging energy, interdot coupling and cross gate coupling strength, making them an ideal way to probe charge rearrangements in QDs. Fig. \ref{Figure 38}(a) shows the evolution of the charge stability diagram of a large GDQD (200 nm in diameter) for applied perpendicular magnetic fields ranging from 4 to 10 T \cite{Chiu2015a}. The field-dependent changes in the dimensions of the honeycomb (highlighted by the dotted hexagonal outlines) indicate the variations in the capacitances $C_{g1}$ and $C_{g2}$ and thus the changes in the charging energies of both dots [see Eqs. (\ref{eqn 36}), (\ref{eqn 37}), (\ref{eqn 44}) and (\ref{eqn 45})]. The QD charging energies vary from $E_{C1}$$\approx$3 meV and $E_{C2}$$\approx$6 meV at \textit{B}=4 T to $E_{C1}$$\approx$2.2 meV and $E_{C2}$$\approx$3.5 meV at 10 T. Note that the subscript 1(2) in $C_{g1(2)}$ and $E_{C1(2)}$ denotes the gate-dot capacitance and the charging energy for QD$_{1(2)}$. These results suggest that the `effective sizes' of both dots increase at high \textit{B}-fields, which is reflected on the decreasing charging energies. A schematic model shown in Fig. \ref{Figure 38}(b), (c) and (d) serve as a qualitative explanation for this observation \cite{Chiu2015a}. Consider a varying background potential \textit{V} in a model QD, as shown in Fig. \ref{Figure 38}(b), where \textit{V} fluctuates from positive (blue) to negative (red), passing through \textit{V}=0 (green). If \textit{V} varies slowly, in each region of a large dot, the energy bands will approximately correspond to the shifted energy bands of 2D graphene with the Fermi energy set to zero. A band gap is introduced to represent the quantum confinement effects of the dot, such that in the \textit{V}=0 (green) region, the density of states (DOS) is very low or 0, whereas in the \textit{V}$<<$0 (\textit{V}$>>$0) regions, it gives rise to the electron (hole) puddles with a high DOS, as shown in Fig. \ref{Figure 38}(c). The DOS in the dot changes dramatically at high \textit{B}-fields, where the lowest LL (LL$_{0}$) is well developed, with the consequent closing of the band gap. Thus, in the \textit{V}=0 region, the DOS is expected to increase, resulting in the development of non-chiral channels connecting the puddles [the yellow region in Fig. \ref{Figure 38}(d)], whereas in the \textit{V}$<<$0 (\textit{V}$>>$0) regions, the DOS decreases due to the more energetically separated LLs in high \textit{B}-fields. At the same time, the other LLs begin developing together with the chiral magnetic edge channel, as indicated in Fig. \ref{Figure 38}(d) by the red (blue) arrows for the electron (hole) puddles. Since in this regime the DOS decreases in the bulk of the puddles while it increases at their edges, electron transport through the dot is not confined to a particular puddle but can be delocalized in the dot by flowing through both the chiral edge channels (red or blue arrows) and non-chiral channels (yellow region). In this sense, the current is delocalized in the dot, and charge rearrangement can be observed compared with the case of low \textit{B}-fields.

\subsection{3.3.2. Charge pumping}
\begin{figure}[!t]	
\includegraphics[scale=0.7]{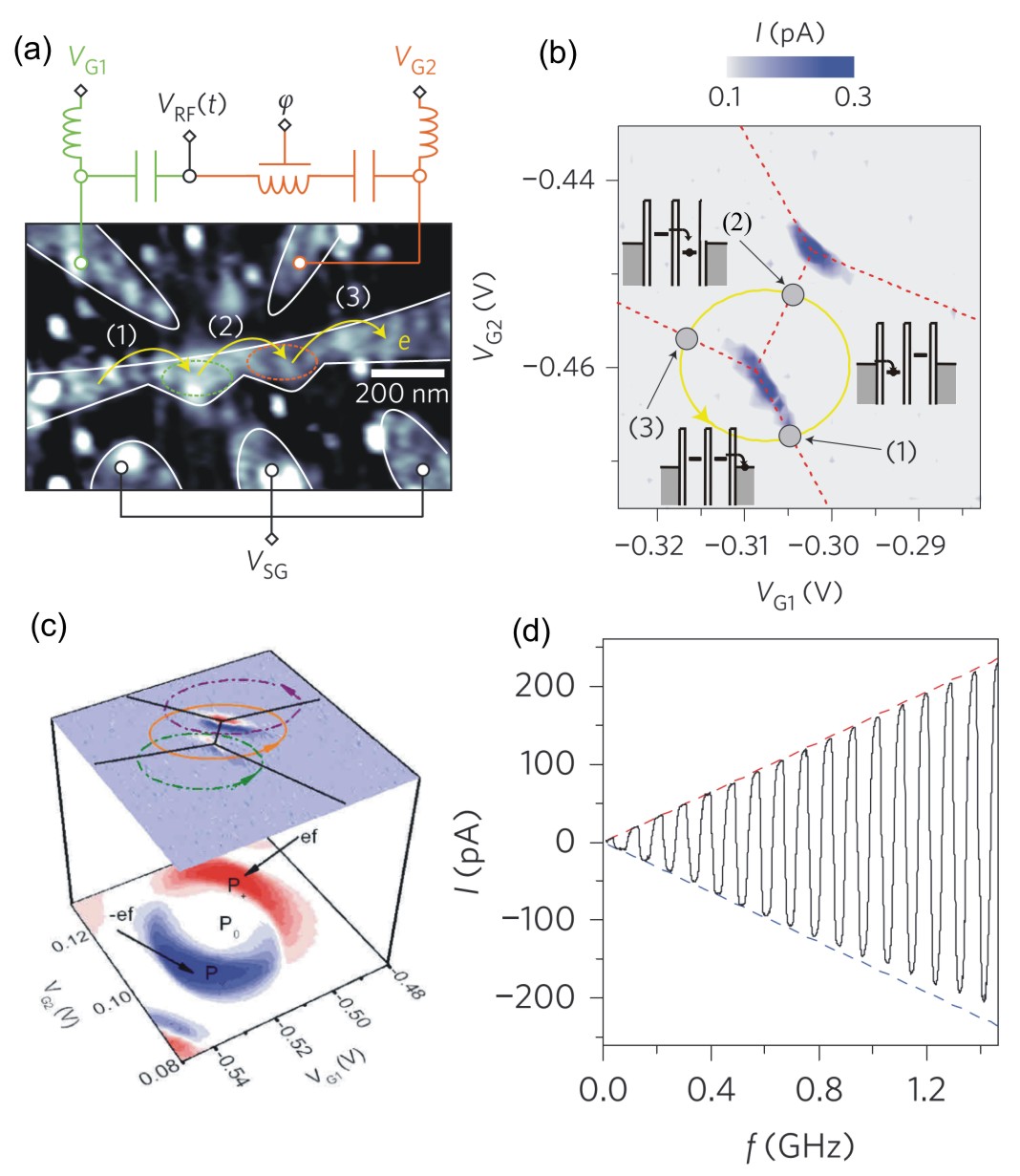}
\caption{(a) Atomic force micrograph of the device that shows the gates used to generate the pumped current in a GDQD device. An oscillating voltage $V_{RF}$($t$) is added to the DC voltages $V_{G1}$ and $V_{G2}$. A phase difference $\varphi$ is added to $V_{RF}$ before being added to one of the gates, which describes a circular trajectory (yellow circle) shown in (b). (b) Source-drain current as a function of $V_{G1}$ and $V_{G2}$ with an applied bias $\leq$1 $\mu$V. The trajectory (yellow) that encircles a triple point, passing through the sequence of transitions (1)$\rightarrow$(2)$\rightarrow$(3), as indicated in both (a) and (b). The insets denote different configuration of QD's energy level. (c) Plot showing a direct comparison between the DC (top) and AC (bottom) current behavior with $f$ = 12 MHz and $P$ = -25 dBm. Regions $P_{+}$, $P_{-}$ and $P_{0}$ refer to the positive, negative and zero pumped current, respectively. (d) Pumped current as a function of frequency at a power of $P$ = -15 dBm. Adapted with permission from ref. \cite{Connolly.2013a}. Copyright 2013 Nature Publishing Group.}  
\label{Fig7}
\end{figure}

Charge pumping, which refers to a device that can shuttle $n$ electrons per cyclic variation of control parameters to give the quantised current $I$ $\equiv$ $nef$, provides an exquisite way to link the electrical current to the elementary charge $e$ and frequency $f$ \cite{Geerligs1990,Kouwenhoven1991,Pothier1992}. Such quantized charge transport can be realized when out-of-phase RF signals are applied to the plunger gates of a DQD, as discussed in section 2.2 \cite{Fuhrer2007b,Chorley2012,Connolly.2013a}. Fig. \ref{Fig7}(a) shows a schematic of the measurement circuit and AFM image of a GDQD device used for charge pumping. The AC voltages $V_{RF}$($t$) on both plunger gates with a phase difference $\varphi$ between them drives the DQD into different charge states around the triple point. When $\varphi$ = 90, it effectively forms a circular pump loop through three charge states in the stability diagram: (1) loading an electron from source reservoir into the left dot, (2) electron transfer from left dot to the right dot and (3) unloading an electron from the right dot to the drain reservoir, as shown in Fig. \ref{Fig7}(a) and (b). When a cycle is complete, a single charge has been transferred from source to drain reservoir and establishes a current. The frequency $f$ of $V_{RF}$ determines the value of the quantized pumped current $I$ = $ef$, and the amplitude of $V_{RF}$ determines the size of area in gate space where pumped current is generated. Depending on the type of triple point that the pumping circle encloses, it generates a different direction of current. Thus, the current recorded around two nearby triple points will present a circular shape with equal values but different signs. Fig. \ref{Fig7}(c) shows a direct comparison of the locations in gate space around a pair of triple points without RF (top) and with RF (bottom) voltages applied to the plunger gates. If the pump loop only encloses one triple point (green and purple loop), it results in a flat regions, labeled $P_{+}$ and $P_{-}$, with a quantized pumped current $P_{+,-}$ = $\pm$$ef$ in the stability diagram. However, when the pump loop encloses a pair of triple points (orange loop), it leads to repeatedly increasing and decreasing the occupancy of each QD without any net transfer of electrons from source to drain. Thus, there is a central region (labeled $P_{0}$) where $I$ $\approx$ 0, giving rise to the crescent shape of pumped current as shown in Fig. \ref{Fig7}(c). Unambiguous confirmation of quantized charge pumping is shown in Fig. \ref{Fig7}(d), which plots the pumped current as a function of $f$ with the DC gate voltages fixed at the center of the $P_{+}$ region. The oscillatory behavior is introduced because of a frequency-dependent phase shift in the RF circuit. The pumped current follows the quantized value $I$ = $\pm$$ef$ over a range of frequencies up to gigahertz, an order of magnitude faster than the traditional metallic pump \cite{Keller1996}. The pumping frequency in graphene is characterized by the RC time constant of the tunnel barriers, where R and C are the effective resistance and capacitance of the GNRs. The two-dimensional nature of graphene leads to a small C and results in a large pump frequency set by the tunnel rate of tunnel barriers (GNRs) \cite{Connolly.2013a}.

To summarize chapter 3, we have reviewed the transport properties of graphene single dots and double dots fabricated on SiO$_{2}$ substrates. Additional relevant reviews of research on GQDs on SiO$_{2}$ substrates can be found in ref. \cite{Molitor2011,Stampfer2011,Guttinger2012,Neumann2013}. We conclude the main observations with the summary of references given in Table I. In the next chapter, we review graphene nanostructures fabricated on hBN and discuss how the transport properties change with the reduced influence of substrate disorder.

\ \\
\begin{center}
\begin{tabular}	  {|p{2cm}||p{6cm}|} 
 \hline
 \multicolumn{2}{|c|}{TABLE I. References for the main observations} \\
 \hline
 References& Main observations of graphene single quantum dots (GSQDs) \\
 
 \hline
 \cite{Schnez2009} & Observation of excited states \\
 \cite{Gutinger2008} & Charge detection \\
 \cite{Guttinger2009,Chiu2012} & Fock-Darwin spectrum in the few-electron and many-electron regimes    \\
 \cite{Guttinger2010} &  Zeeman splitting of spin states  \\
 \cite{Allen2012} & First suspended GQDs \\
 \cite{Goossens2012,Allen2012,Muller2014} & Bilayer GQDs defined by top gates \\
 \cite{Droscher2012,Volk2013} & High-frequency gate manipulation on GQDs \\
 \cite{Klimov2012} & GQDs defined $via$ strain engineering \\

 \hline

References& Main observations of graphene double quantum dots (GDQDs)\\
\hline
\cite{Molitor2010,Liu2010,Volk2011}    &Observation of excited states \\
\cite{Volk2011}    &Zeeman splitting \\
\cite{Volk2011,Fringes2012}    &Bilayer graphene double dot \\
\cite{Liu2010}    &GDQDs defined by gated GNRs \\
\cite{Roulleau2011}    &Electron-phonon coupling \\
\cite{Wei2013}    &Metal gate tuning \\
\cite{Connolly.2013a}&   Charge pumping  \\
\cite{Chiu2015a}&   Charge redistribution in magnetic fields \\
\cite{Deng2015}&   RF sensing of the number of charges \\
\hline

\end{tabular} 
\end{center}

\maketitle
\section{4. Graphene nanostructures on hBN substrate}

While the behavior of graphene nanostructures fabricated on SiO$_{2}$ is clearly influenced by localized states, it remains an open question whether they originate predominantly from substrate disorder or edge roughness. In the following sections, we review the studies of graphene nanodevices fabricated on hBN substrate. These devices, with reduced substrate disorder potential, are expected to enable the influence of substrate and edge disorder to be studied separately.

\subsection{4.1. Graphene nanoribbons on hBN}
\begin{figure*}[!t]	
\includegraphics[scale=0.63]{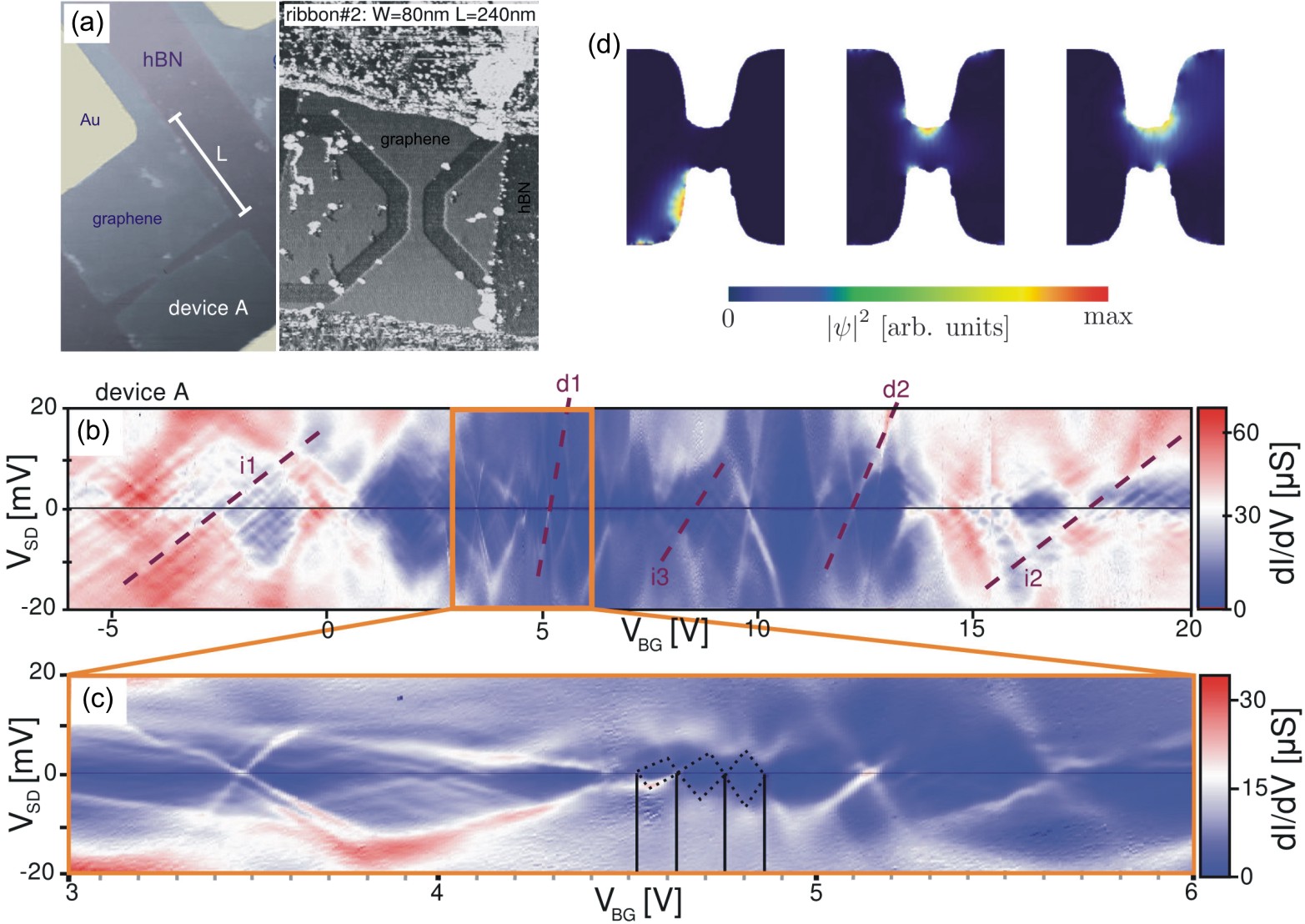}
\caption{(a) Scanning force microscopy images of two GNR/hBN devices with different geometries. (b) Differential conductance of device A [left panel in (a)] as a function of applied back-gate and bias voltage. (c) Closeup of a region highlighted by yellow rectangle in (b). (d) Results for the tight-binding simulation of a 30$\times$30 nm GNR connected to open leads of width 140 nm. $\psi$ in the color bar denotes the wavefunction. Each panel corresponds to an eigenstate which is localized in different area of device. Adapted with permission from ref. \cite{Bischoff2014a}. Copyright 2014 American Physical Society.}  
\label{Fig10}
\end{figure*}

GNRs fabricated on hBN substrate with micrometer and a few tens of nanometer scales have been reported in two experimental studies \cite{Bischoff2012,Bischoff2014a}. While the micron-sized GNRs on hBN indeed showed an improved mobility (55000 cm$^{2}$V$^{-1}$s$^{-1}$) and reduced disorder density (below 10$^{10}$ cm$^{-2}$), no major differences were observed for long GNRs [$W$=80 nm, $L$=240 nm, as shown in the right panel of Fig. \ref{Fig10}(a)], compared to their counterparts on SiO$_2$ \cite{Bischoff2012}. In both cases, electrical transport is characterized by the percolation process through localized charge puddles formed along the GNRs. It leads to an assumption that the edges - which are expected to be similar for reactive-ion-etched ribbons on both SiO$_{2}$ and hBN - dominate charge transport in those long graphene nanostructures. However, the situation changes when a relatively short GNR (30 nm $\times$ 30 nm), as shown in the left panel of Fig. \ref{Fig10}(a), was studied on hBN substrates \cite{Bischoff2014a}. Fig. \ref{Fig10}(b) shows the differential conductance as a function of applied DC bias and back-gate voltage for this device, where a series of Coulomb diamonds in the transport gap indicate the existence of localized states, as can be also seen in the GNRs on SiO$_2$ (see section 3.1). However, a close inspection on the size of smallest observed diamonds, as shown in Fig. \ref{Fig10}(c) [a zoom-in of a region highlighted by orange rectangle in Fig. \ref{Fig10}(b)], shows that the size of the localized states are in fact larger than expected. Using $C_{BG, loc} = e/\Delta V_{BG}$, the smallest diamonds [dotted lines in Fig. \ref{Fig10}(c)] spanning about 0.1 - 0.2 V in back-gate voltage corresponds to a capacitance of 1.6 - 0.8 aF. Employing the plate capacitor model corrected with a factor of 1.5 for stray fields, the area of a site of localized charge can be estimated using A$\approx$$\frac{ed}{\epsilon\epsilon_{0}\Delta V_{BG}}\times\frac{1}{1.5}$, where $\epsilon$ and $d$ are the dielectric constant and thickness of SiO$_{2}$ \cite{Bischoff2014a}. The area associated with a diamond spanning 0.1 V in back-gate, for example, is estimated to be 9000 nm$^{2}$ and is ten times larger than the geometrical constriction size ((30 nm)$^{2}$). 

To address this observation, a tight-binding simulation was performed to calculate the wave function in a 30$\times$30 nm GNR connected to 140 nm wide graphene leads, as shown in Fig. \ref{Fig10}(d). Note that the random fluctuations of the boundary with an amplitude of 2 nm is introduced to address the edge roughness. As can be seen, states localized mostly in the constriction (middle panel) will contribute to the Coulomb blockade and are likely to result in wide diamonds. On the contrary, some states are strongly localized at an edge outside of the constriction (left panel) which do not contribute to the transport. In between of these two cases, there are a number of strongly localized states, as shown in the right panel of Fig. \ref{Fig10}(d), that localize along the rough edge of the constriction and extend along the edge also into the leads of the device. These states are potentially responsible for the smaller diamonds observed experimentally [Fig. \ref{Fig10}(c)]. While the consideration of graphene leads is the key factor for the extension of wavefunction, two conditions are also crucial for this effect to be seen. One is that the substrate disorder has to be much weaker than the edge disorder, and the other is the edge-to-bulk ratio of device has to be large enough for the edge to play an important role. In the next section, a transport work on GQD on hBN will show that the edge is the dominant source of disorder for QD size less than 100 nm in diameter. Therefore, for wider GNRs (small edge-to-bulk ratio) or GNRs on SiO$_{2}$ (strong substrate disorder), localization along the edge still happens but bulk contributions are expected to dominate transport and result in the small localized states, and thus, the wide Coulomb diamonds.

\subsection{4.2. Graphene quantum dots on hBN}

\begin{figure*}[!t]	
\includegraphics[scale=0.83]{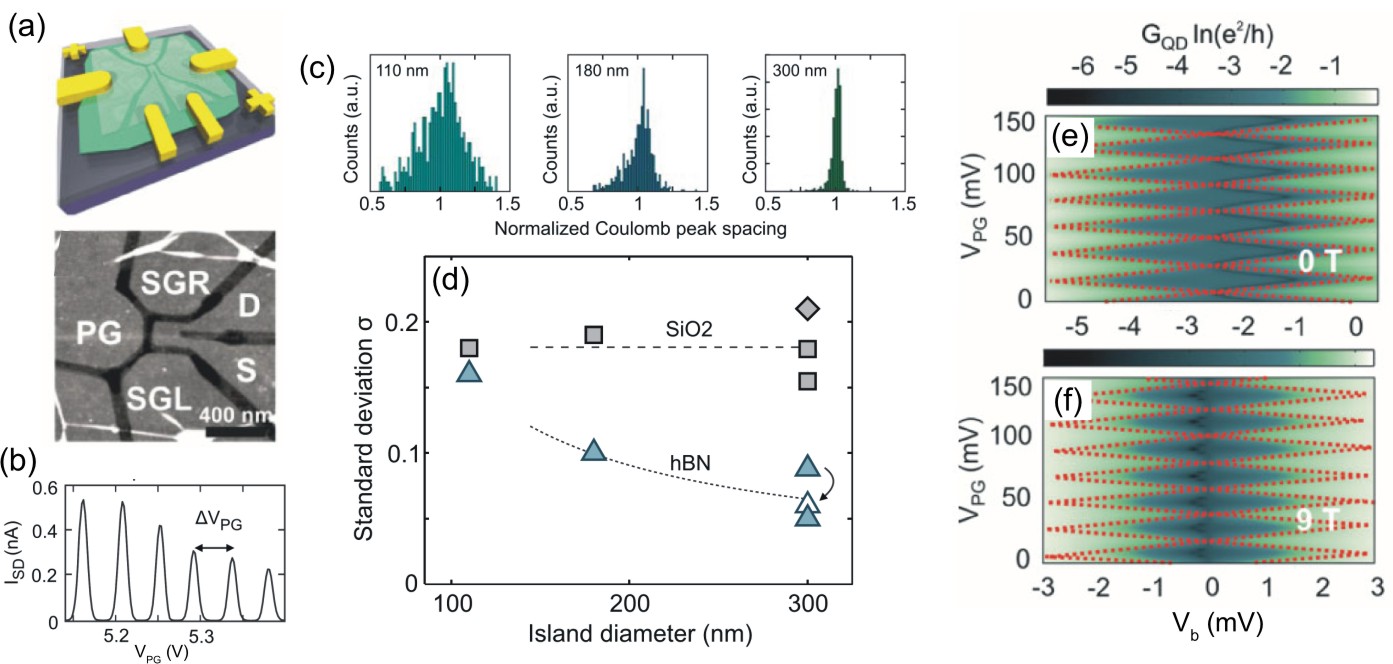}
\caption{(a) Top panel: Schematic illustration of a graphene SET on hBN. Bottom panel: Atomic force micrograph of an etched GQD on hBN with a diameter of 180 nm. (b) Source-drain current $I_{SD}$ as a function of $V_{PG}$ for the device shown in (a). (c) Normalized peak-spacing distribution for GQDs on hBN with diameters of $d$=110 nm (left panel), $d$=180 nm (middle panel), and $d$=300 nm (right panel). (d) Summary plot of the standard deviation $\sigma$ of the normalized peak-spacing distribution for different sized GQD on a SiO$_{2}$ (rectangular data points) and hBN (triangular data points) substrate. (e) Coulomb Diamond measurement of a GQD ($d$=300 nm) on hBN at a perpendicular B-field of 0 T. (f) The same measurement as (e) but at $B$=9 T. (a, b, c, d) adapted with permission from ref. \cite{Engels2013}. Copyright 2013 American Institute of Physics. (e, f) adapted with permission from ref. \cite{Epping2013}. Copyright 2013 John Wiley and Sons.}  
\label{Fig11}
\end{figure*}

GQDs with different diameters ranging from 100 to 300 nm have been fabricated on hBN substrates for transport characterization, as noted in a few literature \cite{Engels2013,Epping2013}. The sizes of the dots are close to the order of the expected size of charge puddles in bulk graphene on hBN ($\approx$ 100 nm in diameter) \cite{Xue2011}, so the substrate disorder is expected to play less important role. Fig. \ref{Fig11}(a) shows the schematic illustration of such a device (top panel) and the atomic force micrograph of an etched GQD on hBN with a diameter of 180 nm (bottom panel). The QD levels are tuned by a plunger gate (PG) while two side gate (SGR and SGL) are used to tune the resistance of the tunnel barrier GNRs. In a regime where the two barriers are pinched-off, the current $I_{SD}$ as a function of plunger gate voltage, as shown in Fig. \ref{Fig11}(b), confirms that the QD is operating in the Coulomb blockade regime. For a more detailed comparison between GQDs resting on hBN and SiO$_{2}$, the distribution of the Coulomb-peak spacing $\Delta V_{PG}$, i.e., the spacing between two subsequent Coulomb peaks, are statistically studied among dots with different sizes fabricated both on hBN and SiO$_{2}$ substrates. The normalized Coulomb peak spacings $\Delta V_{PG}$/$\overline{\Delta V_{PG}}$ for GQDs on hBN are reported as histograms in Fig. \ref{Fig11}(c), for QD diameter $d$=110 nm (left panel), $d$=180 nm (middle panel) and $d$=300 nm (right panel), respectively. The same type of measurements are also performed for GQDs on SiO$_{2}$, and the results are summarized in Fig. \ref{Fig11}(d), where the standard deviation of the normalized peak spacing distribution ($\sigma$) as a function of the QD diameter is presented for both hBN and SiO$_{2}$ substrates. A clear difference can be seen between these two cases. The standard deviation for GQDs on hBN shows a clear decreasing dependence from 0.16 for the dot with $d$=110 nm to 0.05 for the dot with $d$=300 nm, while in the case of GQDs on SiO$_{2}$ it is independent of $d$. The standard deviation $\sigma$, which can be considered as the strength of peak-spacing fluctuations, may result from (i) the fluctuations of single particle level spacing $\Delta$, (ii) fluctuations of the charging energy $E_{C}$ (i.e., fluctuations in the size of the dot), or (iii) fluctuations of the lever arm $\alpha$ (i.e., the position of the dot). The single-particle level spacing in GQD is $\Delta$($N$)=$\hbar v_{F}/(d\sqrt{N}$), where $N$ is the number of charge carriers on the dot and $v_{F}$ is the Fermi velocity \cite{Schnez2009}. If $N$ is the only variable, the single particle level spacing $\Delta$($N$) gives an upper limit at the order of 0.03 to $\sigma$ for $N$= 600 (the number of peaks studied), and should be independent of the dot size and substrate. This is not in agreement with the data shown in Fig. \ref{Fig11}(d) and leads to the assumption that the remaining two sources are responsible for the variability in peaks spacing. The standard deviation for GQDs on hBN can be represented as $\sigma$ $\approx$ $\sigma^{hBN}$ + $\sigma^{edge}$/$d$ $\approx$ 0.01 + 16/$d$ [nm], where $\sigma^{hBN}$ represents the substrate-induced disorder (independent of dot size) and $\sigma^{edge}$ represents the edge-induced disorder (scale with size as the edge-to-bulk ratio changes). Note that both values are obtained from the fit of the dotted line in Fig. \ref{Fig11}(d). By contrast, the standard deviation for GQDs on SiO$_{2}$ is independent of dot size and reads $\sigma^{SiO_{2}}$ $\approx$ 0.18. This suggests that the potential landscape in the dot on SiO$_{2}$ is dominated by substrate induced disorder, while contributions due to edge roughness, which are expected to scale with the size of the sample, play a minor role. These $\sigma$ values also lead to the conclusions that (i) the substrate-induced disorder in GQDs on hBN is reduced by roughly a factor 10 as compared to SiO$_{2}$ ($\sigma^{SiO_{2}}$=0.18 to $\sigma^{hBN}$=0.01), (ii) edge roughness is the dominating source of disorder for GQDs with diameters less than 100 nm. The latter conclusion agrees with the observations in small GNRs on hBN, as already discussed in the previous section.

The reduced substrate disorder of GQDs on hBN can also reflect on the magneto-transport. If the magnetic length of the electrons on the GQDs is on the order of the disorder potential length scale, the electrons can accumulate in different charge puddles, leading to charge redistribution in the dot thus changing the charging energy (as discussed in 3.3.1). However, as a result of the reduced bulk disorder, this effect is assumed not to occur for GQDs on a hBN substrates. Fig. \ref{Fig11}(e)(f) show the comparison of Coulomb diamond measurements of a $d$=300 nm GQD on hBN at $B$=0 T and $B$=9 T. It can be seen by the similar Coulomb diamonds (with a charging energy $E_{C}$ $\approx$ 3 meV) for both magnetic fields, that the QD is stable and well-defined at 9 T, supporting the notion that the effective size of GQD is not affected in high magnetic fields.

\ \\
In summary, we have reviewed the transport properties of GNRs and GSQDs on hBN substrates. The effect of reduced substrate disorder can be observed both in GNRs (smaller Coulomb diamond) and GQDs (mean Coulomb peak spacing fluctuating with QD's size). Here we note that, although the reduced substrate disorders for GQDs on hBN can in principle suppress some possible sources for fast spin relaxation, the spin-related transport phenomena (such as spin blockade in DQDs) are still expected to suffer from the edge states in the etched GNR tunnel barriers. 

\maketitle
\section{5. Nanostructures of Transition Metal Dichalcogenides}
\begin{figure*}[!t]	
\includegraphics[scale=0.72]{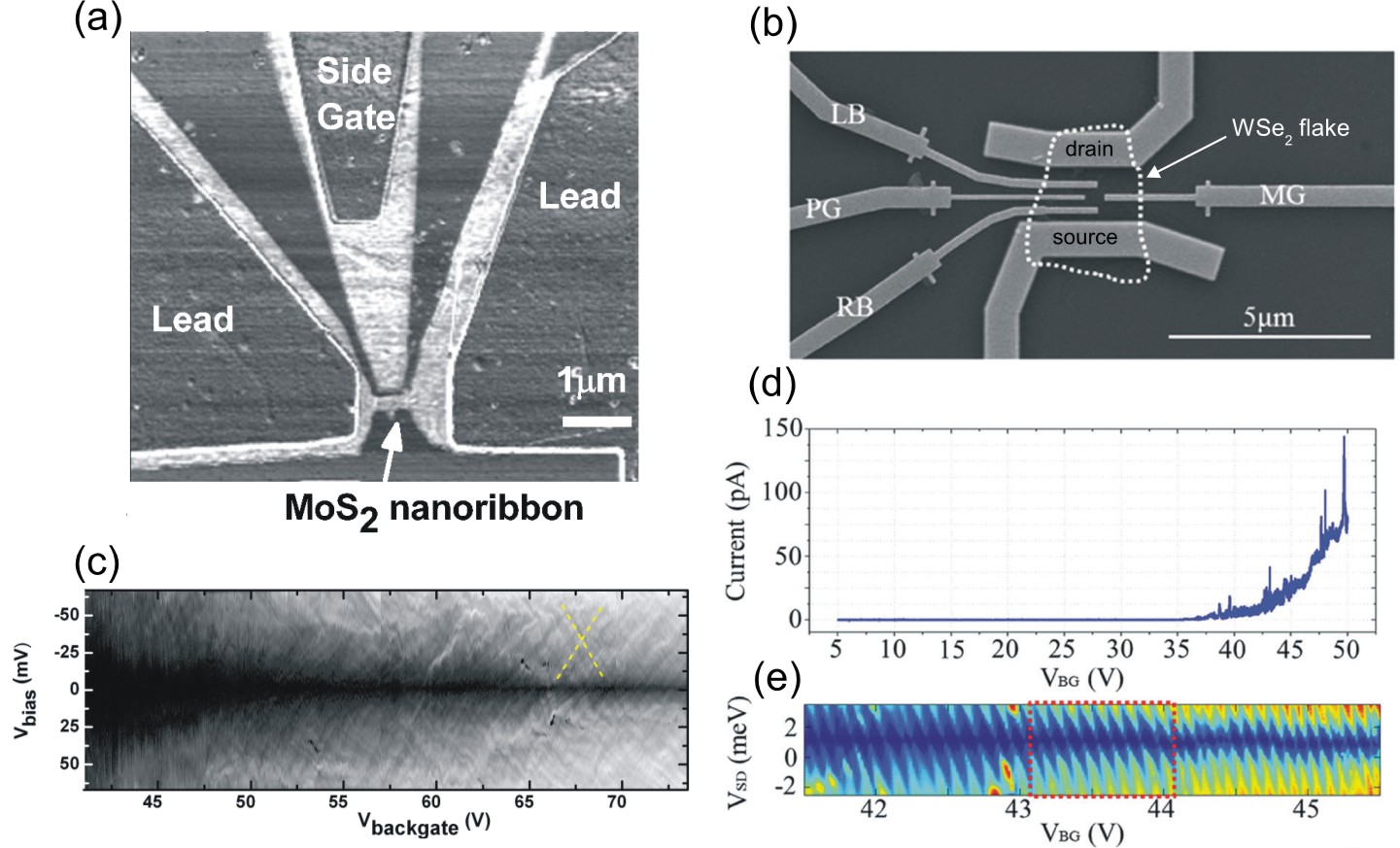}
\caption{(a) AFM image of the MoS$_2$ nanoribbon. (b) SEM image of the WSe$_2$ quantum dot device. The WSe$_2$ flake (4.5 nm in thickness) is highlighted by the white dotted line. The WSe$_2$ flake was directly contacted by the source/drain electrodes, and was separated from the four top gates (LB, PG, RB and MG) by a 40 nm layer of Al$_2$O$_3$ grown by atomic layer deposition (ALD). (c) Differential conductance of the MoS$_2$ nanoribbon as a function of DC bias voltage $V_{bias}$ and back-gate voltage $V_{backgate}$, with the side gate floated. (d) Source-drain current flow through the WSe$_2$ device as a function of back-gate voltage ($V_{BG}$). (e) Coulomb diamond measurements for WSe$_2$ QD from $V_{BG}$ = 41.5 V to $V_{BG}$ = 45.5 V, with all other top gate voltages fixed at -2 V. (a, c) adapted with permission from ref. \cite{Li2013}. (b, d, e) adapted with permission from ref. \cite{Song2015}. Copyright 2015 Royal Society of Chemistry.}  
\label{Fig39}
\end{figure*}

The existence of band gaps close to the wavelengths of visible light has earned TMD nanostructures considerable attention in optical studies \cite{Gopalakrishnan2014,Lin2015,Yan2016}. However, it is also the sufficiently large bandgaps that distinguish TMDs from graphene and allows their nanostructures to be defined using electrical gating, as is commonly done in GaAs 2DEG systems. In this manner, the edge roughness created during the lithographic etching process, which is a common case in GQD fabrications, can be avoided. To date, to the best of the author's knowledge, only two studies have emerged in the literature that are relevant to transport in TMD nanostructures, including MoS$_2$ nanoribbons and WSe$_2$ single quantum dots \cite{Li2013,Song2015}. In the former case, the nanoribbons were fabricated using RIE etching [Fig. \ref{Fig39}(a)], whereas in the latter, the QDs were defined using metal top gates [Fig. \ref{Fig39}(b)]. Both devices were fabricated on SiO$_{2}$/Si substrates. The Coulomb diamond measurement for a MoS$_2$ nanoribbon is shown in Fig. \ref{Fig39}(c). The existence of diamonds confirms the presence of small localized sates (or QDs) in the nanoribbon. Moreover, the fact that larger diamonds are formed in the middle of the transport gap, whereas smaller diamonds are located away from the gap, indicating that the size of the localized state is strongly dependent on the Fermi energy (or back-gate voltage), as is also observed for GNRs on SiO$_{2}$/Si substrates. This finding suggests that, in both cases (graphene nanoribbons and MoS$_2$ nanoribbons on SiO$_{2}$/Si substrates), the potential in the nanoribbons can be described as a superposition of the substrate disorder potential and the confinement-induced energy gap, as already discussed in section 3.1. The back-gate sweep for the WSe$_2$ device is shown in Fig. \ref{Fig39}(d) and exhibits the characteristic behavior of an n-doped semiconductor with a transport gap for V$_{BG}$ $\leq$ 35 V. The Coulomb blockade regime for a single dot can be achieved by tuning the WSe$_2$ flake into the conducting regime (V$_{BG}$ $>$ 35 V) while keeping the area below the top gates in an insulating state (V$_{MG}$$=$V$_{PG}$$=$V$_{LB}$$=$V$_{RB}$ $=$ $-$2 V), such that the Coulomb diamonds can be measured as a function of V$_{BG}$, as shown in Fig. \ref{Fig39}(e). In this study, the charging energy $E_C$ was estimated to be approximately 2 meV, which corresponds to a QD radius of $r$ $=$ 260 nm (the plate capacitance model $E_{C}$=$e^{2}/8\epsilon\epsilon_{0}r$ is used, where $\epsilon$ is the relative permittivity of WSe$_2$) and is in reasonable agreement with the area defined by the top gates. Although the the 0D behavior was demonstrated, no magnetic field dependence of the Coulomb resonances was reported in this work. 

Despite a lack of experimental studies on the topic, here, we briefly discuss how the Coulomb resonances of TMD QDs should evolve with magnetic fields. The Hamiltonian for a QD in a two-dimensional semiconducting TMD defined by electrostatic gates under the influence of a perpendicular magnetic field can be written as follows (note that the model assumes that the system is n-doped and that the Hamiltonian describes the conduction band at the $K$ ($K'$) point) \cite{Kormanyos2014}:

\begin{eqnarray}
H_{dot} &=& H^{\tau,s}_{el} + H^{intr}_{SO} + H^{\tau}_{vl} + H_{sp} + V_{dot} \nonumber \\ 
&=& \frac{\hbar^2\hat{q}_+\hat{q}_-}{2m^{\tau,s}_{eff}} + \frac{1 + \tau}{2} sgn(B_z)\hbar\omega^{\tau,s}_c + \tau\Delta_{CB}s_z + \frac{\tau}{2}g_{vl}\mu_BB_z + \frac{1}{2}\mu_Bg^{\bot}_{sp}s_zB_z + V_{dot}
\label{eqn 55}
\end{eqnarray} The wave numbers $q_{\pm}$ = $q_{x}$ $\pm$ $iq_y$ are measured from the $K$ and $K'$ valley points of the TMD; therefore, the band dispersion is parabolic and isotropic (note that at zero $B$-field, $q_+q_-$ = $q^2_x$ + $q^2_y$). Here, the cyclotron energy $\hbar\omega^{\tau,s}_c$ = $e\left|B_z\right|/m^{\tau,s}_{eff}$ where $B_z$ is the applied magnetic field in z-axis and $m^{\tau,s}_{eff}$ denote the effective masses for bands with different valley ($\tau$ = $\pm$ 1) and spin ($s$ = $\pm$ 1) indices. $H^{intr}_{SO}$ = $\tau\Delta_{CB}s_z$ denotes the intrinsic spin-orbit coupling in the TMD, where $s_z$ is the spin Pauli matrix and $\Delta_{CB}$ determines the coupling strength. The next term, $H^{\tau}_{vl}$ = $\frac{\tau}{2} g_{vl}\mu_BB_z$, breaks the valley symmetry of the Landau levels ($g_{vl}$ is the valley $g$-factor, and $\mu_B$ is the Bohr magneton) and describes how the valley states move in the magnetic field. Finally, $g^{\bot}_{sp}$ = $g_e$ + $g^{\bot}_{so}$ is the total $g$-factor, where $g_e$ is the free-electron $g$-factor and $g^{\bot}_{so}$ is the out-of-plane effective spin $g$-factor addressing the SOC. $V_{dot}$ is the confinement potential for a QD of radius $R_d$ and describes the hard-wall boundary conditions: $V_{dot}$($r$) = 0 for $r \leq R_d$ and $V_{dot}$($r$) = $\infty$ for $r \geq R_d$. To solve the band dispersion for this Hamiltonian, we follow the route of solving the Fock-Darwin spectrum for the GQD, as depicted in section 2.1. We set $V_{dot}$ = 0 and find the eigenvalue and eigenfunction of Eq. (\ref{eqn 55}), and the bound-state solutions of the QD can then be determined from the condition that the wave function must vanish at $r$ = $R_d$. For further details on this formalism, refer to ref. \cite{Kormanyos2014}.

\begin{figure*}[!t]	
\includegraphics[scale=0.74]{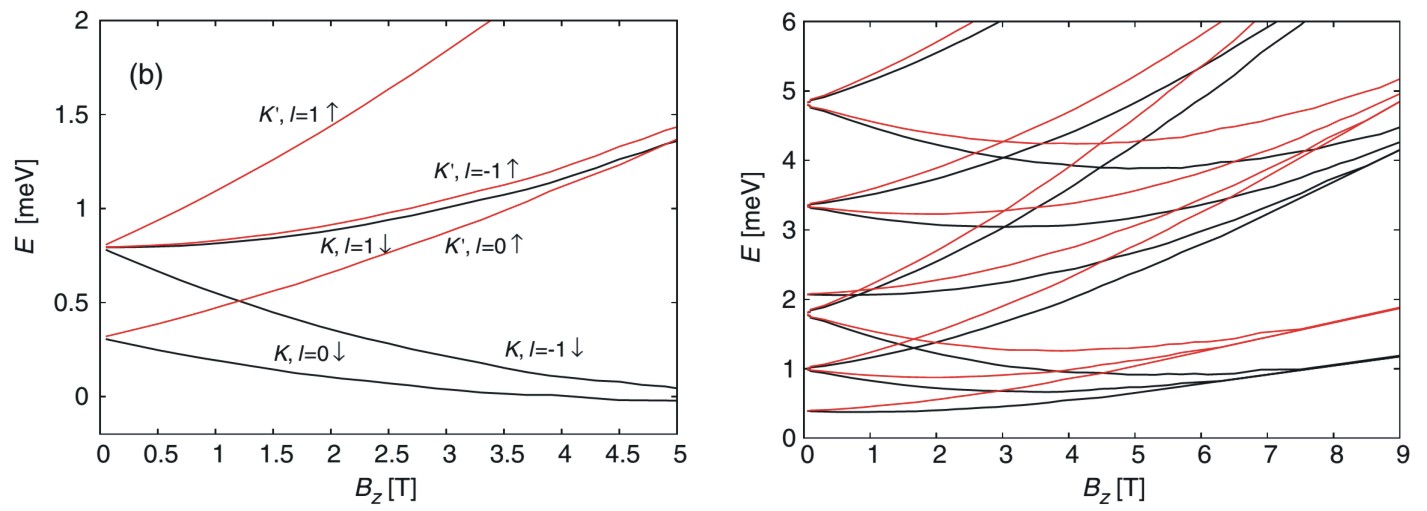}
\caption{(a) Fock-Darwin spectrum of a MoS$_2$ QD of radius $R_d$ = 40 nm. Labels show the valley ($K$ or $K'$), the orbital quantum number $l$, and the spin state ($\uparrow$ or $\downarrow$) for each level. (b) Fock-Darwin spectrum of a WS$_2$ QD of radius $R_d$ = 40 nm. Black (red) lines represent the spin $\uparrow$($\downarrow$) states from the $K$($K'$) valley. Adapted with permission from ref. \cite{Kormanyos2014}. Copyright 2014 American Physical Society.}  
\label{Fig40}
\end{figure*}

The numerically calculated spectra for QDs of $R_d$ = 40 nm in MoS$_2$ and in WS$_2$ are shown in Fig. \ref{Fig40}(a) and Fig. \ref{Fig40}(b), respectively. Note that the parameters $m^{\tau,s}_{eff}$, $g_{vl}$ and $g^{\bot}_{sp}$ used in the simulations for MoS$_2$ and WS$_2$ are different \cite{Kormanyos2014}. As can be seen in Fig. \ref{Fig40}(a), this spectrum mimics the one that we derived for a 2DEG QD [Fig. \ref{Figure 36}(a)] because of the quadratic dispersion in the model. At zero magnetic field, states with an angular momentum of $\pm l$ within the same valley are degenerate, due to the effective time-reversal symmetry acting within each valley. For a finite magnetic field, all levels are both valley and spin split, which is different from Fig. \ref{Figure 36}(a) where spin is not considered. For large magnetic fields, when $l_B$ $\leq$ $R_d$, the dot levels merge into Landau levels, as also shown in Fig. \ref{Figure 36}(a) and (b). From Fig. \ref{Fig40}(a), one can see that the spin and valley degrees of freedom are locked together (meaning that spin $\downarrow$ ($\uparrow$) electrons only reside in the $K$ ($K'$) valley), suggesting that TMD QDs can be used as simultaneous valley and spin filters for single electrons. The spectrum for WS$_2$, as shown in Fig. \ref{Fig40}(b), is similar to that for MoS$_2$, but the level spacing at zero field is larger because of the smaller effective mass in WS$_2$ (the mean level spacing can be approximated as $\delta$ = $\frac{2\pi\hbar^2}{m_{eff}A}$, where $A$ is the area of the dot \cite{Kormanyos2014}). By contrast, at a finite $B$-field, the splitting between states belonging to different valleys (and also spins, as spin and valley are locked) is significantly larger for the MoS$_2$ than for the WS$_2$ because of the different signs of $\Delta_{CB}$ and, consequently, the different spin polarizations of the lowest levels in the two materials. The valley and spin pairs could, in principle, serve as valley or spin qubits, but the crossing between levels with different quantum numbers $l$ in finite magnetic fields may add complication to its realization. One way to circumvent this problem may be to use the lowest Kramers pairs [$\left|l = 0, K', \uparrow\right\rangle$ and $\left|l = 0, K, \downarrow\right\rangle$; see Fig. \ref{Fig40}(a)] at a low $B$-field as a combined spin-valley qubit. Note that the above results apply not only for monolayer TMDs but also for TMDs with an odd number of layers, in which the inversion symmetry is broken.


\maketitle
\section{6. Nanostructures of topological insulators}

Topological surface states are ideal for certain applications in low-power electronics \cite{Zutic2004} and quantum computing devices \cite{Nayak2008,Moore2009}. However, electron-transport measurements on the TI surface states sometimes involve the unavoidable residual bulk carriers. TI nanostructures offer several advantages to overcome this problem and enable studying the fundamental nature of the topological surface states. First, their large surface-to-volume ratio greatly reduces the bulk carrier contribution in the overall electron transport. Second, TI nanostructures have well-defined nanoscale morphology, ideal for quantum interference experiments such as Aharonov-Bohm (AB) oscillations. Third, the field-effect gating to tune the Fermi level in TI nanostructures is relatively easy, due to the fact that they are only tens of nanometers thick. Thus, the design of nanostructures on TI is attractive both in probing the nature of confined topological modes \cite{Bardarson2010,Zhang2010} as well as the possibility of using them as building block for quantum information applications \cite{Akhmerov2009,DeFranceschi2010}. 

\begin{figure*}[!t]	
\includegraphics[scale=0.71]{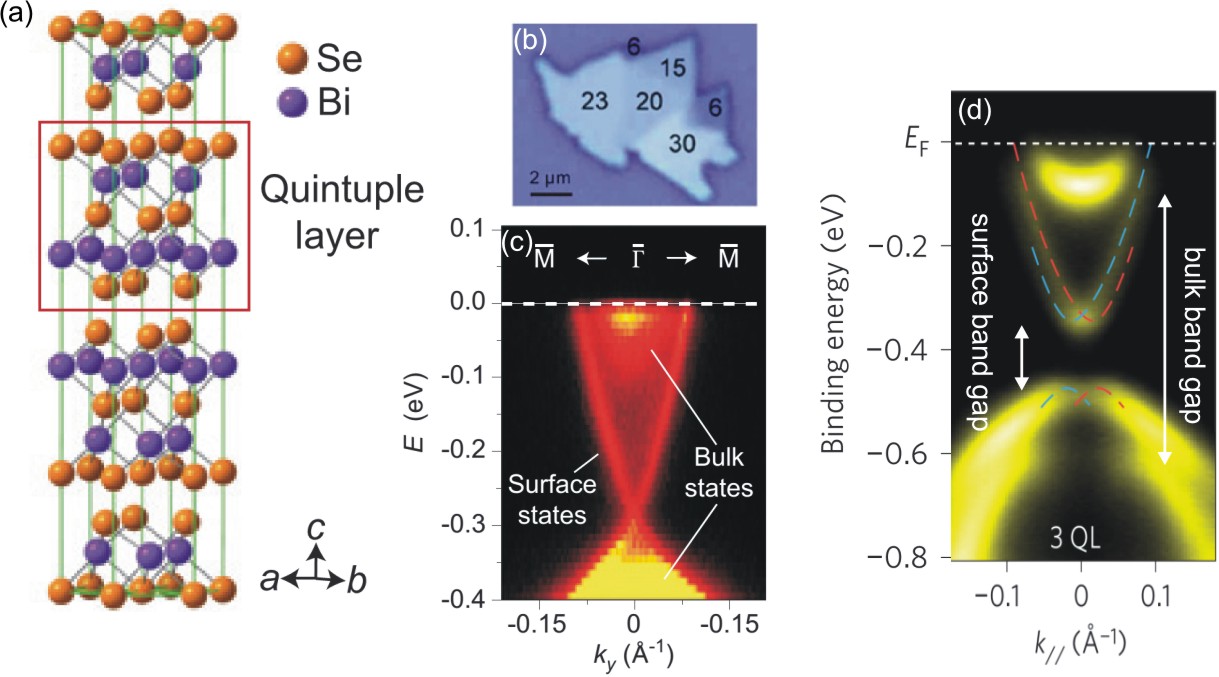}
\caption{(a) Layered crystal structure of 3D topological insulators Bi$_{2}$Se$_{3}$. Each layer consists of five atomic sheets (a quintuple layer), which are bonded together by van der Waals interactions along the c-axis. (b) Optical micrograph of Bi$_{2}$Se$_{3}$ thin flakes produced by mechanical exfoliation. The overlaid numbers indicate the number of quintuple layers of Bi$_{2}$Se$_{3}$. (c) Angle-resolved photoemission spectroscopy (ARPES) measurements of surface electronic band dispersion on Bi$_{2}$Se$_{3}$ along the $\overline{\Gamma}$-$\overline{M}$ momentum-space cut. (d) ARPES spectrum (along the $\overline{\Gamma}$-$\overline{K}$ direction) of a three quintuple layer (QL) Bi$_{2}$Se$_{3}$ slab shows a gap opening in the surface states. (a, c) adapted with permission from ref. \cite{Kong2011}. Copyright 2011 Nature Publishing Group. (b) adapted with permission from ref. \cite{Checkelsky2011}. Copyright 2011 American Physical Society. (d) adapted with permission from ref. \cite{Zhang2010a}. Copyright 2010 Nature Publishing Group.}  
\label{Fig41}
\end{figure*}

Bi$_{2}$Se$_{3}$ is a 3D topological insulator, in which Se-Bi-Se-Bi-Se are bonded covalently into a quintuple layer (QL), and a crystal is formed $via$ van der Waals force connecting each QL as shown in Fig. \ref{Fig41}(a). Unlike Bi$_{1-x}$Sb$_{x}$ alloy, Bi$_{2}$Se$_{3}$ is stoichiometric compound which can be mechanically exfoliated from bulk [see Fig. \ref{Fig41}(b)] at higher purity \cite{Moore2009,Moore2010}. In addition, the large enough bulk band gap (0.3 eV) equivalent to 3,600 K enables the topological phase to be seen at room temperature \cite{Xia2009}. Fig. \ref{Fig41}(c) shows an angle-resolved photoemission spectroscopy (ARPES) measurement performed on Bi$_{2}$Se$_{3}$, where only a single surface state with electronic dispersion almost the same as an idealized Dirac cone is present. In order to induce quantum confinement in Bi$_{2}$Se$_{3}$, and thus, to define nanostructures, a band-gap has to be introduced in the gapless surface states. This can be achieved by thinning Bi$_{2}$Se$_{3}$ into a thin slab. The tunnel coupling between the top and bottom surface induces a symmetry-breaking and creates a thickness-dependent surface band gap \cite{Zhang2010a,Cho2011}, as shown in Fig. \ref{Fig41}(d), where an opening of band gap of surface electrons in a 3 QL Bi$_{2}$Se$_{3}$ slab is presented. The other approach is to place magnetic materials proximity to a TI, in this way the broken time-reversal symmetry can result in a surface band gap at the interface \cite{Fu2008,Qi2008,Ji2012,Wei2013a}. In the following sections, we will focus on Bi$_{2}$Se$_{3}$ nanostructures (including nanowires and quantum dots), in which synthesis methods and their transport properties will be reviewed.

\subsection{6.1. Bi$_{2}$Se$_{3}$ nanowires}
\begin{figure}[!t]	
\includegraphics[scale=0.51]{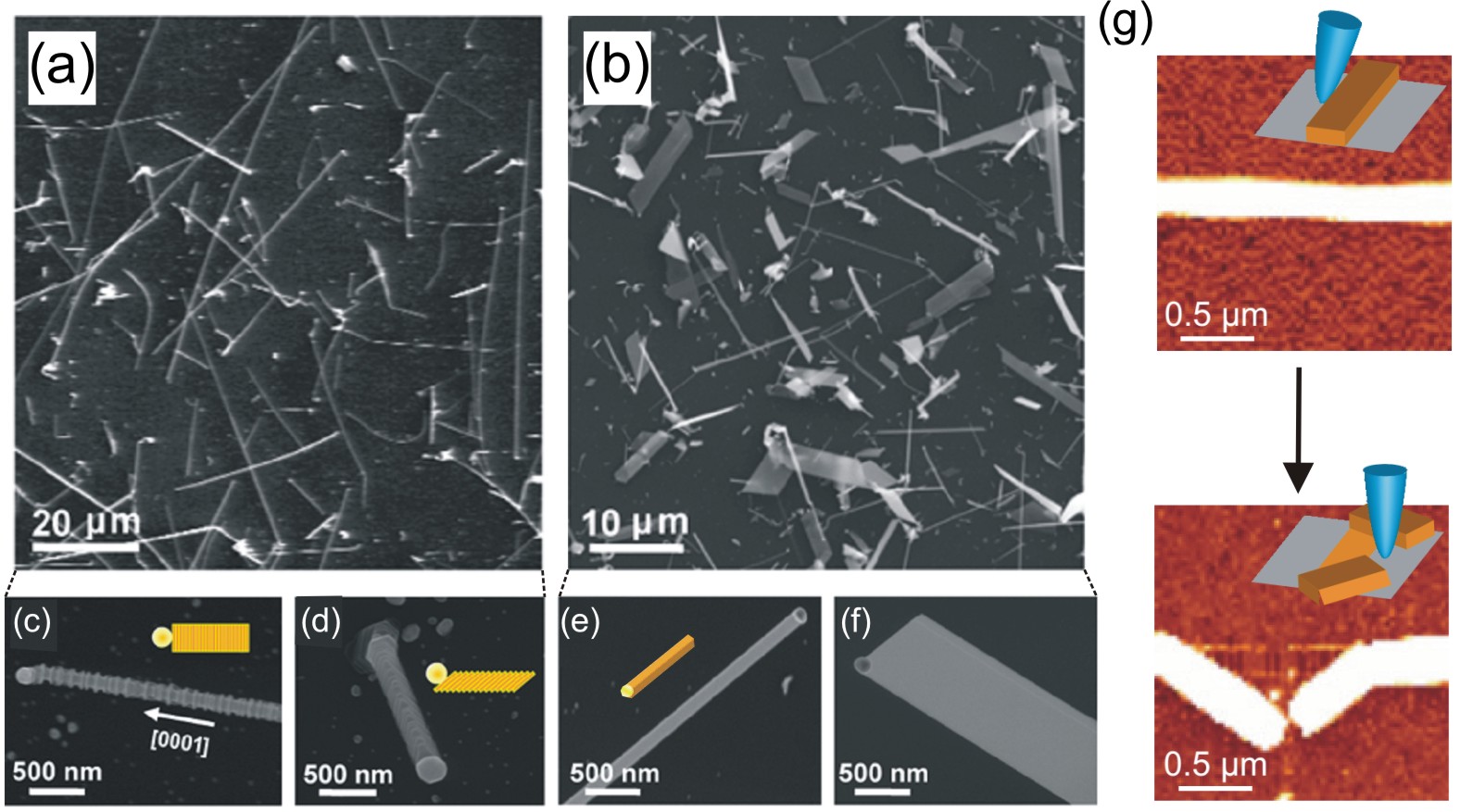}
\caption{Scanning electron microscope (SEM) images of Bi$_{2}$Se$_{3}$ nanostructures grown via VLS method. (a) SEM image of Bi$_{2}$Se$_{3}$ nanowires grown on a warm substrate. (b) SEM image of Bi$_{2}$Se$_{3}$ nanoribbons grown on a cool substrate. (c) Nanowire grown along c-axis. (d) Nanowire grown off c-axis. (e) Quasi-1D narrow nanoribbon. (f) Sheetlike wide nanoribbon. (g) Exfoliation of Bi$_{2}$Se$_{3}$ nanoribbons using an AFM, in which multiple layers of the materials are ‘knocked off’ by the tip. (a, b, c, d, e, f) adapted with permission from ref. \cite{Kong2010}. Copyright 2010 American Chemical Society. (g) adapted with permission from ref. \cite{Hong2010}. Copyright 2010 American Chemical Society.}  
\label{Fig14}
\end{figure}

Methods for producing topological insulator nanomaterials can generally be categorized as bottom-up synthesis and top-down exfoliation \cite{Kong2011}. The top-down exfoliation is the so-called scotch tape method, which is commonly used to exfoliate graphene from natural graphite. As for the bottom-up approach, topological insulator nanoribbons are made by Au-catalyzed vapor-liquid-solid (VLS) growth in a tube furnace \cite{Kong2010,Peng2010}. In such a growth, the source material Bi$_{2}$Se$_{3}$ powder is placed in the hot center of the furnace while Si substrates coated with 20 nm Au nanoparticles are placed in the downstream side of the furnace. The furnace is then heated to high temperature in the range of 450-580 degrees and is kept at the high temperature for 1-5 h, followed by a natural cool-down period. The SEM images of the as-grown Bi$_{2}$Se$_{3}$ nanowires and nanoribbons are shown in Fig. \ref{Fig14}(a) and (b), respectively. 
At the warm zone of the furnace, the growth of quasi-one-dimensional materials is dominant [see Fig. \ref{Fig14}(c), (d), (e)] while at the cool zone the nanoribbons with lateral dimensions of several micrometers are the dominant growth product [see Fig. \ref{Fig14}(f)]. The use of Au nanoparticle as catalyst induces the nucleation and growth of the nanomaterial, their presence at the end of each nanowire and nanoribbon can be viewed as an evidence of VLS growth mechanism. The thickness of these ribbons is roughly determined by the size of the Au nanoparticles (few nanometers), and the width of the ribbon varies from 50 nm to tens of micrometers. Another approach to fabricate thin Bi$_{2}$Se$_{3}$ nanoribbons down to few nanometers is shown in Fig. \ref{Fig14}(g), where an atomic force microscope tip is used to sweep off the extra layers of Bi$_{2}$Se$_{3}$ to form a thinner ribbon \cite{Hong2010}.

\begin{figure*}[!t]	
\includegraphics[scale=0.78]{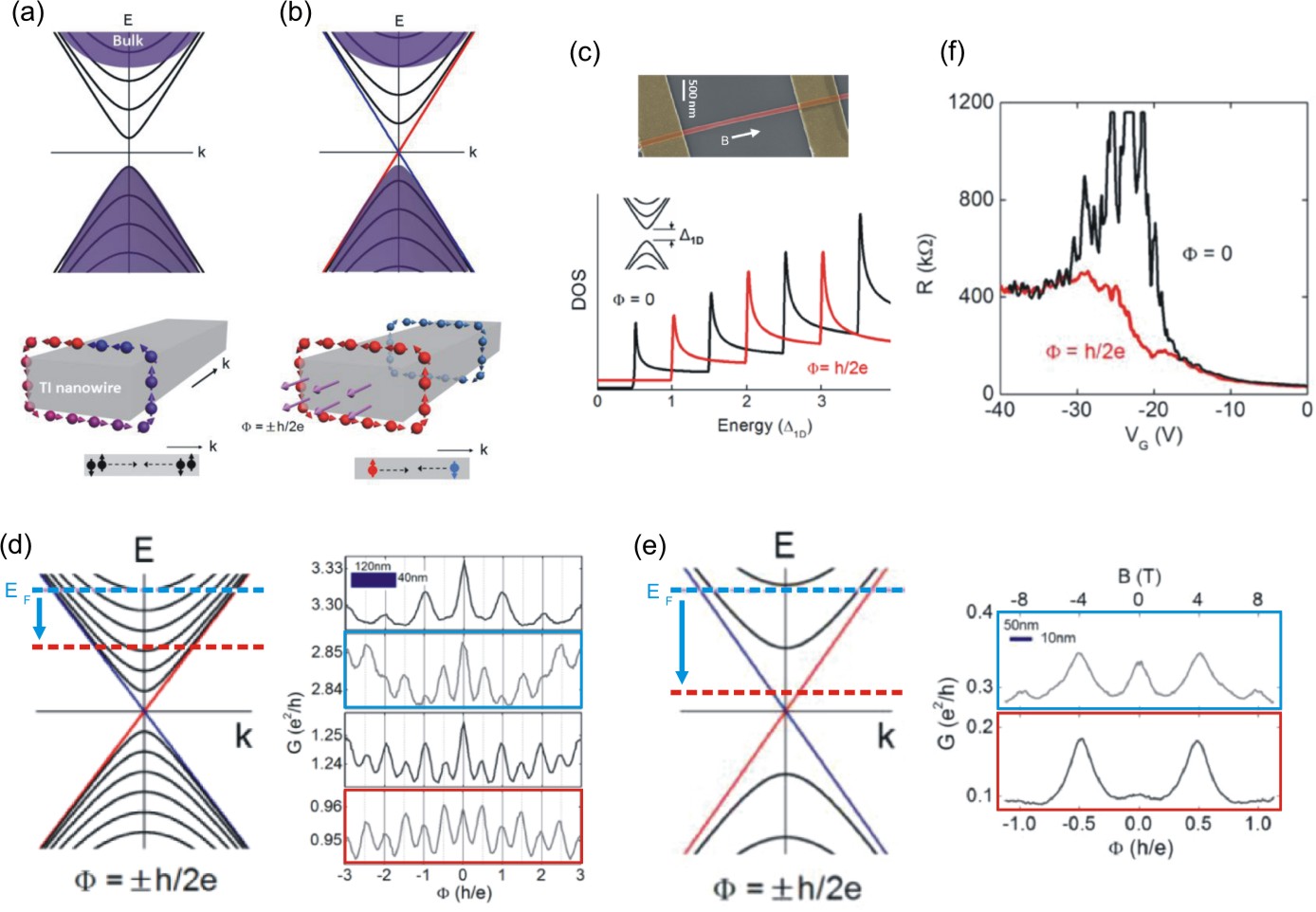}
\caption{(a) Top: Band structure of a topological insulator nanowire in the absence of a magnetic field. The black solid lines represent the discrete 1D subbands with spin degeneracy. Bottom: Cartoon illustrating the additional Berry phase $\pi$ obtained from a 2 $\pi$ rotation of electrons (red to blue color transition). (b) Top: Band structure of a topological insulator nanowire in a magnetic field corresponding to $\Phi$ = $\pm$$h$/2$e$. The gapless bands (red and blue) are not spin degenerate and are referred as 1D helical modes. Bottom: The same as the cartoon in (a) but with $\Phi$ = $\pm$$h$/2$e$, which results in helical 1D channels (blue and red). (c) Top: A false-color scanning electron microscopy image of a Bi$_{2}$Se$_{3}$ core (40 nm height $\times$ 120 nm width) and an amorphous Se shell (2 nm) nanowire. Bottom: Densities of states (DOSs) of the surface electron 1D modes, as a function of the 1D band gap ($\Delta_{1D}$), for $\Phi$ = 0 and $\Phi$ = $h$/2$e$. (d) Left: Schematic band structure of the 1D modes in a TI nanowire with a larger perimeter ($L$ = 300 nm) for a flux value of $\Phi$ = $\pm$$h$/2$e$. The location of the Fermi level ($E_{F}$) is tuned $via$ gating. Right: Magnetoconductance oscillations of the device shown in (c) at four different gate voltages (from top to bottom $V_{G}$= -40 V, -45 V, -85 V, and -95 V). The color of each rectangle corresponds to that of the $E_{F}$ in the left panel. (e) Left: Schematic band structure of the 1D modes in TI nanowires with a smaller perimeter ($L$ = 100 nm) for a flux value of $\Phi$ = $\pm$$h$/2$e$. Right: Magnetoconductance of another nanowire device (10 nm height $\times$ 50 nm width) at $V_{G}$ = -25 V (top) and $V_{G}$ = -45 V (bottom). (f) Resistance ($R$) vs gate voltage ($V_{G}$) graph for a nanowire (10 nm height $\times$ 40 nm width), at $\Phi$ = 0 (black) and $\Phi$ = $h$/2$e$ (red). Adapted with permission from ref. \cite{Hong2014}. Copyright 2014 American Chemical Society.}  
\label{Fig15}
\end{figure*}

TI nanowires are ideal for studying the quantum interferences of topological surface states because of their well-defined nanoscale morphology. Aharonov-Bohm (AB) oscillations are particularly suitable for probing the surface states since a quantum phase is introduced when surface electrons complete a closed loop along the perimeter of a TI nanowire. However, the AB oscillations in TI nanowires in fact suggest a different type of $B$-filed modulation of the electronic band structure, as will be discussed below. When the surface electrons' mean free path exceeds the perimeter of the TI nanowire, the conical band dispersion of 2D Dirac fermions transforms into discrete 1D subbands [see Fig. \ref{Fig15}(a)] due to the quantum confined periodic boundary condition along the perimeter direction. This behavior mimics the transition from graphene to a carbon nanotube, except for the fact that the TI surface states are spintextured (see section 1.3). Interestingly, these 1D subbands can be periodically modulated by a magnetic flux $\Phi$, as described by $E (n,k,\Phi)$ = $\pm$$hv_{F}$$\left[k^{2}/4\pi^{2} + (n + 1/2 - \Phi/\Phi_{0})^{2}/L^{2} \right]$$^{1/2}$, where $n$ is the subband index (integer), the half-integer (1/2) term is from the spin Berry phase, $\Phi_{0}$=$h$/$e$, and $L$ and $v_{F}$ are the nanowire perimeter and the Fermi velocity, respectively \cite{Bardarson2010,Hong2014}. This 1D band dispersion gives rise to magnetoconductance oscillations with a periodicity of $h$/$e$, in contrast to the $h/2e$ periodicity of the oscillations in metallic cylinder conductors (known as Altshuler-Aronov-Spivak (AAS) oscillations). This $B$-field dependency originates from the Dirac nature of the TI surface electrons, which adds a Berry phase ($\pi$) when surface electrons complete a 2$\pi$ rotation along the nanowire circumference. At certain values of the magnetic flux ($\Phi$ = $\pm$$h$/2$e$, $\pm$3$h$/2$e$.., etc.), the phase shift induced by the magnetic flux cancels the spin Berry phase ($\pi$), resulting in a single, gapless 1D mode. Thus, all the 1D subbands of the surface electrons are gapped at zero magnetic field but can be restored to a gapless 1D state when the magnetic flux is $\Phi$ = $\pm$$h$/2$e$, $\pm$3$h$/2$e$.., etc. [see Fig. \ref{Fig15}(b)]. The resulting 1D subband densities of states (DOSs) for $\Phi$ = 0 (or $nh$/$e$) and $\Phi$ = $h$/2$e$ (or $nh$/$e$ + $h$/2$e$) are shown in the bottom panel of Fig. \ref{Fig15}(c).

Fig. \ref{Fig15}(c) (top panel) shows a false-color scanning electron microscopy image of a heterostructure nanowire used to study AB magnetoconductance oscillations. This nanowire consisted of a Bi$_{2}$Se$_{3}$ core (40 nm height $\times$ 120 nm width) and an amorphous Se shell (2 nm), which were synthesized using VLS growth, as discussed above. A magnetic field was applied along the length of the nanowire to introduce a magnetic flux encircled by electrons moving along the perimeter of the nanowire. AB interference measurements recorded at different applied back-gate voltages are shown in the right panel of Fig. \ref{Fig15}(d), while the associated Fermi levels are indicated in the left panel. Two oscillations with periodicities of $h$/2$e$ and $h$/$e$ are observed, of which the $h$/2$e$ periodicity is from AAS effect and the $h$/$e$ periodicity (AB-like peaks) is from the periodic modulation of the 1D subbands by magnetic flux. Depending on where the Fermi level (determined $via$ gate tuning) intersects with the DOS [bottom panel of Fig. \ref{Fig15}(c)], the AB-like oscillation can exhibit conductance peaks at either $\Phi$ = $nh$/$e$ or $\Phi$ = $nh$/$e$ + $h$/2$e$. By contrast, the AAS peaks that appear at every half value of the quantum flux ($nh/2e$) should be independent of the gate tuning. The AB-like peaks in the first (top) and 3rd panels of Fig. \ref{Fig15}(d) are located at $\Phi$ = 0, $\pm h/e$, $\pm 2h/e$.., whereas those in the 2nd and 4th panels are located at $\Phi$ = $\pm h/2e$, $\pm 3h/2e$.., etc. AAS peaks with a relatively weak amplitude coincide with the AB-like peaks and appear in all panels of Fig. \ref{Fig15}(d) [e.g., see the top panel of Fig. \ref{Fig15}(d); at $\Phi$ = $nh$/$e$, AB-like and AAS peaks both exist, whereas the peaks at $\Phi$ = $nh$/$e$ + $h$/2$e$ can be attributed purely to the AAS effect].

The evidence for quantum oscillations from the 1D subband model is the electrical gate modulation. When the Fermi energy is far away from the Dirac point [see the blue dashed line and rectangle in Fig. \ref{Fig15}(d)], the Fermi level crosses multiple 1D subbands. The change in the DOS induced by the magnetic flux (one subband's DOS) is much smaller than the total DOS (entire subbands' DOS), resulting in a small oscillation amplitude at $\Phi$ = $\pm$$h$/2$e$, $\pm$3$h$/2$e$.., etc. The oscillation amplitude at half values of the quantum flux can be enhanced when the Fermi energy is tuned closer to the Dirac point [see the red dashed line and rectangle in Fig. \ref{Fig15}(d)], where the Fermi level crosses fewer 1D subbands, and the change in the DOS induced by the magnetic flux becomes pronounced at $\Phi$ = $\pm$$h$/2$e$, $\pm$3$h$/2$e$.., etc. This gating effect on the magnetoconductance oscillations is even stronger in a thinner nanowire (10 nm height $\times$ 50 nm width), where the 1D band gap is now sufficiently large and its thinner nature enables the Dirac point to be reached $via$ gating. The magnetoconductance oscillations of this device at two different back-gate voltages are shown in the right panel of Fig. \ref{Fig15}(e), with the the associated Fermi levels shown in the left panel. As can be seen, when the Fermi level is at the Dirac point (see the red dashed line and red rectangle), no 1D electronic state exists at zero magnetic field, resulting in the absence of a peak at $\Phi$ = 0 (the remaining weak signal at $\Phi$ = 0 is due to the AAS contribution). However, when an additional magnetic flux restores the gapped 1D band to the gapless 1D state, an oscillation peak is found at $\Phi$ = $\pm$$h$/2$e$. The different natures of the gapped 1D mode ($\Phi$ = 0) and the gapless 1D mode ($\Phi$ = $\pm$$h$/2$e$) can be clearly demonstrated by a gate voltage sweep, as shown in Fig. \ref{Fig15}(f). In the case of zero magnetic flux ($\Phi$ = 0), the resistance begins to diverge near the Dirac point ($V_{G}$= -22 V) due to the gapped 1D subband. By contrast, in the case of $\Phi$ = $\pm$$h$/2$e$, the resistance converges near the Dirac point, indicating the Fermi level is tuned across the gapless (helical) 1D state. A more detailed study of the gapless state at $\Phi$ = $\pm$$h$/2$e$ has revealed that it is robust against additional impurities but can be easily destroyed under time-reversal symmetry breaking (i.e., under perpendicular magnetic fields) \cite{Hong2014}.

\subsection{6.2. Bi$_{2}$Se$_{3}$ quantum dots}

\begin{figure}[!t]	
\includegraphics[scale=0.47]{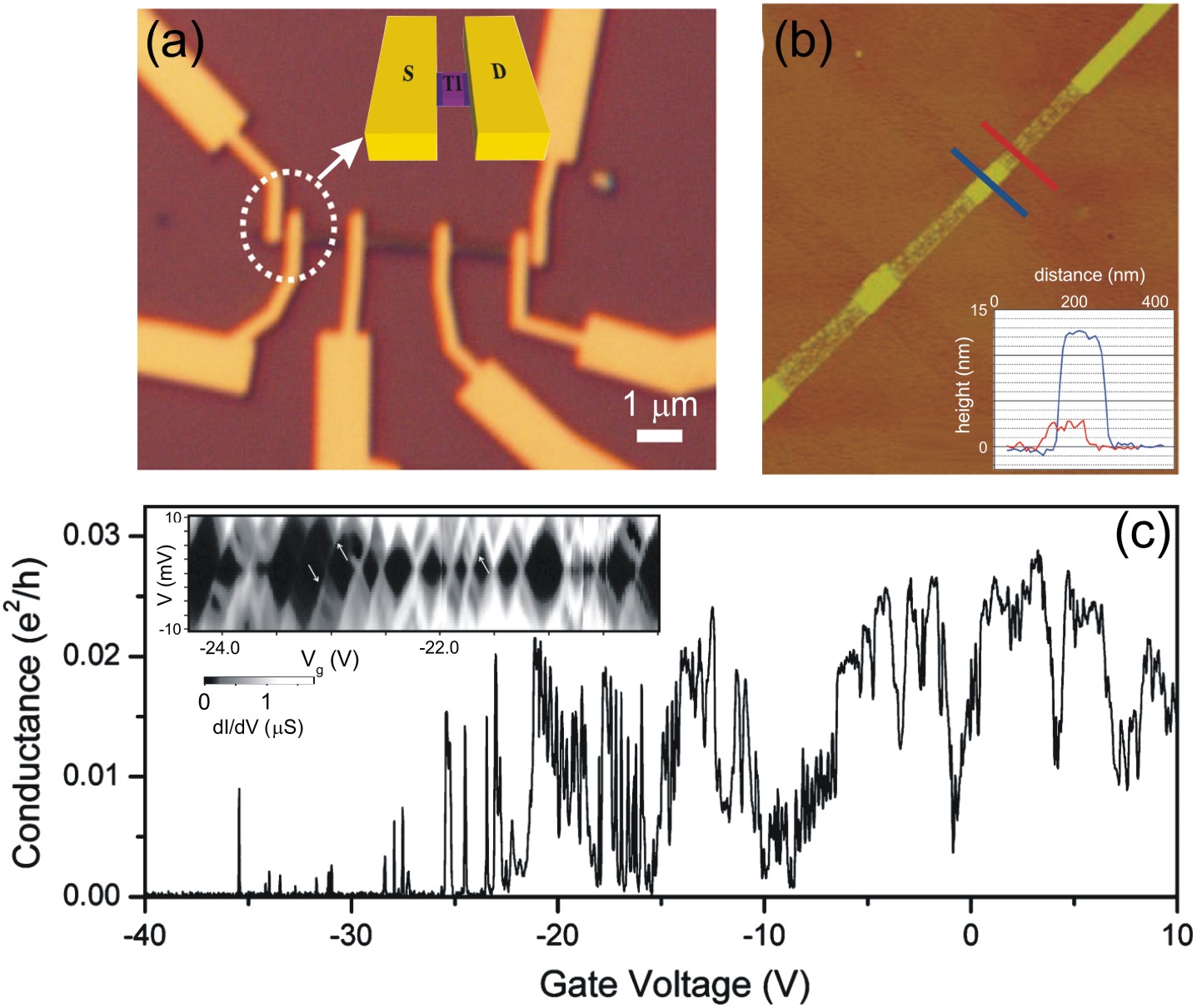}
\caption{(a) Optical micrograph of a 7 nm thick, mechanically exfoliated Bi$_{2}$Se$_{3}$ nanoribbon contacted by Cr/Au (2 nm/28 nm) contacts. Inset: Schematic of a Bi$_{2}$Se$_{3}$ quantum dot forms between two contacts. Dashed circle shows the device studied in ref. \cite{Cho2012}. (b) Atomic force micrograph of a 12 nm thick nanoribbon after etching with PMMA mask. Inset: Line traces of topographic data along blue line (unetched area) and red line (etched area). (c) Gate voltage dependence of differential conductance ($G=dI/dV$) shows a current pinch-off regime with quasi-periodic narrow conductance peaks. Inset: Coulomb diamond measurements on the device. Arrows indicate the excited states of the QD energy levels. Adapted with permission from ref. \cite{Cho2012}. Copyright 2012 American Chemical Society.}  
\label{Fig13}
\end{figure}

A TI quantum dot is an interesting platform in which single electron transport could be spin polarized and can be potentially used as a single-spin generator. Fig. \ref{Fig13}(a) shows an optical micrograph of a Bi$_{2}$Se$_{3}$ quantum dot device. A Bi$_{2}$Se$_{3}$ nanoribbon with a width of 200 nm and a thickness of 7 nm was mechanically exfoliated on Si/SiO$_{2}$ substrate. In order to create quantum confinement, parts of the Bi$_{2}$Se$_{3}$ nanoribbon were etched (by N$_{2}$ plasma) into a thin slab to induce a tunnel coupling between top and bottom surfaces, and thus open a band gap. Fig. \ref{Fig13}(b) shows an AFM image of the nanoribbon after etching, where the etched region with a thickness of 3-5 QLs [inset in Fig. \ref{Fig13}(b)] will be contacted by Cr/Au (2 nm/28 nm) electrodes. The etched Bi$_{2}$Se$_{3}$ thin films act as gate-tunable tunnel barriers, thus a quantum dot can be defined between two adjacent etched regions (i.e., between two electrodes), as shown in the inset of Fig. \ref{Fig13}(a). Fig. \ref{Fig13}(c) shows the differential conductance $G=dI/dV$ as a function of back-gate voltage $V_{g}$ at zero source-drain bias $V_{b}$ = 0 for the device shown in Fig. \ref{Fig13}(a). The current pinch-off region at large negative $V_{g}$ indicates the n-doped Bi$_{2}$Se$_{3}$ thin films are tuned at the band-gap, in which the transport is suppressed for $V_{g}\leq$-24 V (transport gap regime). Within the transport gap, quasi-periodic narrow conductance peaks were observed, indicating the device is operating in the Coulomb blockade regime. This could be further confirmed by the Coulomb diamond measurements, where $G=dI/dV$ as a function of both $V_{g}$ and $V_{b}$ was shown in the inset of Fig. \ref{Fig13}(c). The diamonds with charging energies $E_{C}$ ranging from 5-10 meV suggest a variation in dot size, possibly due to the inevitable gate modulation of tunnel barriers therefore alter the size of the dot. The charging energy of the smallest diamond (5 meV) corresponds to an effective dot radius $r=$ 113 nm if the disc plate capacitance model $E_{C}$=$e^{2}/8\epsilon\epsilon_{0}r$ is used, in agreement with the dimension of the dot A= $L\times W$= (200 $\times$ 200) nm$^{2}$. The excited states were resolved at the edges of the diamond at around 1 meV, as indicated by the arrows in the inset of Fig. \ref{Fig13}(c). Although the spin-texture of surface states may enable TI quantum dots to be used as single-spin generators, non spin-polarized bulk contribution cannot be ruled out. Thus, future experiments are needed to detect spin polarization of charges through TI QDs for further applications.

\maketitle
\section{7. Summary and perspective}
In conclusion, we have provided the theoretical background for and a historical review of a number of electron transport experiments performed on nanostructures fabricated from graphene, transition metal dichalcogenides and topological insulators. The transport properties of graphene nanodevices (GNRs, GSQDs and GDQDs) fabricated on SiO$_{2}$ and hBN substrates were both reviewed. GNRs fabricated on SiO$_{2}$ and hBN both show the presence of strongly localized states, but in the case of the former, they predominantly originate from substrate disorder, whereas in the latter case, they can be attributed to the edge roughness if the edge-to-bulk ratio is large. GSQDs fabricated on SiO$_{2}$ and hBN show a distinct difference in their Coulomb blockade peak-spacing fluctuations, indicating that edge roughness is the dominant source of disorder for QDs with diameters of less than 100 nm. We have also described the power of pulsed gating for probing electron relaxation times in a GSQD and generating frequency-dependent pumped currents in a GDQD. For TMD nanostructures, despite the rather limited number of experimental studies on transport in such devices, we discussed the Fock-Darwin spectra of TMD QDs for future reference. Finally, we reviewed a few topological insulator nanodevices and discussed their exotic topological surface states. The discrete 1D subbands of a Bi$_{2}$Se$_{3}$ nanowire can be periodically modulated by a magnetic field applied along its length, resulting in a $h/e$ periodicity in its magnetooscillations whose amplitude is tunable by gate voltage. We also reviewed a Bi$_{2}$Se$_{3}$ single quantum dot, in which the single-electron tunneling regime can be manifested.

The absence of spin blockade in GDQDs \cite{Moriyama2009,Molitor2010,Liu2010,Volk2011,Wang2012,Wei2013,Chiu2015} and the fact that the measured spin relaxation times in 2D graphene flakes are shorter than expected \cite{Tombros2007,Han2010a,Han2011,Maassen2012,Drogeler2014} suggest that there are extrinsic effects that govern the spin relaxation dynamics in graphene. These effects maybe related to the scattering of electrons off magnetic impurities originating from carbon atom vacancies or graphene edge roughness. Since graphene constrictions have been used as tunnel barriers in most GQDs reported thus far, alteration of the electron spin by the constriction edge during transport is inevitable. One possible solution to this problem is to use an electrical-field-induced bandgap in bilayer graphene to define GQDs \cite{Goossens2012,Allen2012}. However, the small induced energy gap ($\approx$ 200 meV \cite{Zhang2009b}) may limit the available energy range for quantum dot operation. In this sense, the large bandgaps of few-layer TMDs can be advantageous for defining QDs $via$ electrical gating. The spin- and valley-polarized states in TMD QDs also make them an interesting candidate for quantum information processing. Nevertheless, gate-defined DQDs have not been reported on either material thus far, nor have spin blockade or spin relaxation time measurements. Another approach of reducing the edge effect of GNRs is to use functionalized graphene, such as fluorinated graphene (FG), to simultaneously define the quantum confinement and passivate the graphene edges. Nano-patterning of fluorinated graphene has been achieved by employing scanning probe lithography and electron beam irradiation \cite{Withers2011,Lee2011}; in each case, graphene nanochannels are surrounded by insulating FG to form constrictions. However, to date, no functionalization-defined GQDs have been reported experimentally.

Vertical tunneling to GQDs using hBN as a tunnel barriers may also serve as a promising solution for minimizing the edge effect from graphene constrictions \cite{Ashoori1993}. Vertical GQD tunneling devices offer several advantages over their lateral counterparts. Tunneling can continue to occur even if the graphene bulk is in an insulating state (for example, at high B-fields where Landau levels are formed), and is expected to provide better resolution in Fock-Darwin spectrum since the B-field modulation on the GNR tunnel barriers is absent \cite{Chiu2012}. Recently, a vertical stack of a GNR and a 2D graphene sheet (separated by a 13-nm-thick hBN layer) has been reported, in which the Coulomb resonances of the GNR was used to probe the local density of states of the graphene sheet \cite{Bischoff2015}. In addition, a stack of two GNRs vertically separated by a 12-nm-thick hBN layer has shown the characteristic features of a capacitively coupled double-dot structure in the charge stability diagram \cite{Bischoff2015a}. However, experimental studies on vertical tunneling to graphene nanostructures through thin hBN as a tunnel barrier are still lacking. Further efforts are encouraged to build such a system and investigate the spin relaxation dynamics in it.  

The spin-momentum locking nature of TI surface states offers the possibility of using TI QDs as single spin generators. To date, no transport studies have been reported in which the spin polarization in a TI QD has been detected. More sophisticated designs for the TI quantum dots are needed to study the controllability of the confined charges and their spin-related transport properties.

\maketitle
\section{6. Acknowledgments}
The authors would like to acknowledge Malcolm Connolly, Alessandro Cresti, Edbert Sie, Valla Fatemi, Seung Sae
Hong, Yi Cui and Andor Kormanyos for helping with preparation of the manuscript and proof read.

\end{document}